\documentclass[prd,twocolumn,preprintnumbers,superscriptaddress,showpacs,floatfix,amsmath,amssymb,showkeys,nofootinbib]{revtex4}
\usepackage{epsfig}
\usepackage{hyperref}
\usepackage{amssymb}
\usepackage{dsfont}

\newcommand{\bea}{\begin{eqnarray}}
\newcommand{\eea}{\end{eqnarray}}
\newcommand{\bite}{\begin{itemize}}
\newcommand{\eite}{\end{itemize}}

\begin{document}

\preprint{DESY 12-065, Edinburgh 2012/04, HU-EP-12/14, Liverpool LTH 919}

\title{
Wilson loops to 20th order numerical stochastic perturbation theory
}

\author{R.~Horsley}
\affiliation{School of Physics, University of Edinburgh, Edinburgh EH9 3JZ, UK}

\author{G.~Hotzel \footnote{Present address: Institut f\"ur Physik, Humboldt-Universit\"at zu Berlin, D-12489 Berlin, Germany}}
\affiliation{Institut f\"ur Theoretische Physik, Universit\"at Leipzig,  D-04109 Leipzig, Germany}

\author{E.-M.~Ilgenfritz}
  \affiliation{Institut f\"ur Physik, Humboldt-Universit\"at zu Berlin, D-12489 Berlin, Germany}
  \affiliation{Joint Institute for Nuclear Research, VBLHEP, 141980 Dubna, Russia}

\author{R.~Millo}
\affiliation{Theoretical Physics Division, Department of Mathematical Sciences,
             University of Liverpool,  Liverpool L69 3BX, UK}

\author{H.~Perlt}
\affiliation{Institut f\"ur Theoretische Physik, Universit\"at Leipzig,  D-04109 Leipzig, Germany}

\author{P.~E.~L.~Rakow}
\affiliation{Theoretical Physics Division, Department of Mathematical Sciences,
             University of Liverpool,  Liverpool L69 3BX, UK}

\author{Y.~Nakamura}
\affiliation{RIKEN Advanced Institute for Computational Science, Kobe, Hyogo 650-0047, Japan}

\author{G.~Schierholz}
\affiliation{Deutsches Elektronen-Synchrotron DESY, D-22603 Hamburg, Germany}

\author{A.~Schiller}
\affiliation{Institut f\"ur Theoretische Physik, Universit\"at Leipzig,  D-04109 Leipzig, Germany}

\begin{abstract}
We calculate perturbative contributions of Wilson loops of various sizes up to order 20 in
$SU(3)$ pure lattice gauge theory at different lattice sizes
for Wilson gauge action
using the technique of numerical stochastic perturbation theory.
This allows us to investigate 
the perturbative series for various Wilson loops at high orders of perturbation theory.
We observe differences in the behavior of those series
as function of the loop order $n$.
Up to $n=20$ we do not find evidence for the factorial growth of the expansion
coefficients often assumed to characterize an asymptotic series. 
Based on the actually observed behavior we sum the series in a model 
parametrized by hypergeometric functions.
For Wilson loops of moderate sizes the summed series in boosted perturbation theory 
reach stable plateaus already for moderate orders in perturbation theory.
The coefficients in the boosted series become much more stable in the result of smoothing 
the coefficients of the original series effected by the hypergeometric model.
We introduce generalized ratios of Wilson loops of different sizes.
Together with the corresponding Wilson loops from 
standard Monte Carlo measurements
they enable us to assess their non-perturbative parts.
\end{abstract}

\keywords{Lattice gauge theory, stochastic perturbation theory}

\pacs{11.15.Ha, 12.38.Gc, 12.38.Cy,12.38.Aw}

\maketitle

\section{Introduction}

Since the non-perturbative gluon condensate 
has been introduced by Shifman, Vainshtein and Zakharov~\cite{Shifman:1978bx} 
there have been many attempts to obtain reliable numerical 
values for this quantity. 
It has become clear very soon that lattice gauge theory provides a promising tool to
calculate the gluon condensate from first principles using Wilson loops $W_{NM}$ of 
various sizes $N \!\times\! M$. 
The perturbative expansion of the Wilson loop -- which does not depend on an external scale -- 
is especially simple since it cannot depend on logarithms.
In ~\cite{Banks:1981zf,Di Giacomo:1981wt} the plaquette was used whereas larger Wilson loops 
have been investigated in~\cite{Kripfganz:1981ri,Ilgenfritz:1982yx}. In all cases it turned 
out, that the knowledge -- as precisely as possible -- of the  
large order perturbative tail of the Wilson loops is crucial.
In the last decade, the application of numerical stochastic perturbation theory (NSPT)
\cite{Alfieri:2000ce} has pushed the perturbative order of the plaquette up to order $n=10$
\cite{DiRenzo:2000ua} and even $n=16$~\cite{Rakow:2005yn}.

Apart from the desired evaluation of the gluon condensate, there is a general interest in the 
behavior of perturbative series in QCD (for an investigation see~\cite{Meurice:2006cr}). 
In perturbation theory observables can be written as series of the form
\begin{equation}
  \mathcal{O} \sim\, \sum_{n} c_n \lambda^n\,,
\end{equation}
where $\lambda$ denotes some generic coupling, e.g. $\alpha_s$. 
It is generally believed that these series are asymptotic ones, 
and it is often assumed that for large $n$ the leading growth of the coefficients $c_n$
can be parametrized as~\cite{ZinnJustin}
\begin{equation}
  c_n \sim \,C_1\,(C_2)^n\,\Gamma (n+C_3)
  \label{eq:coefffactorial}
\end{equation}
with some constants $C_1,C_2,C_3$, i.e., they show a factorial behavior. 

Using the technique of NSPT one can reach loop
orders of perturbation theory where a possible set-in of this assumed
behavior becomes testable. In~\cite{Narison:2009ag}
Narison and Zakharov discussed the difference between short and long 
perturbative series and 
its impact on the determination of the gluon condensate.

In this paper we present perturbative calculations of Wilson loops in NSPT
for the Wilson gauge action (with $\beta={6}/{g^2}$)
\begin{equation}
  S_W[U]=\beta \sum_P \left[ 1 - \frac{1}{6} \mathrm{Tr} \left( U_P + U_P^{\dagger} \right) \right]
  \label{eq:WGAction}
\end{equation}
up to order $n=20$ for lattice sizes $L^4$ with $L=4,6,8,12$.
The computation for $L=12$ were performed on a NEC SX-9 computer of RCNP 
at Osaka University, all others on Linux/HP clusters at Leipzig University.

The paper is organized as follows. In Section \ref{sec:NSPT} we explain 
how the loop order expansion of Wilson loops has been obtained in NSPT. 
In Section \ref{sec:ptseries} we discuss a model which allows us to sum up 
completely the obtained Wilson loops series on finite lattices.
As an alternative we apply boosted perturbation theory consisting 
in a rearrangement of the series such that already for a summation up to 
relatively low loop number good convergence of the
summed series can be achieved. These results are used to estimate
the gluon condensate in Section \ref{sec:GG}. Finally we draw our conclusions.

Some preliminary results have been presented in recent lattice proceedings~\cite{Ilgenfritz:2009ck,Horsley:2010af}.
In the present work we give the computational details of the Langevin calculation for the final statistics reached
and significantly extend the analysis part using boosting and series summation as well as 
adding new aspects to the analysis of Wilson loops of moderate size.

\section{NSPT and Wilson loops  up to 20 loops}
\label{sec:NSPT}

\subsection{The strategy of NSPT}
\label{subsec:strategy}

Numerical stochastic perturbation theory -- based on stochastic quantization~\cite{Parisi:1980ys} -- allows 
perturbative calculations on finite lattices up to finite but high loop order $n$,
unrivalled by the standard diagrammatic approach in lattice perturbation theory.
Practical limits are set only by computer time, storage limitations and machine precision.
For instance, in order to calculate in the $n$-loop order 
in the simplest realization of NSPT in the Euler scheme, one has to keep simultaneously 
links corresponding to roughly $2n$ gauge field configurations for a given lattice size.
If one wants to keep for practical reasons also the gauge fields (vector potentials) besides the 
gauge field links 
themselves, the storage requirement is even doubled. In addition, the computer time of the Langevin simulation scales quite severely,
we found it roughly goes like $n^3$.

The algorithm of NSPT has been introduced and discussed in detail in~\cite{Alfieri:2000ce, Di Renzo:2004ge}.
For convenience, we will here repeat the main points for pure $SU(3)$ lattice gauge theory.
The stochastic  evolution of the gauge field links $U_{x,\mu}$, located at the link between lattice sites 
$x$ and $x+\hat\mu$, occurs in an additional ``Langevin time'' $\tau$. This process is described
by the Langevin equation
\begin{equation}
  \frac{\partial}{\partial \tau} U_{x,\mu}(\tau;\eta) = {\rm i}\; \bigl\{
  \nabla_{x,\mu} S_W[U]- \eta_{x,\mu}(\tau) \bigr\} \;
  U_{x,\mu}(\tau;\eta) \,.
  \label{eq:Langevin_I}
\end{equation}
The so-called drift term is given by the 
variation of the Euclidean gauge action $S_W[U]$: it is written in terms of the left Lie derivative
$\nabla_{x,\mu}$ which keeps the links in the $SU(3)$ group manifold. 
The process is made stochastic by additive white noise $\eta_{x,\mu}(\tau)$. 
In the limit of large $\tau$ the distribution of subsequent,
simultaneous gauge link fields converges to the Gibbs measure 
$P[U] \propto {\rm{exp}}(- S_W[U])$.

As in any numerical approach one needs to discretize the ``Langevin time'' as a sequence 
$\tau \to k \epsilon$, with running step number $k$. It is known that, in order to extract 
correct equilibrium physics, one needs to perform the double 
extrapolation $k \to \infty$ and $\epsilon \to 0$, the latter in order not to violate detailed balance.
For the numerical solution of the Langevin 
equation we adhere to a particular version of the Euler scheme that guarantees all the link 
matrices $U_{x,\mu} \in SU(3)$ to stay in the group manifold:
\begin{equation}
  U_{x,\mu}(k+1; \eta)  =  \exp \bigl( {\rm i}~F_{x,\mu}[U,\eta] \bigr) \; U_{x,\mu}(k; \eta)
  \label{eq:iteration}
\end{equation}
with the force term for the update of the gauge links $U_{x,\mu}(k; \eta)$
in the form
\begin{eqnarray}
  F_{x,\mu}[U, \eta]  =  \epsilon~\!\nabla_{x,\mu} S_W[U] + \sqrt{\epsilon}~\eta_{x,\mu}\,,
  \label{eq:force}
\end{eqnarray} 
$\eta $  being a traceless $3 \!\times\! 3$ noise matrix.
In case of the Wilson gauge action that force term 
takes the form
\begin{eqnarray}
   F_{x,\mu} &=&  
   \frac{ \beta \epsilon}{12} \sum_{U_P\supset U_{x,\mu}} \left[ \left(U_P -U_P^{\dagger} \right) \right.
   \nonumber
  \\
    && \left.
   - \frac{1}{3} \mathrm{Tr} \left(U_P -U_P^{\dagger} \right) \, \mathds{1}\right] 
   +  \sqrt{\epsilon}~\eta_{x,\mu}   \,.
  \label{eq:forcewilson}
\end{eqnarray}

We expand each link matrix at any time step  
in the bare coupling constant $g$ around the trivial vacuum $U_{x,\mu}=\mathds{1}$.
Since $\beta=6/g^2$, the expansion reads
\begin{equation}
  U_{x,\mu}(k; \eta) \to  \mathds{1} + \sum_{m \ge 1} \beta^{-m/2} U_{x,\mu}^{(m)}(k; \eta) 
  \,.
  \label{eq:expansion_of_U}
\end{equation} 
If one rescales the time step to $\varepsilon = \beta \epsilon$, the expansion 
(\ref{eq:expansion_of_U}) converts the Langevin equation (\ref{eq:iteration}) into a system 
of simultaneous updates in terms of the expansion coefficients 
of $U_{x,\mu}^{(m)}(k; \eta)$ and of similar expansion coefficients for the force 
$F_{x,\mu}$ in (\ref{eq:forcewilson}), but free of adjustable constants.
While the random noise $\eta$ enters only the lowest order equation, higher orders 
are rendered stochastic by the noise propagating from lower 
to higher order 
terms. The system is usually truncated according to the maximal order of the perturbative 
gauge link fields one is interested in.

For NSPT it is indispensable to perform stochastic gauge fixing by using a variant of
gauge transformations
\begin{equation}
  U_{x,\mu}^G = G_x \, U_{x,\mu} \, G^\dagger_{x + \hat \mu}  
  \label{eq:gtrafo}
\end{equation}
with $G_x$ derived from the Landau gauge and
expanded in powers of $1/\sqrt\beta \sim g$.
A convenient solution for the gauge transformation $G$ comes with the choice 
\begin{equation}
  G_x= \exp \left\{ - \alpha \, 
  \sum_\nu \left(\frac{ A_{x+\hat\nu/2,\nu}-A_{x-\hat\nu /2,\nu}}{a} \right)\,
    \right\} \,, 
  \label{eq:localstep}
\end{equation}
where the series variant of the expression has to be taken.
Here the (antihermitean) vector potential $A_{x+\hat{\mu}/2,\mu}$ is related to the link 
matrices $U_{x,\mu}$ via
\begin{equation}
  A_{x+\hat{\mu}/2,\mu}  =  \log U_{x,\mu} \,,
  \label{eq:log_mapping}
\end{equation}
and an expansion similar to (\ref{eq:expansion_of_U}) 
taking values in the algebra $su(3)$ is applied for the potential.

The need for stochastic gauge fixing comes from the fact that the diffusion of the 
longitudinal component of the $A_{\mu}$ fields is unbounded and hence their norms 
would diverge in the course of the stochastic process.
Although gauge-invariant quantities are in principle not affected by these divergences, 
the performance eventually runs into trouble due to loss of accuracy. 
It turns out that one step of (\ref{eq:gtrafo}) using (\ref{eq:localstep}) alternating with
the Langevin step (\ref{eq:iteration}) is sufficient 
to keep fluctuations under control, if
$\alpha$ is chosen of order $\alpha \sim \varepsilon$.

The influence of zero modes of the gluon field on the performance of 
the Langevin process has been critically discussed in~\cite{Di Renzo:2004ge}.
Since zero modes (constant modes) of the gauge fields do not
contribute to the discretized divergence present in (\ref{eq:localstep}),
they would not be subtracted by performing the gauge transformation.
We take the simplest prescription of subtracting zero modes at every order by hand.
This completes the specification how NSPT is used in our calculations.

Let us remark that, whenever we speak about contributions of some order 
to an observable
constructed out of links,
this has to be understood in the sense of
an expansion
\begin{equation}
  \langle {\cal O} \rangle \to \sum_{m \ge 0} \beta^{-m/2} \langle {\cal O}^{(m)} \rangle \ , 
  \label{eq:expansion_of_O}
\end{equation}
and the expansion coefficient $\langle {\cal O}^{(m)} \rangle$ are extracted out of
the expanded r.h.s.
of (\ref{eq:expansion_of_U})
by comparing coefficients of equal powers $\beta^{-m/2}$ (or $g^m$).
In the notation of (\ref{eq:expansion_of_O}) even integers $m$ correspond to genuine loop 
contributions (with loop order $m/2$). In the computer implementation of NSPT we practically measure 
observables for various small but finite values of $\varepsilon$. 
The final result is then obtained by performing the extrapolation to $\varepsilon \rightarrow 0$
for the observables in each loop order.

\subsection{NSPT results for Wilson loops in high order perturbation theory}
\label{subsec:nsptres}

In lattice gauge theory the Wilson loop as a gauge invariant quantity built
only out of gauge field links is defined as the trace of a
product of link fields along a closed path $C$
\begin{equation}
  W_C[U]= \frac{1}{3} \, \mathrm{Tr}\prod\limits_{(x,\mu) \in C} U_{x,\mu} \,.
  \label{eq:wloopdef}
\end{equation}
Having at our disposal the expansion of the links (at finite Langevin step size) 
close to the trivial vacuum $U_{x,\mu}^{(0)}\equiv \mathds{1}$
to all orders in $g \propto 1/\sqrt{\beta}$\footnote{From now we use as expansion parameter 
the gauge coupling $g$, using the same notation for the coefficients $U_{x,\mu}^{(m)}$.}
\begin{equation}
  U_{x,\mu} \equiv \sum_{m \ge 0}  U_{x,\mu}^{(m)} \, g^m \, ,  
  \label{eq:expansion_of_UU}
\end{equation}
we construct perturbative Wilson loops within a 
given ``Langevin configuration'' (at fixed ``Langevin time'').
Inserting the expansion~(\ref{eq:expansion_of_UU}) for the links in~(\ref{eq:wloopdef}) 
we collect terms of equal power in $g$ on 
the right hand side and identify these with the $n$-th loop order contribution $W_{C}^{(n)}$ on the
left hand side
\bea
  && \sum\limits_{n=0,1/2,1,3/2,\dots} W_{C}^{(n)}\,  g^{2n} =
  \nonumber \\
  &&
  \frac{1}{3} \,
  {\rm Tr} \prod\limits_{(x,\mu) \in C} \Bigl[ \sum\limits_{m_{x,\mu}\ge0} U_{x,\mu}^{(m_{x,\mu})} \, g^{m_{x,\mu}} \Bigr]  \, . 
  \label{eq:Wloop}
\eea
The final result involves averaging over different configurations obtained during the 
Langevin evolution and the extrapolation to $\varepsilon \to 0$.

Here we consider rectangular Wilson loops $C$ of size $N \!\times\! M$, where we restrict the 
maximal side length of the Wilson loop 
to half of the lattice size $L/2$ for a lattice $L^4$.
Therefore, we identify the general perturbative loop order expansion of the Wilson loop $W_{NM}$ in terms of the
bare lattice coupling $g$ as
\begin{equation}
  W_{NM} = \sum_{n=0,1/2,1,3/2,\dots} \,W_{NM}^{(n)}\, g^{2n}
  \label{eq:WNMPTSeries}
\end{equation}
with the Wilson loop expansion coefficients  $W_{NM}^{(n)}$  ($W_{NM}^{(0)} \equiv 1$).
The integer powers $n=1,2,\dots$ in the series~(\ref{eq:WNMPTSeries}) denote the perturbative loop orders 
as in diagrammatic perturbation theory.

In addition, following (\ref{eq:Wloop}) we measure analogues of the loop coefficients $W_{NM}^{(n)}$ 
also for half-integer $n=3/2,5/2,\dots$ (Due to the color trace the coefficient with $n=1/2$ is identically equal to zero).
Averages over coefficients with those half-integers -- which describe {\it non-loop} contributions --
should vanish numerically 
after averaging over a sufficient number of measurements and 
define some level of ``noise'' for finite statistics to be compared to the {\it loop} contributions.
While higher loop order contributions decrease fast with the loop number,
the ``noise'' does not decrease sufficiently fast, staying near zero.  
Therefore, we adopt here 
the criterion that we can take the expansion coefficients 
for a given loop order $n$ for granted (``reliable'') only 
if they can be clearly distinguished numerically from the noisy results 
for adjacent non-loop contributions of orders $n-1/2$ and $n+1/2$.
We do not rule out the possibility of an extrapolation to zero Langevin step size crossing
in a systematic way the noise region near zero from a positive/negative coefficient at large $\varepsilon$
to a negative/positive coefficient at smallest $\varepsilon$.
The coefficient extrapolated to $\varepsilon=0$  might be as small as the noise of the adjacent 
non-loop contributions.

Let us add some details of the perturbative Langevin simulation:
Instead of having one link configuration as in usual Monte Carlo studies, we have to handle 40 link configurations 
building our ``perturbative'' configuration for each $g$ order
to reach loop order 20 at each Langevin step.
So, unavoidably, the different orders in $g$ are correlated, since we have to use 
a correlated system of Langevin equations for each order.

Any simulation for a chosen Langevin step size $\varepsilon$ starts
from a link configuration, where the zeroth order in $g$ of the
expanded links is put equal to one (and remains so during all the
evolution), whereas all non-zero orders in $g$ are set initially to zero (a ``cold'' start).
So any loop contribution is by construction vanishing at the beginning.
Starting from here with the Langevin process including the noise term, the non-zero $g$ orders of the links 
iteratively obtain non-zero values at each link position starting from the lowest order in $g$.
Therefore, the highest order in $g$ needs the highest minimal number of Langevin steps to reach equilibrium.
With decreasing step size $\varepsilon$ that minimal number also increases.

As a criterion to reach the equilibrium of the Langevin process, we studied the behavior of the perturbative plaquette.
By monitoring the highest order of the plaquette at the lowest chosen step size $\varepsilon=0.01$, we observed that 
equilibrium is reached after roughly 2000 Langevin steps.
To be on the safe side we have discarded the first 5000 Langevin steps
after a ``cold'' start before we began measurements of the perturbative
Wilson loops. To increase statistics, we also created new ``parallel''
Langevin trajectories (keeping the same parameter $\varepsilon$) starting
from a configuration already in equilibrium (given in replicas representing
all orders in $g$) after changing the seeds for the white noise.
Only in these cases the strategy of averaging over independent realizations
of noise has been followed. Otherwise, subsequent sequences of noise are
considered as independent.

We have observed that the autocorrelations increase on one side with increasing loop order 
and on the other side with increasing Wilson loop size.
The perturbative Wilson loops have been measured after each $20^{th}$ Langevin step to reduce autocorrelations.
The integrated autocorrelation times are included in the error estimate of the measured quantities.
Typically for the $1\!\times\!1$ Wilson loop the estimated autocorrelation was $O(1)$ at the lowest loop-orders 
and increased up to $O(10)$ at the highest loop orders.
So the relative errors significantly increase with the loop order.
As a result, we have collected the following statistics in measuring the perturbative Wilson loops 
for the different chosen finite Langevin steps sizes and lattice volumes as shown in 
Table~\ref{tab:statistics1}.
\begin{table}[!htb]
  \begin{center}
    \begin{tabular}{|c|c|c|c|c|}
     \hline
  $\varepsilon$ & $L=4$ & $L=6$ & $L=8$  &$L=12$  \\
  \hline
    0.010       & 19522 & 16390 & 21000 & 5672  \\
    0.015       & 12182 & 13366 & 18500 & ---   \\
    0.020       & 11186 & 12726 & 18750 & 5464  \\
    0.030       & 10120 & 10210 & 17500 & 5334  \\
    0.040       &  9620 &  9466 & 17500 & 5200  \\
    0.050       &  9500 &  8500 & 16500 & ---   \\
    0.070       &  9500 &  8500 & 16250 & 5476  \\
  \hline
     \end{tabular}
  \end{center}
  \caption{Number of Wilson loops measurements up to loop order 20 at various lattice volumes $L^4$ 
           and Langevin time steps $\varepsilon$.}
  \label{tab:statistics1}
\end{table}
The statistics has to be understood as follows: The thermalization is not included, 
e.g. 21000 measurements at lattice volume $L^4$ with $L=8$ and $\varepsilon=0.01$ in the Table 
corresponds to 420000 Langevin steps in equilibrium. Those measurements are performed for all orders in $g$, 
the reached results are shown in the Figures below.

Let us first discuss the accuracy and some problems 
we have met in performing the extrapolation to vanishing
Langevin step size $\varepsilon$.
Having several different 
expansion coefficients $W_{NM}^{(n)}(\varepsilon)$ for various 
$\varepsilon$ values 
at a given loop order $n$ available, 
we perform the extrapolation to the
coefficient $W_{NM}^{(n)}$ corresponding to
zero step size by a linear plus quadratic fit ansatz
\begin{equation}
   W_{NM}^{(n)}(\varepsilon) = W_{NM}^{(n)} + A_{NM}^{(n)} \, \varepsilon + B_{NM}^{(n)} \, 
   \varepsilon^2 \,.
  \label{eq:fitWNMeps}
\end{equation}

The $\varepsilon$ behavior depends on the loop order $n$ and the 
Wilson loop size $N \!\times\! M$, as well as on the lattice volume. 
To illustrate the overall behavior we present here results for the 
plaquette $W_{11}$ and the Wilson loop $W_{33}$ for lattice size $L=12$.

The measured perturbative plaquette values $W_{11}^{(n)}$ at all 
integer loop orders $n>0$ (remember that $W_{NM}^{(0)}\equiv 1$) 
behave in a similar way: they are all negative and tend to values different from zero
which can be determined with very good accuracy. 
Except for $n=2$ the zero Langevin step size limit is approached from below with decreasing 
step size $\varepsilon$. 
The clearly non-vanishing fit results decrease monotonically in magnitude with increasing loop order.
This is demonstrated in the left of Figure~\ref{fig:orderW11}, see also the Tables in the Appendix.
\begin{figure*}[!htb]
  \begin{tabular}{cc}
     \includegraphics[scale=0.63,clip=true]{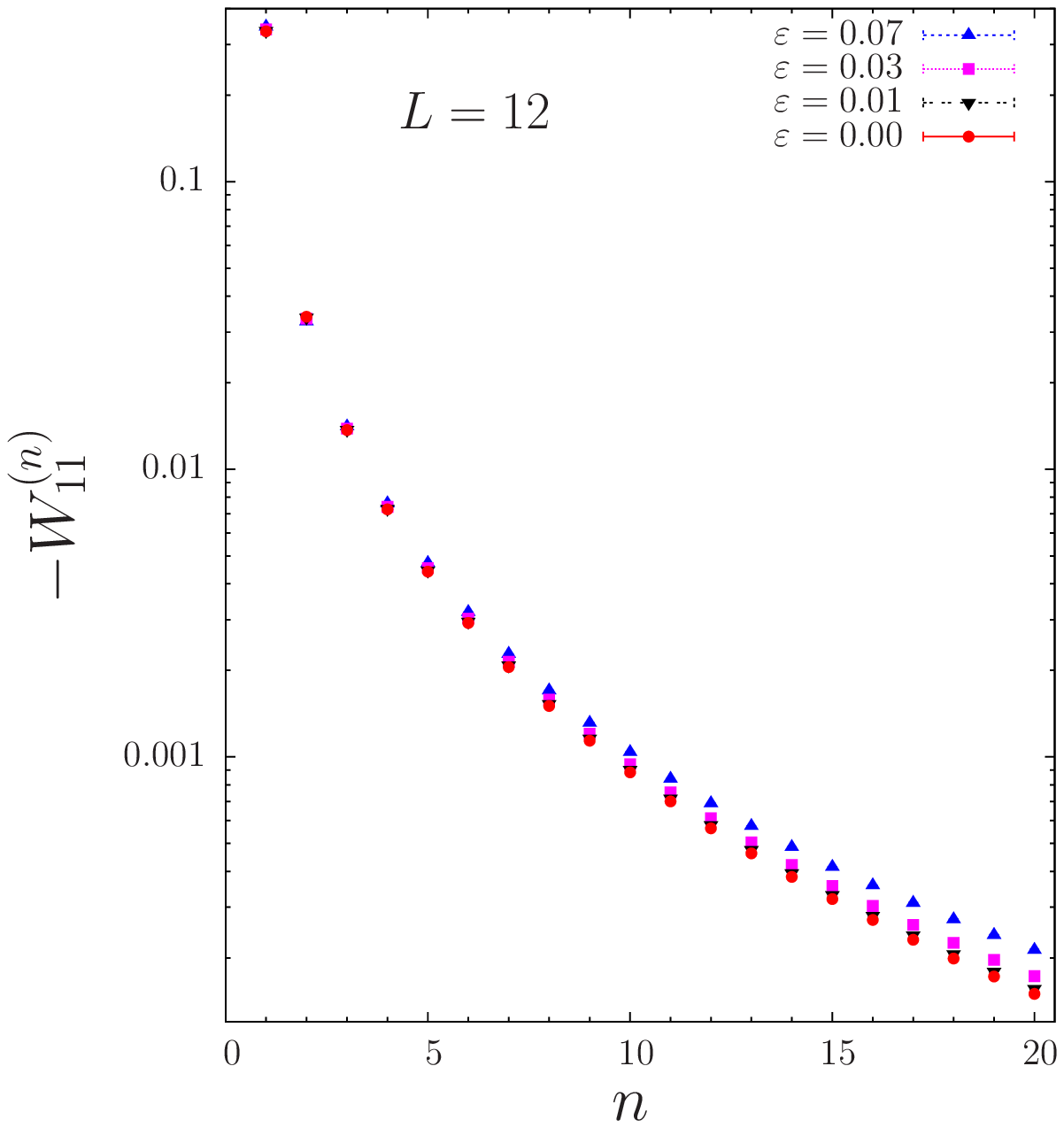}
     &
     \includegraphics[scale=0.63,clip=true]{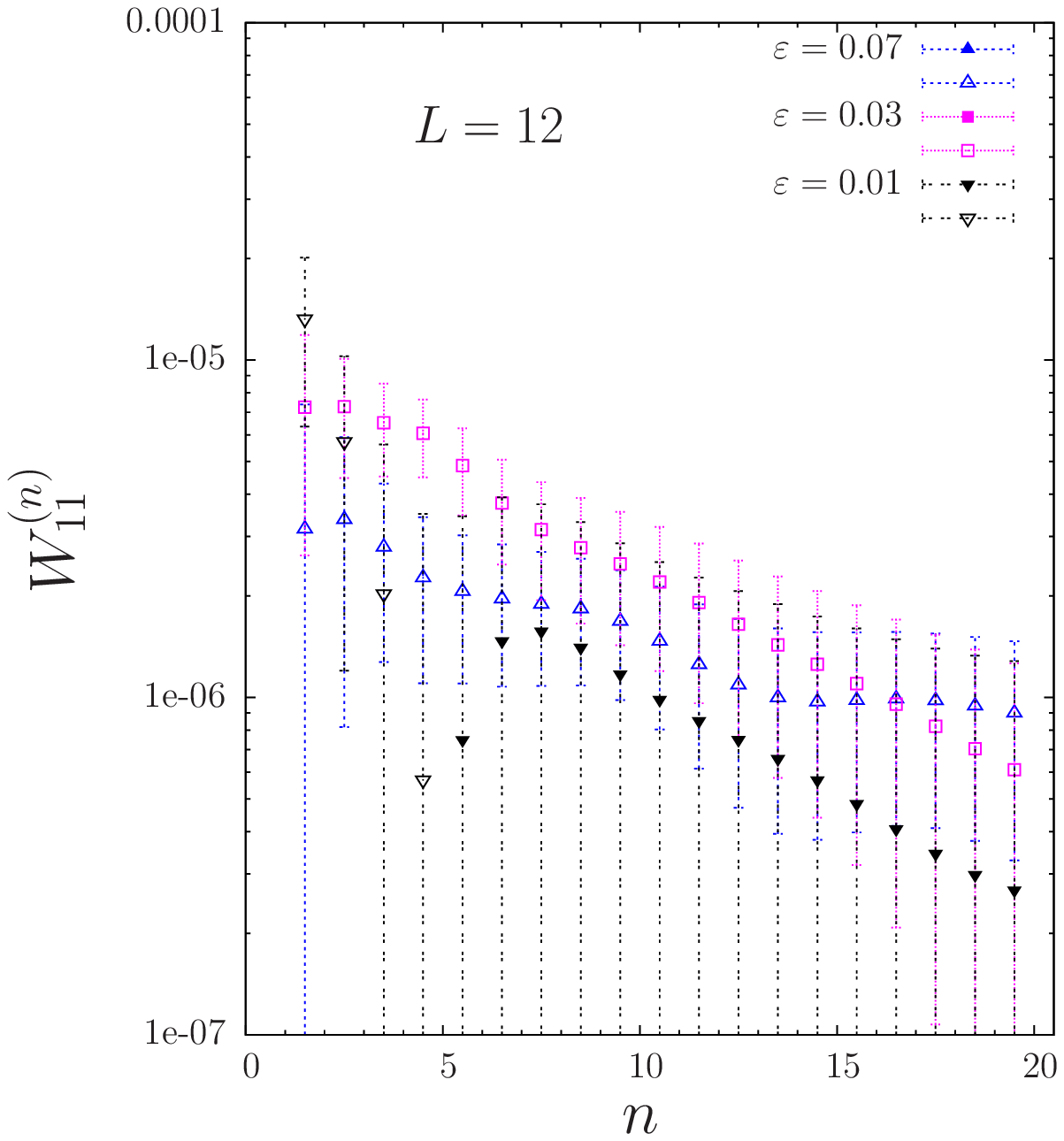}
  \end{tabular}
  \caption{Plaquette expansion coefficients $W_{11}^{(n)}$ at some finite $\varepsilon$  at $L=12$ versus loop number $n$.
           On the left-hand panel the loop expansion coefficients for integer $n$ (signal) are shown
           together with their extrapolations   $\varepsilon \to 0$.
           The right-hand panel shows the non-loop coefficients for half-integer $n$  which are purely noise,
           and 2 or 3 orders of magnitude smaller than the integer coefficients.
           Open/full symbols denote positive/negative numbers.}
  \label{fig:orderW11}
\end{figure*}
The coefficients of odd powers of $g$ should be zero, because the action
is unchanged under $g \leftrightarrow -g$. These non-loop coefficients
are shown in the right-hand panel of Figure~\ref{fig:orderW11}. We observe that these coefficients
are indeed orders of magnitude smaller than the coefficients for even powers of $g$.
To show the quality of the $\varepsilon\to 0$ extrapolation we zoom into the small and large loop number behavior of the expansion 
coefficients. This is demonstrated in Figure~\ref{fig:orderW11zoom}. For better 
visibility, part of the expansion coefficients
at low loop numbers $n$ are multiplied by factors given in the Figure.
\begin{figure*}[!htb]
  \begin{tabular}{cc}
     \includegraphics[scale=0.63,clip=true]{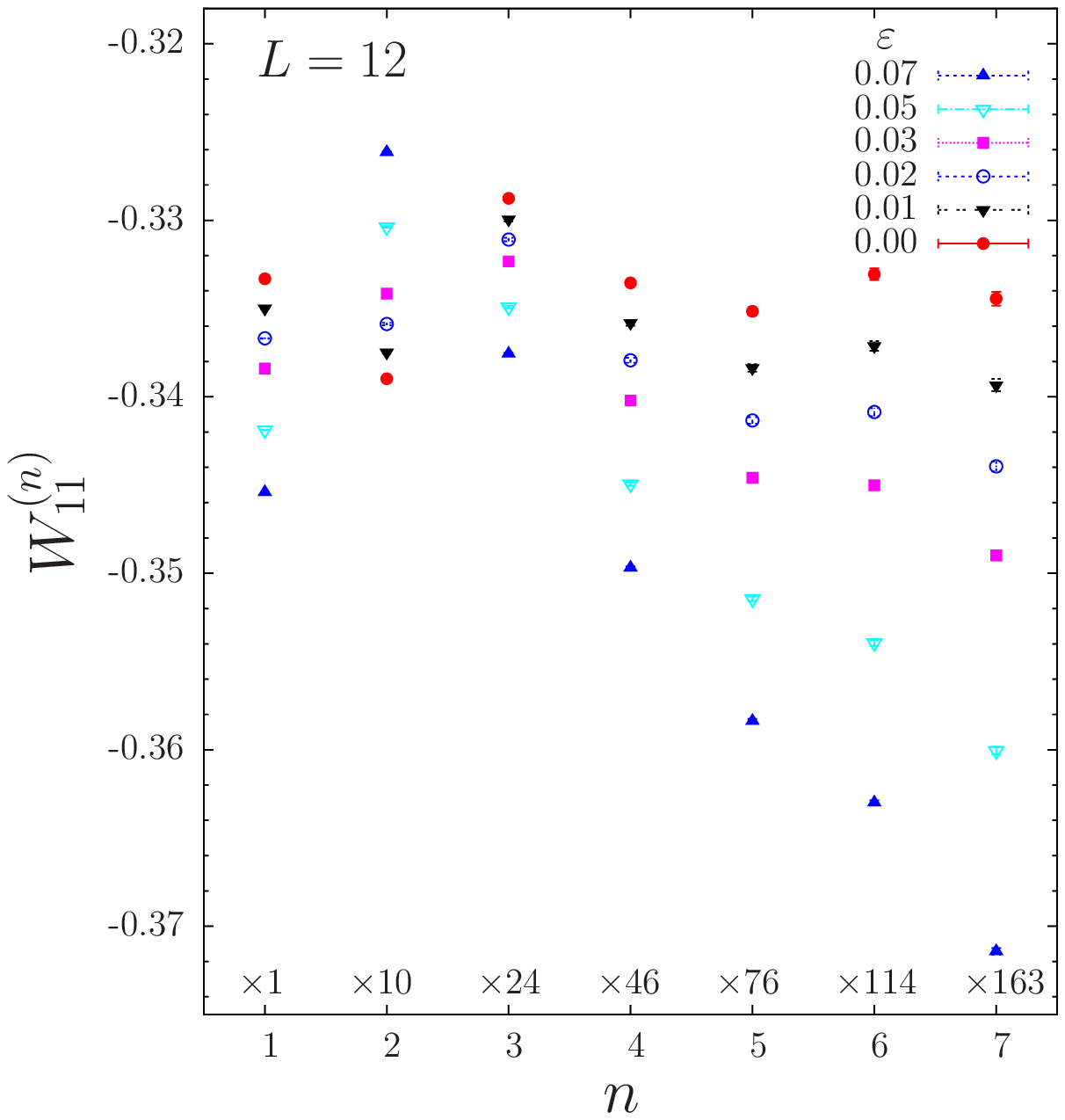}
     & 
     \includegraphics[scale=0.63,clip=true]{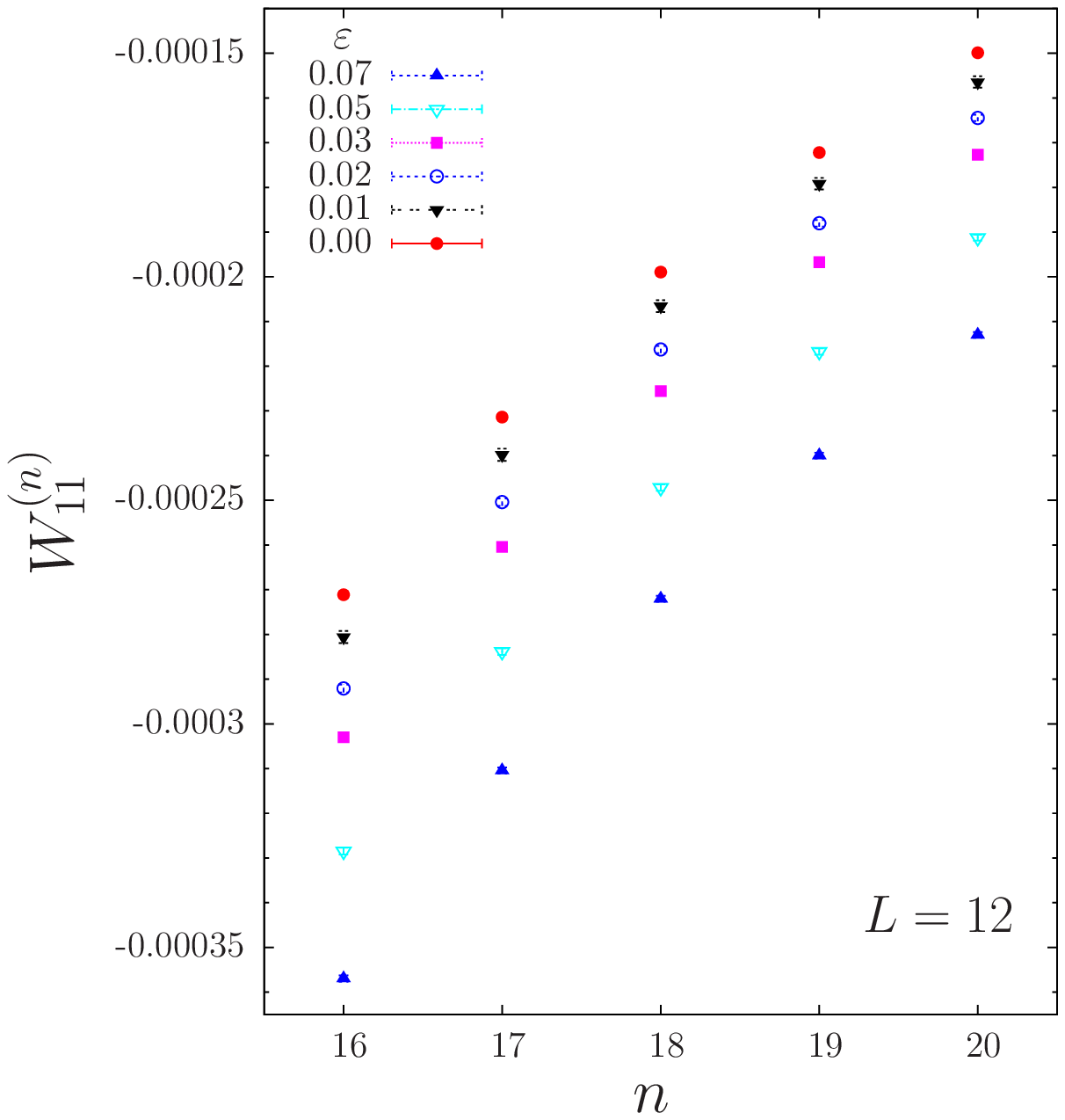}
  \end{tabular}
  \caption{Zoom of left Figure~\ref{fig:orderW11} into small (left) and large (right) loop number region $n$ for all finite $\varepsilon$. 
           The full circles in red are the extrapolated $\varepsilon \to 0$ values. 
           The coefficients in the left Figure at different orders $n$  are multiplied by factors to make them comparable in
           size to those at $n=1$.}
  \label{fig:orderW11zoom}
\end{figure*}

\begin{figure*}[!htb]
  \begin{tabular}{cc}
     \includegraphics[scale=0.63,clip=true]{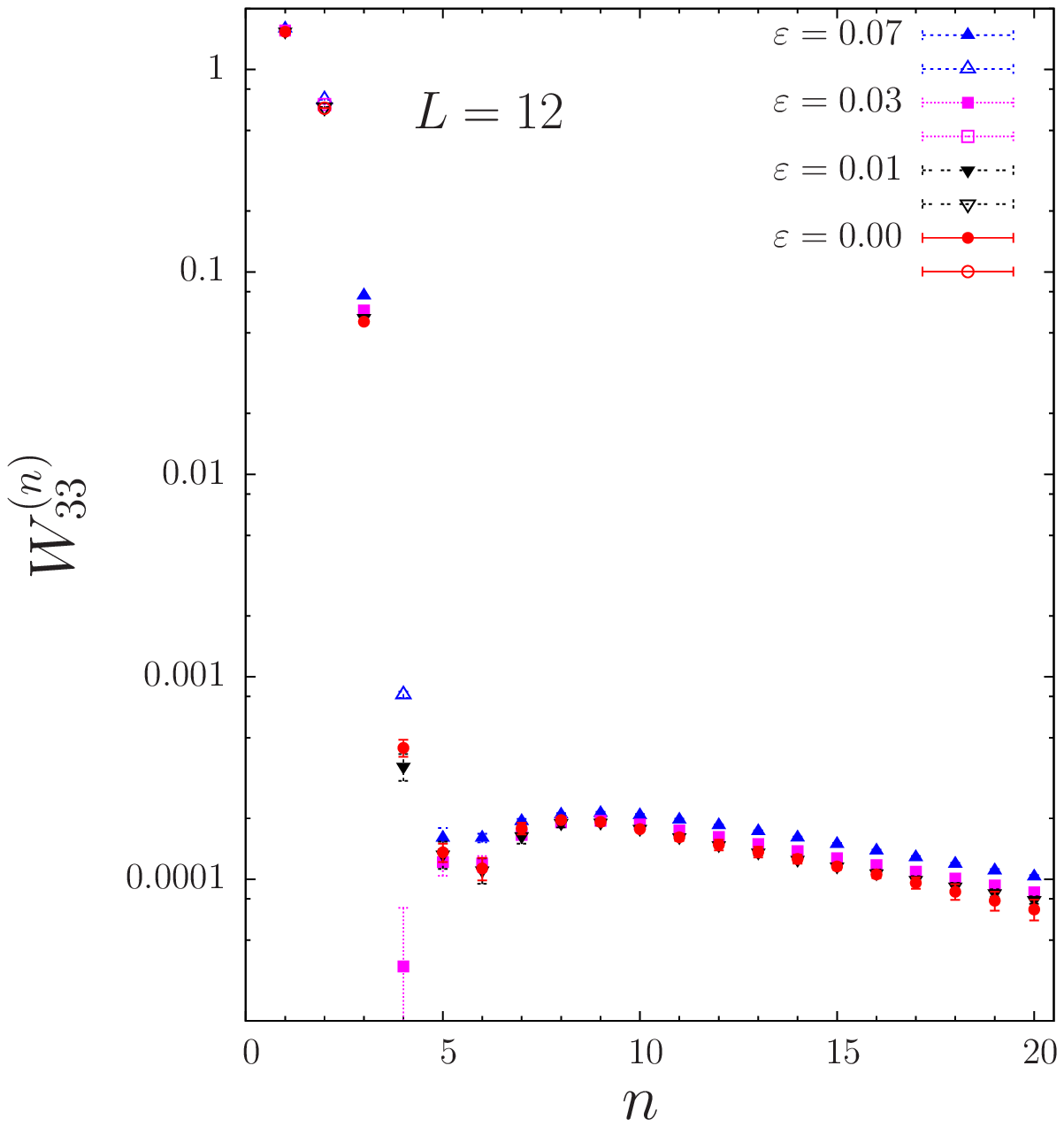}
     &
     \includegraphics[scale=0.63,clip=true]{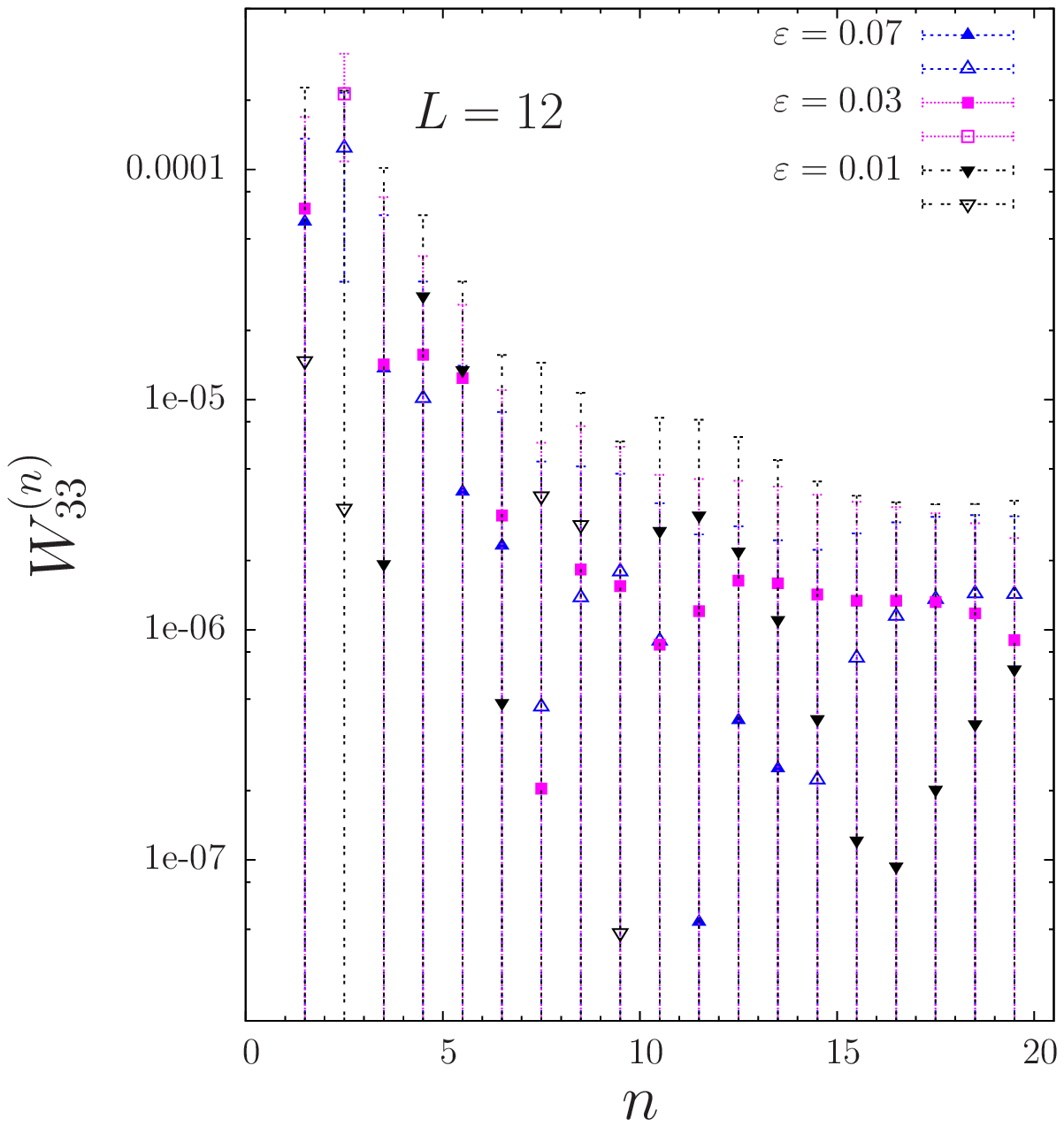}
  \end{tabular}
  \caption{Same as in Figure~\ref{fig:orderW11} but for the Wilson loop expansion coefficients  $W_{33}^{(n)}$.}
  \label{fig:orderW33}
\end{figure*}

Now we consider the Wilson loop $W_{33}$. 
In Figure~\ref{fig:orderW33} we show how the 
loop and non-loop 
expansion coefficients for various Langevin step sizes behave as function of $n$. 
We observe that the noise of the non-loop coefficients
is much larger than in the plaquette case, which has to be expected for Wilson loops with larger areas.
For the smallest half-integer $n$ the 
magnitude of the noise is larger than
the actual (integer) loop results at much larger $n$. But still our 
criterion is fulfilled that a 
Wilson loop coefficient at a given loop order $n$ 
should be larger than the magnitude 
of the noise for the 
adjacent $n-1/2$ and $n+1/2$ non-loop contributions.

\begin{figure*}[!htb]
  \begin{center}
  \begin{tabular}{cc}
     \includegraphics[scale=0.63,clip=true]{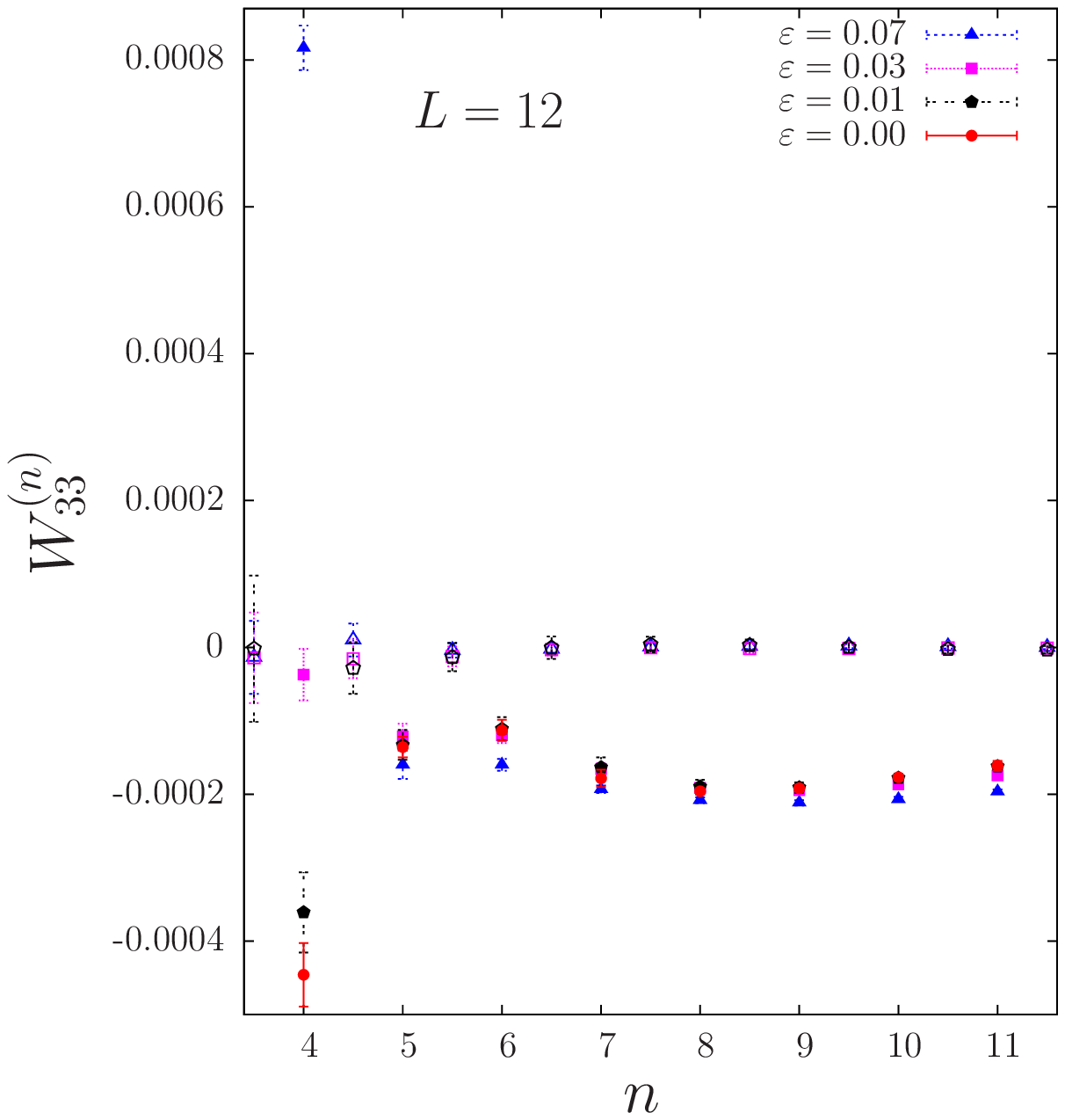}
     &
     \includegraphics[scale=0.63,clip=true]{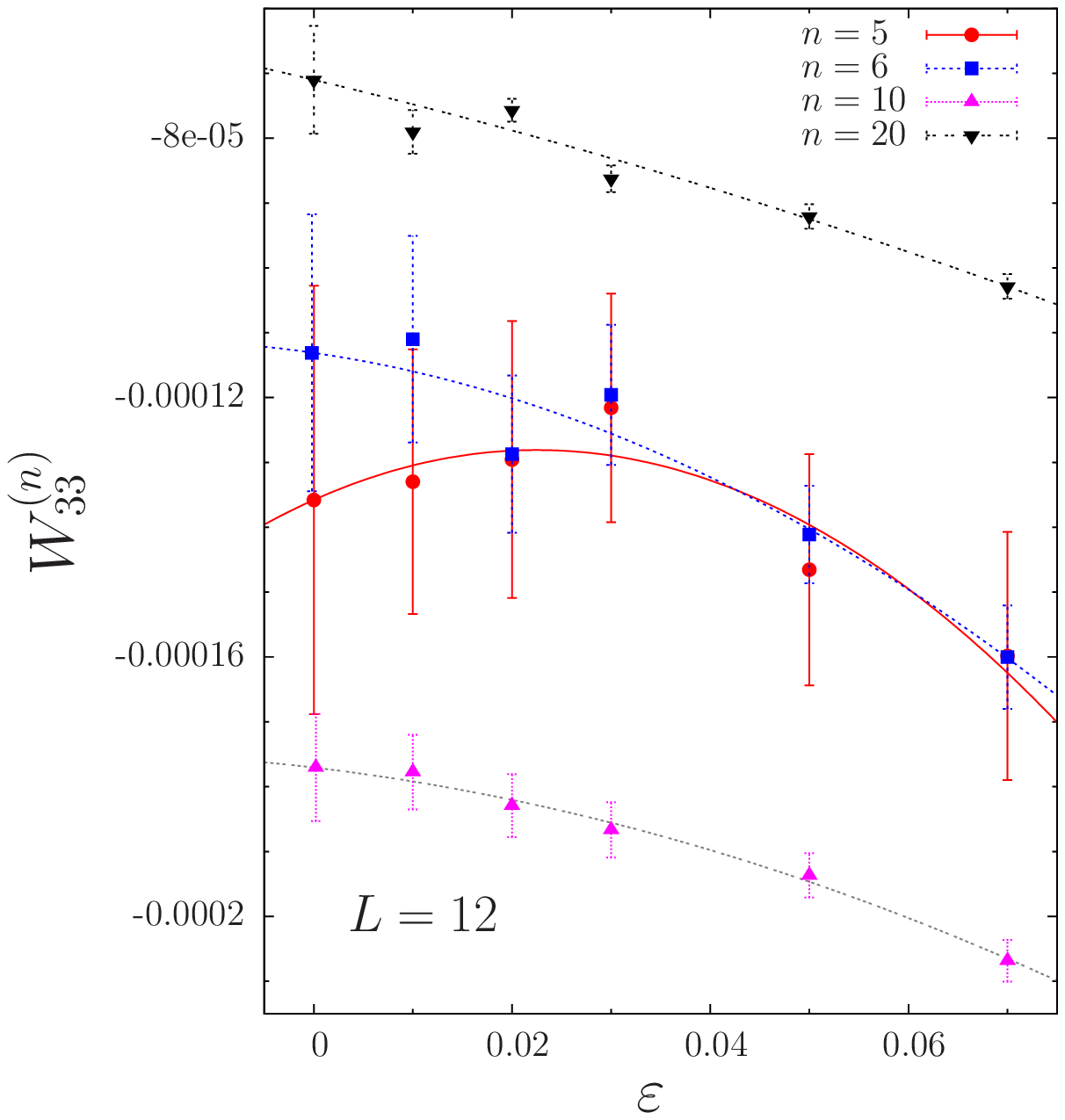}
  \end{tabular}
  \end{center} 
  \vspace{-2mm}
  \caption{The extrapolation to zero Langevin step for Wilson loop expansion coefficients  $W_{33}^{(n)}$: 
           Left: Zoom of both Figs.~\ref{fig:orderW33} in the region [3.5,11.5]. 
           Right: Detailed extrapolation to $\varepsilon\to 0$ for selected loop numbers.}
  \label{fig:orderW33zoom}
\end{figure*}

Contrary to the plaquette case, the loop expansion coefficients alternate in sign for $n \le 3$.
In absolute value the step-size extrapolation $\varepsilon \to 0$ approaches the extrapolated value from above. 
For loop number $n=4$ the situation 
is different (see left Figure~\ref{fig:orderW33zoom}):
The extrapolation of the expansion coefficient to zero Langevin step starts at a positive value 
$W_{33}^{(4)}(\varepsilon=0.07)$, crosses ``zero'' with 
decreasing $\varepsilon$ 
and points towards a negative value $W_{33}^{(4)}$ at zero Langevin step.
Remember that near zero we have 
the ``noise'', shown in that Figure as well, 
by the adjacent non-loop contributions $3.5$ and $4.5$.
The magnitude of that noise is comparable to $W_{33}^{(4)}$ for $\varepsilon=0.03$ only, and a reliable almost linear 
extrapolation to zero Langevin step is possible.
For the next higher loop numbers $n>4$ the extrapolation to zero 
Langevin step becomes clearly non-linear as 
shown in more detail in the right of Figure~\ref{fig:orderW33zoom} for 
some loop numbers $n$.
The extrapolated zero step size results are still clearly distinguishable from the adjacent non-loop expansion coefficients.
Therefore, according to our criterion, those extrapolations can be considered as reliable. 
For larger loop numbers $n \ge 10$ the $\varepsilon$ dependence becomes less non-linear again.
For those $n$ the expansion coefficients of $W_{33}$ as function of $n$ behave similar to those of the plaquette though
their slope slightly differs.

\begin{figure*}[!htb]
  \begin{center}
     \begin{tabular}{cc}
        \includegraphics[scale=0.63,clip=true]{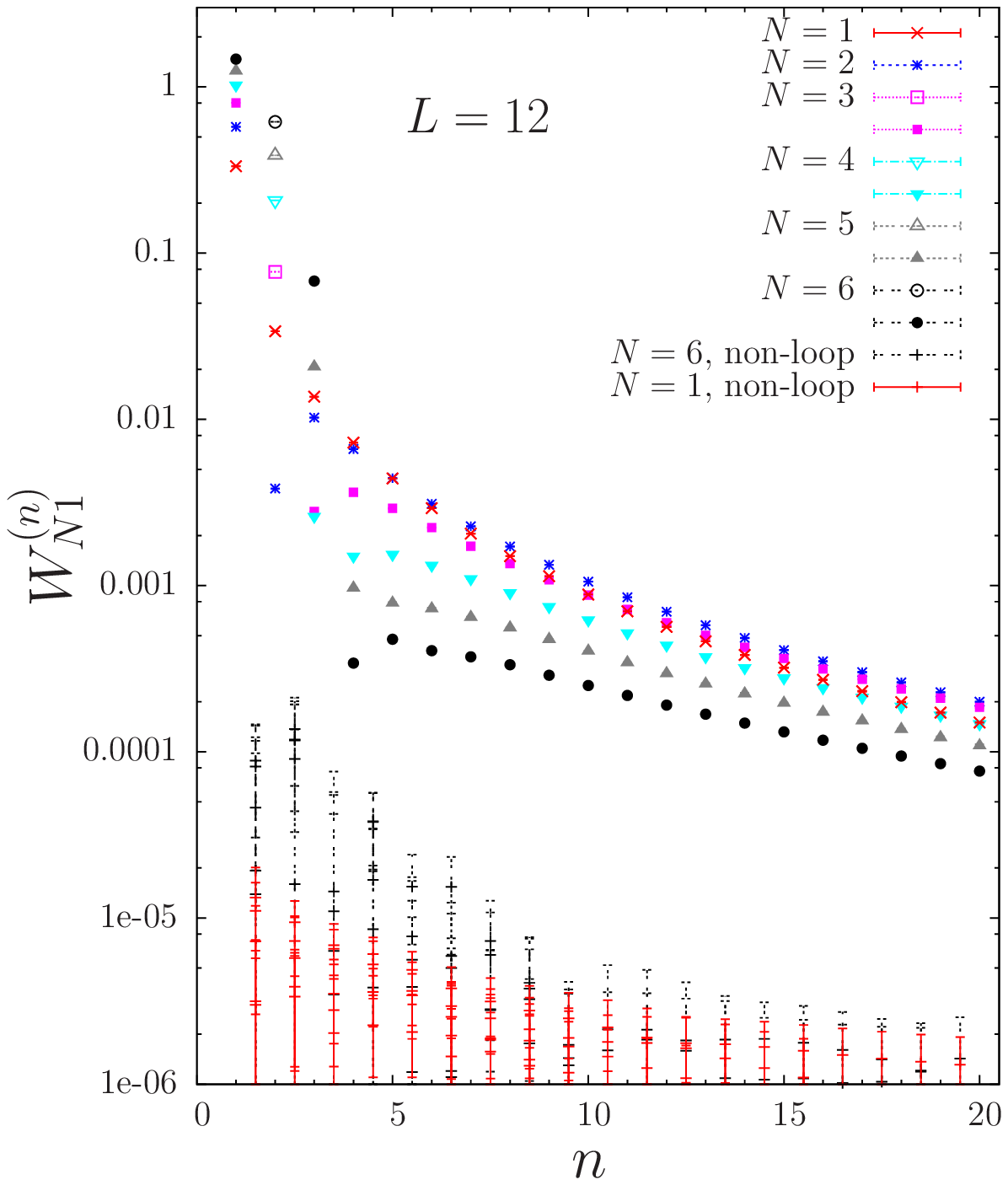}
        &
        \includegraphics[scale=0.63,clip=true]{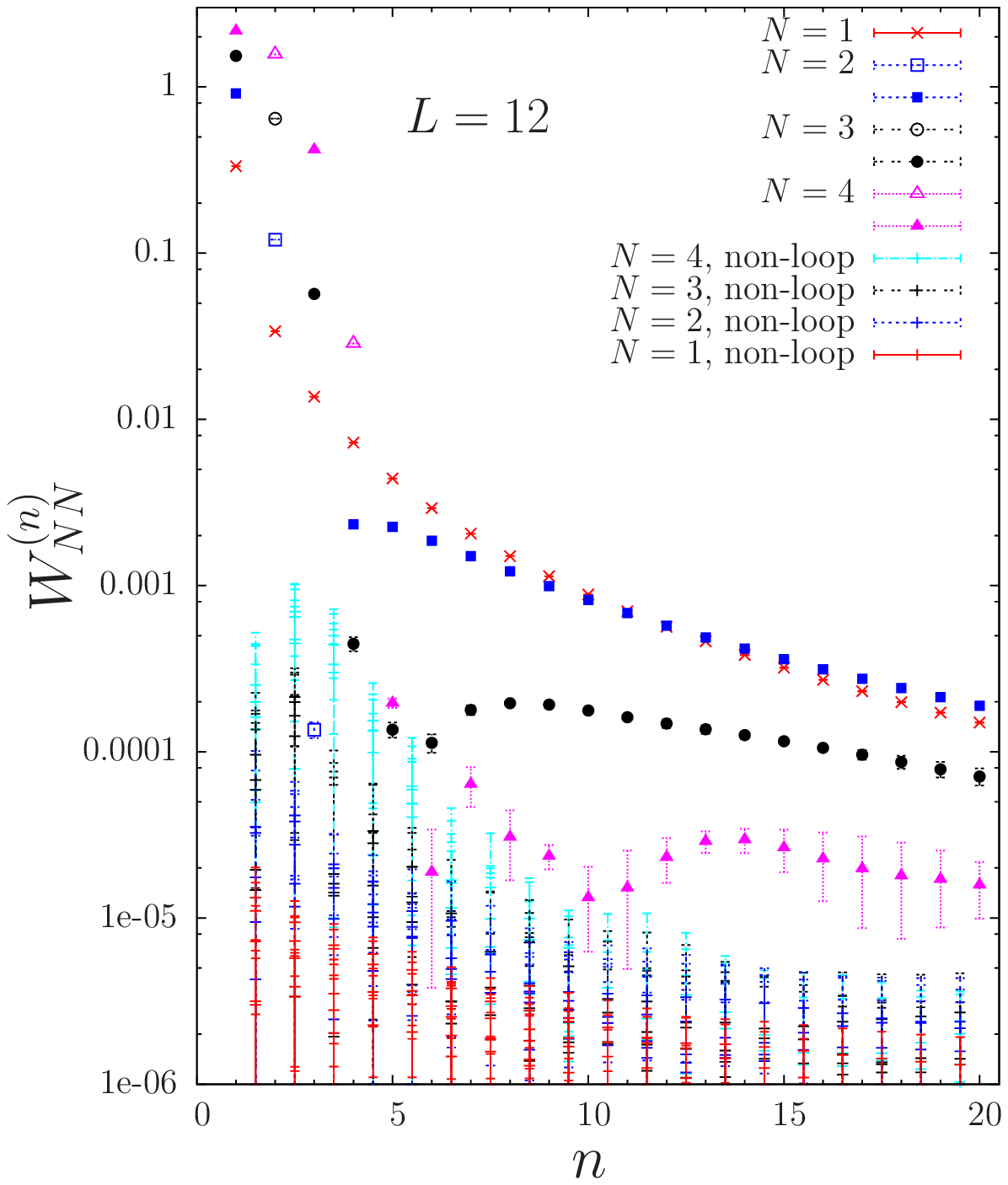}
     \end{tabular}
  \end{center}
  \caption{Selected loop coefficients $W_{NM}^{(n)}$ for $L=12$ versus loop order $n$ together with typical values in magnitude
  of non-loop coefficients.
  Positive/negative signs of the coefficients are given by open/full symbols, all $W_{11}^{(n)}<0,W_{21}^{(n)}<0$. Left: elongated Wilson loops
  $N\!\times\!1$  with $N=1,\dots 6$, right: square Wilson loops $N\!\times\!N$ with $N=1,\dots,4$.}
  \label{fig:WNMwil}
\end{figure*}

In Figure~\ref{fig:WNMwil} 
we show some results for the loop coefficients (extrapolated to $\varepsilon=0$) of elongated 
($W_{N1}^{(n)}$, left) and square ($W_{NN}^{(n)}$, right) Wilson loops for various size $N$ as function of loop order $n$
for a $12^4$ lattice and compare them to the noise.
At larger $n$, 
a behavior without sign changes is observed for all considered Wilson loops that could be interpreted 
as ``asymptotic''.
We note that the precision of the extrapolated loop coefficients for the larger Wilson loops
drops down and also the signal to noise ratio decreases. Still, the signal for the shown Wilson 
loops is clearly above the noise for all orders. For square Wilson loops with $N \ge 4$ (not shown)
or other larger Wilson loops 
the statistics was insufficient to get a clear signal out of the noise for larger orders (see also Appendix). 
In the analysis below we concentrate on the smallest Wilson loops. 

In addition we have to raise the question about the infinite volume limit of the series.
In the perturbative series the leading finite-size correction is expected to be proportional to $1/L^4$.
For additional non-leading corrections we tried the heuristic ansatz
\begin{equation}
  W_{NM,L}^{(n)} = W_{NM,\infty}^{(n)} + a_{NM}^{(n)}\, \frac{1}{L^4}+ b_{NM}^{(n)}\,\frac{\log L}{L^6} \,,
  \label{eq:fitansatz}
\end{equation}
which describes well the $L$-dependence of one- and two-loop coefficients of the perturbative Wilson loops for various loop sizes.  
Those coefficients are known from standard finite volume lattice perturbation theory (\cite{Bali:2009}, the basic formulae have 
been given in~\cite{Heller:1984hx}). 
Note that the one- and two-loop NSPT coefficients reproduce the finite volume lattice perturbation theory reasonably well
as shown in Table \ref{tab:BaliTab} for some examples.
\begin{table*}[!htb]
  \begin{center}
     \begin{tabular} {|c| c| c c| c c|}
        \hline
        &&&&&\\ 
        $W_{NN}$ & $L$ & NSPT (1-loop) &  Bali (1-loop)  & NSPT (2-loop) &  Bali (2-loop) \\
        &&&&&\\ 
        \hline
        &&&&&\\ 
        $W_{22}$ &  $4$   & $-0.87468(13)$   & $-0.87500$  & $0.10404(07)$    & $0.10406$\\ 
                 &  $6$   & $-0.90752(12)$   & $-0.90762$  & $0.11830(10)$    & $0.11837$\\ 
                 &  $8$   & $-0.91147(03)$   & $-0.91141$  & $0.11998(02)$    & $0.11993$\\ 
                 &  $12$  & $-0.91264(02)$   & $-0.91261$  & $0.12043(01)$    & $0.12038$\\[1.2ex]
        $W_{33}$ &  $6$   & $-1.50088(30)$   & $-1.50093$  & $0.60906(34)$    & $0.60866$\\ 
                 &  $8$   & $-1.52849(12)$   & $-1.52803$  & $0.63654(08)$    & $0.63632$\\ 
                 &  $12$  & $-1.53552(06)$   & $-1.53533$  & $0.64388(01)$    & $0.64360$\\[1.2ex]
        $W_{44}$ &  $8$   & $-2.14092(23)$   & $-2.14016$  & $1.52436(28)$    & $1.52331$\\ 
                 &  $12$  & $-2.17001(10)$   & $-2.16922$  & $1.57160(10)$    & $1.57006$\\[0.2ex]
        \hline
      \end{tabular}
    \end{center}
  \caption{Comparison of one- and two-loop results for NSPT and 
           finite volume standard lattice perturbation theory~\cite{Bali:2009}.}
  \label{tab:BaliTab}
\end{table*}

\begin{figure*}[!htb]
  \begin{center}
     \begin{tabular}{cc}
        \includegraphics[scale=0.63,clip=true]{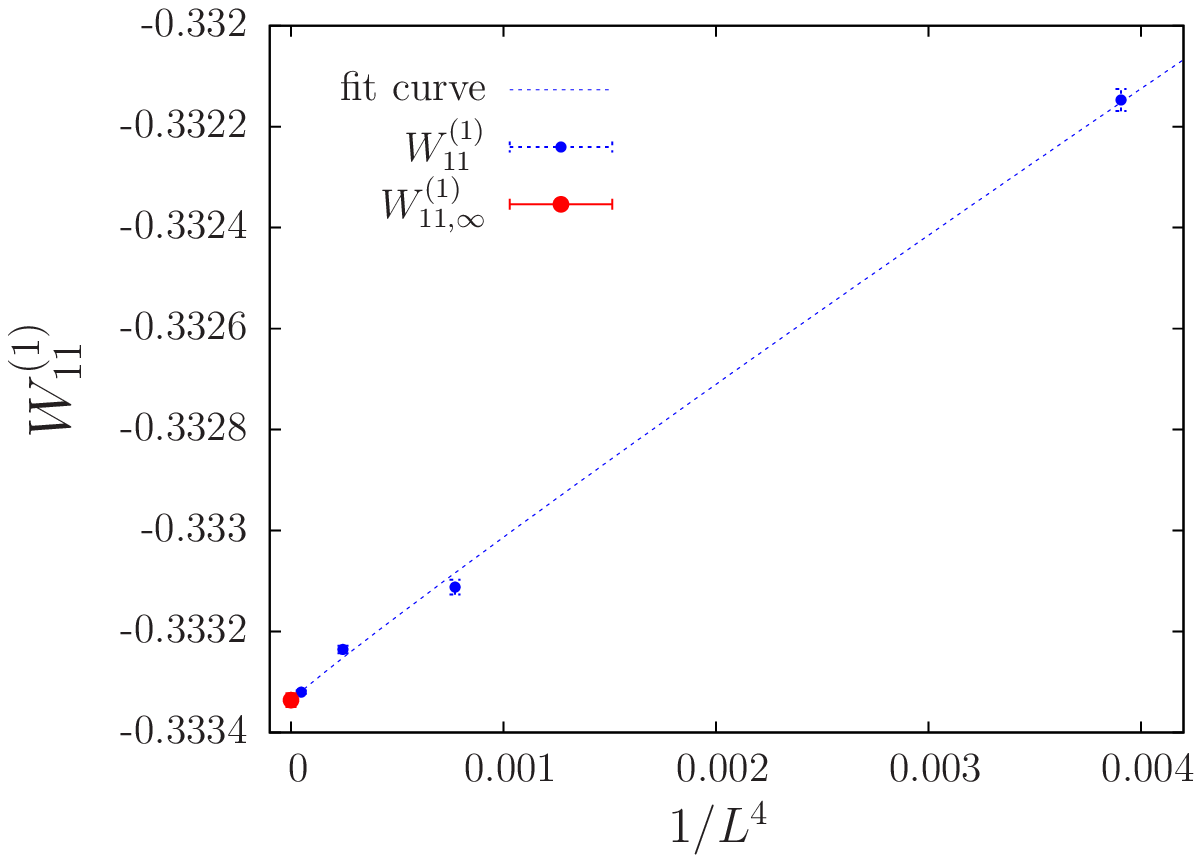}
        &
        \includegraphics[scale=0.63,clip=true]{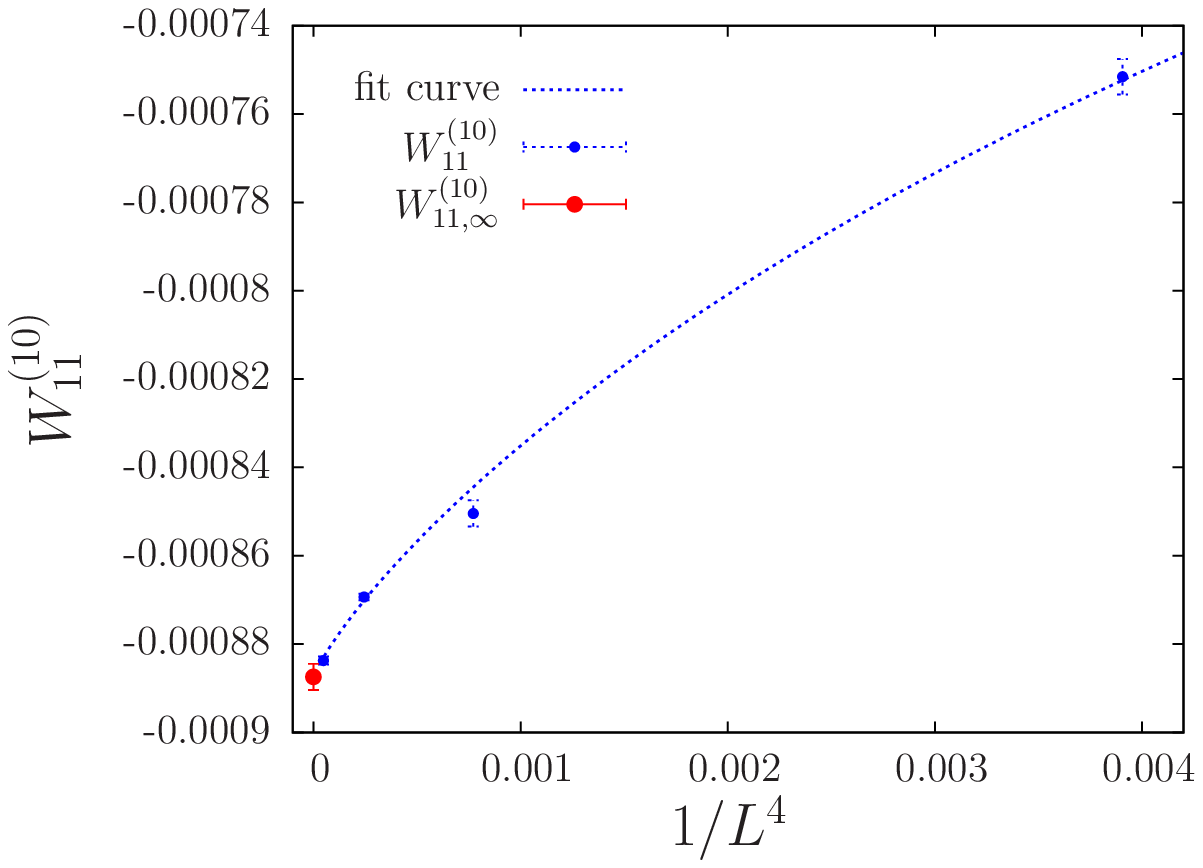}
     \end{tabular}
  \end{center}
  \caption{Extrapolation $L\rightarrow \infty$ for $W_{11}^{(n)}$ for loop orders $n=1$ (left) and $n=10$ (right).}
  \label{fig:W11InfPlaquette}
\end{figure*}

For lower loop orders a simple $1/L^4$ dependence was sufficient in the fits in agreement 
with~\cite{Heller:1984hx}. 
Higher loop coefficients, however, need further
corrections  which we have chosen in the form~(\ref{eq:fitansatz}).
In Figure~\ref{fig:W11InfPlaquette}
we show two selected extrapolations for the one- and ten-loop expansion coefficient. 
From the volume dependence of all orders and sizes of the Wilson loop we conclude that we can treat the lattice volume $12^4$
being already  near to the infinite volume limit.
Therefore, in the subsequent analysis  we use  that lattice size as a reasonable approximation
for volume independent results of the series.
In the Appendix we present the expansion coefficients for all available lattice volumes
and Wilson loop sizes.

\section{Perturbative series of Wilson loops at large orders}
\label{sec:ptseries}

There is plenty of evidence that perturbative series in continuum QCD are divergent,
at best asymptotic. This would mean that, beginning from some perturbative
order $n>n^\star$, the coefficients of the series should grow factorially.
The situation might be different for perturbative series on finite lattices.
Here we have both ultraviolet and infrared cut-offs and the growth could be modified 
significantly. With our computed coefficients 
of the loop expansion up to order $n=20$ we are able to check this to a so 
far inaccessible level.

For finite lattices one could try to use the raw NSPT coefficients for evaluating
the corresponding underlying infinite series. 
This requires to deduce a kind of asymptotic model providing the complete perturbative answer.
Formally, one can use such a model designed for finite lattices also in a version adapted to the
coefficients extrapolated to $L \rightarrow \infty$.
Although the extrapolation seems to yield smooth limits, it is certainly not allowed to sum 
a series based on these extrapolated coefficients up to infinity.
In this case there exist at least two possibilities.
The first consists in taking into account possible renormalon effects and estimating the truncated tail 
of the series (cf. e.g.~\cite{Meurice:2006cr}). 
This procedure, however, strongly depends on whether a 
clear factorial growth of the coefficients in the 
perturbative region under consideration has been identified. 
We will see that this is very difficult to justify from our results. 
A second possibility consists in applying boosting, i.e. a rearrangement of
the series, resulting in a (rather) stable plateau of the 
truncated sum as function of the maximal perturbative order 
$n^{*}$ that is included,
and to use this as the final perturbative result at given coupling.

\subsection{Plaquette} 
\label{subsec:Plaquette}

In 2001, when only the first 10 loops of the plaquette series as expansion in the bare coupling
were known from~\cite{DiRenzo:2000ua}, some of the present authors tried plotting the data
in various ways in order to find a fit ansatz
which could describe the known data
and would be able to predict the unknown higher 
coefficients~\cite{Horsley:2001uy}.
A logarithmic plot of 
$W_{11}^{(n)}$ against $n$ shows a curve with decreasing slope, 
well described by an asymptotic behavior 
\begin{equation} 
  W_{11}^{(n)} \sim n^{-(1+\gamma)} \, u^n,  
  \label{powexp} 
\end{equation} 
i.e.~an exponential in $n$, multiplied by
a power of $n$ (see Figure~\ref{fig:orderW11} and Tables in the Appendix). 
This is a somewhat unexpected result, because 
a series of this type has a finite radius of convergence, 
$g^2 < 1/u$, and sums to give a result with a 
power-law singularity of the form 
\begin{equation}
  (1 - u g^2)^\gamma. 
  \label{powsing} 
\end{equation} 

A more sensitive way of showing the large $n$ behavior
of a series is the Domb-Sykes plot~\cite{DombSykes}. If the series
has the form 
\begin{equation} 
  \sum_n c_n g^{2n} 
\end{equation} 
we calculate $r_n$, the ratio of neighboring coefficients, 
\begin{equation} 
   r_n \equiv \frac{c_n}{c_{n-1}} \, ,
  \label{rn-def}
\end{equation}
and plot it as a function of $1/n$. The intercept as $1/n \to 0$ 
(if the limit exists) gives the radius of convergence. The behavior 
for small $1/n$ (i.e. large $n$) tells us the nature of the 
dominant singularity. 
A function with the power-law singularity~(\ref{powsing}) has the expansion
\begin{eqnarray} 
  && (1 - u g^2)^\gamma = 
  1 - \gamma u g^2 + \cdots 
  \nonumber
  \\  
  &&
  + \frac{ \Gamma(n- \gamma)}{\Gamma(n+1) \Gamma(-\gamma) } (u g^2)^n + \cdots 
  \label{powfit} 
\end{eqnarray}
which leads to the ratio of neighboring coefficients depending on the 
parameters $u$ and $\gamma$
\begin{eqnarray} 
  \frac{c_n}{c_{n-1}} =  u \left( 1 - \frac{1+\gamma}{n} \right) \,.
  \label{rn0} 
\end{eqnarray} 
Therefore, the Domb-Dykes plot $c_n$ versus $1/n$ is a straight-line graph.

\begin{figure*}[!htb]
  \begin{center}
     \begin{tabular}{cc}
        \includegraphics[scale=0.32,clip=true,angle=270]{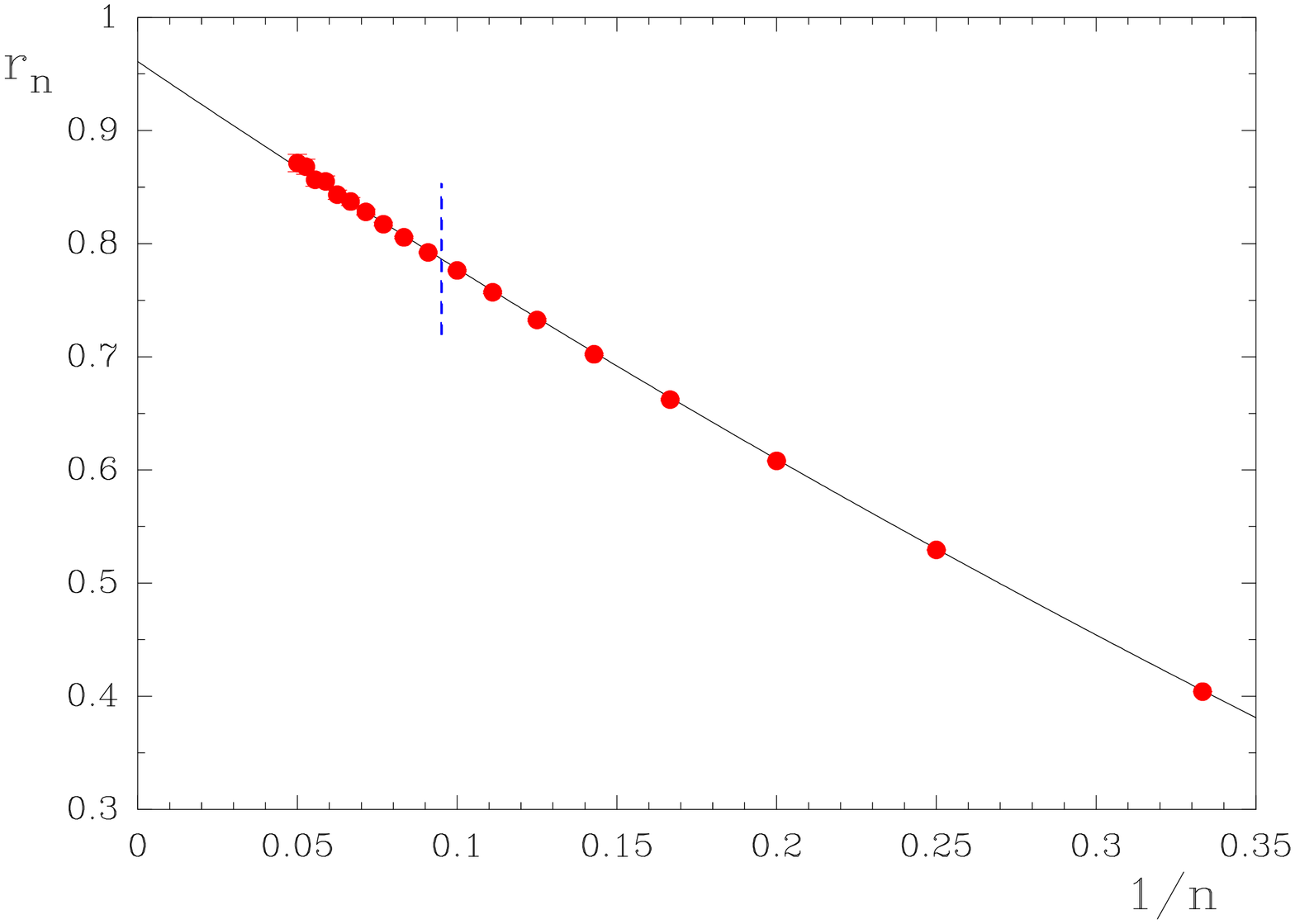}
        &
        \includegraphics[scale=0.32,clip=true,angle=270]{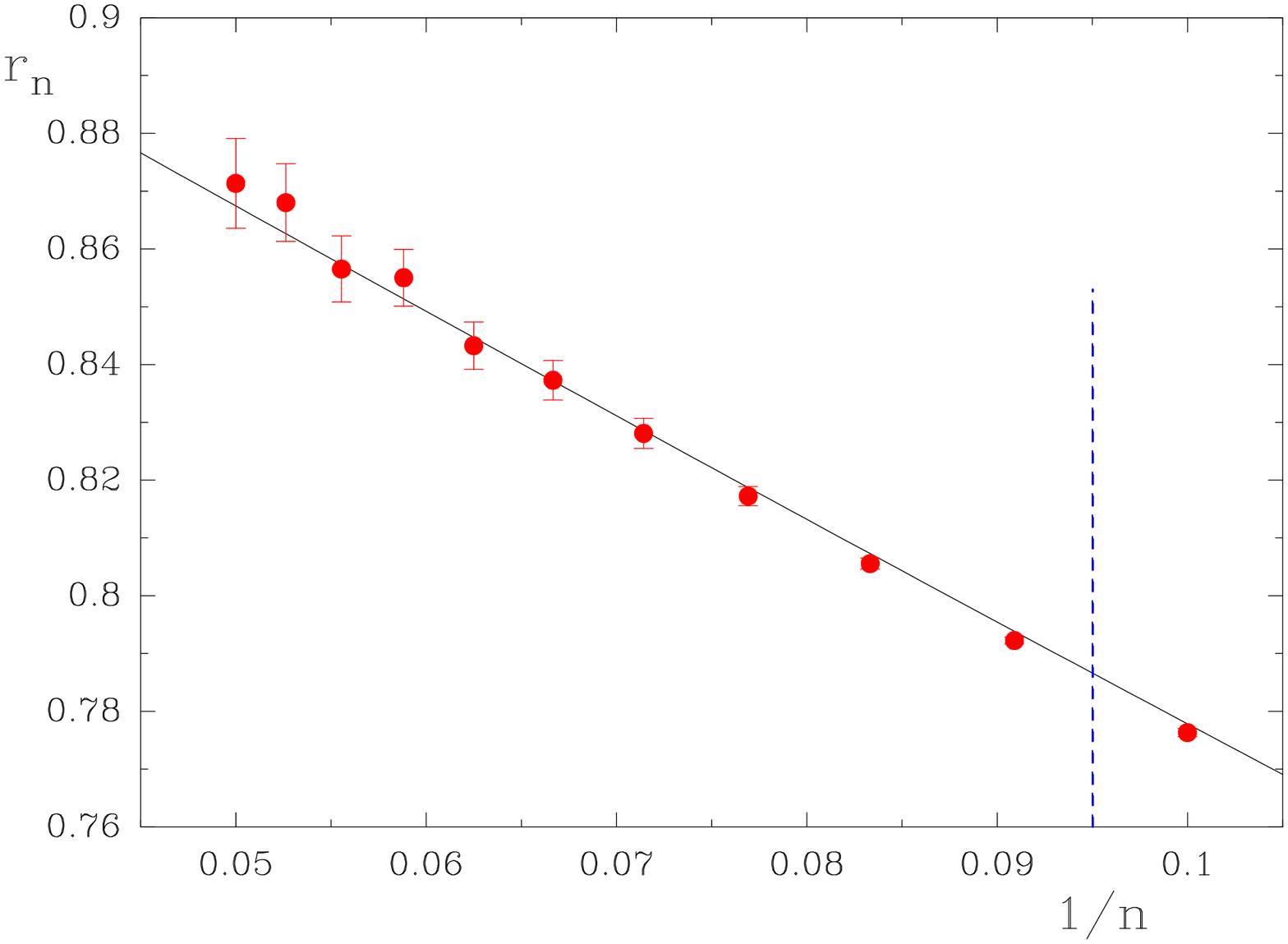}
     \end{tabular}
  \end{center}
  \caption{Current ratio data for the plaquette,
           compared with the prediction of 2001~\cite{Horsley:2001uy},
           plotted with the original parameters. The prediction was based on data with
           $n \le 10$, i.e. to the right of the vertical blue bar.
           The second figure zooms in on the region of new data.
 \label{predict} }
 \end{figure*}

The actual Domb-Sykes plot for the measured perturbative plaquette showed a small curvature. To 
allow for this we added one more parameter and made a fit of the form 
\begin{equation} 
  r_n = u \left( 1 - \frac{1+\gamma}{n+s} \right) \,.
  \label{plaqfit} 
\end{equation} 
This described the data for $ n \in [3,10] $ well, with the 
parameter values~\cite{Horsley:2001uy}  
\begin{equation}
  u = 0.961(9), \quad \gamma = 0.99(7), \quad s= 0.44(10) \;. 
  \label{plaqpar} 
\end{equation}  
We now have 10 more coefficients. How well do the fit parameters
(\ref{plaqpar}) predict the new data? 
In Figure~\ref{predict} 
we compare the current data with the prediction made in 2001.   

The data lie very near the prediction. We have doubled the maximum $n$ value
without seeing any breakdown of the behavior seen at lower $n$. 
In particular, the series still looks like a series with a finite 
range of convergence, $ g^2 < 1.04$. 

\subsection{A model for summing up the Wilson loop series}
\label{subsec:HypMod}

\begin{figure}[!htb]
  \begin{center}
     \includegraphics[scale=0.64,clip=true]{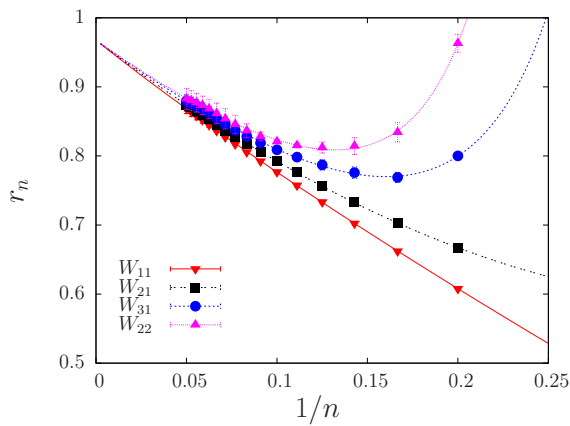}
  \end{center}
  \caption{Domb-Sykes plots for $W_{NM}$ for $n \ge 5$ together with the fit result 
           using (\ref{eq:HRSratio}).}
  \label{fig:DSPlotsWNM}
\end{figure}

Now we have in addition also Wilson loops larger than the plaquette at our disposal.
In Figure~\ref{fig:DSPlotsWNM} 
we show the coefficient ratios $r_n$ for some small size Wilson loops for $n \ge 5$. 
We have seen that at large order $n$ the coefficients in the plaquette 
series have the asymptotic behavior of (\ref{powexp}).
What is the asymptotic behavior of the other Wilson loops? Is it similar? 

\begin{figure}[!htb]
  \begin{center}
    \includegraphics[scale=0.32,clip=true,angle=270]{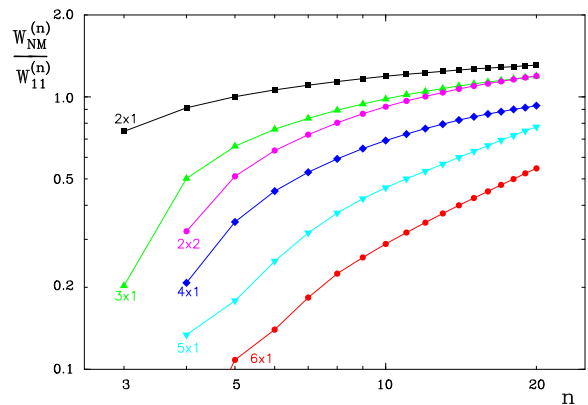}
  \end{center} 
  \caption{A log-log plot of the ratio~(\ref{Wrat}), plotted for different
           sizes of Wilson loops. To guide the eyes, the data points for the loop orders are connected by lines.} 
  \label{WratFig}
\end{figure}

A sensitive way to investigate this is to look at the ratio
between the coefficients of the Wilson loops series and the plaquette 
series. If both have similar behaviors at large order $n$
\begin{equation} 
  \frac{ W_{NM}^{(n)} } { W_{11}^{(n)} } \sim
  \frac{ n^{-(1+\gamma^\prime)} \, (u^\prime)^n }
  { n^{-(1+\gamma)} \, u^n }
  = n^{( \gamma - \gamma^\prime )} \left ( \frac{u^\prime} {u} \right)^n \;.
  \label{Wrat}
\end{equation} 
We plot the ratio~(\ref{Wrat}) for various $NM$ values
in Figure~\ref{WratFig}, as a log-log plot against $n$. 
The plot shows that at large $n$ the ratio scales like a power of $n$, suggesting that the parameter $u$ in
(\ref{powexp}) is the same for all Wilson loops, but the 
power $\gamma$ depends on the size of the loop.
Therefore, $u^\prime = u$ to a very good approximation.
This means that for all Wilson loops the series have the same 
apparent radius of convergence, $g^2 < 1/u$. 
However the curves for different Wilson loops have different 
slopes at large $n$, indicating different asymptotic powers of
$n$, i.e. different values of $\gamma$. 

In Figure~\ref{fig:DSPlotsWNM} one clearly recognizes that for larger loop size the ratios deviate from the almost perfect 
straight line behavior seen for $W_{11}$. 
This deviation can be described rather well by a modification of (\ref{plaqfit}) taking into account 
some curvature, especially for larger loop-sizes $N \!\times\! M$. Parametrizing these effects by an additional parameter
$p$  we make the ansatz
\begin{equation}
  r_n=\frac{c_n}{c_{n-1}} = u \left(1-\frac{1+\gamma}{n}  \right) + 
  \frac{p}{n(n+s)} 
  \label{eq:HRSratio}
\end{equation}
where the first term is the asymptotic form (\ref{rn0}) without curvature.
Relation (\ref{eq:HRSratio}) can be transformed into a recursion relation,
\begin{equation}
  c_n = \left\{ \begin{array}{ll}
       r_n\,  c_{n-1}\,,  & \mbox{if $n > n_0$}\,,\\
        c_{n_0}\,, & \mbox{if $n = n_0$}.\end{array} \right.
  \label{eq:HRSreceq}
\end{equation}
Here $c_{n_0}$ is the input value for 
some lowest measured perturbative coefficient $W_{NM}^{(n_0)}$ at
loop order $n=n_0$ to begin the recursive reconstruction. 
Relation (\ref{eq:HRSreceq}) can be solved to
\begin{eqnarray}
  c_{n,hyp} &=&
  d_{n_0}\,\frac{\left(\sigma-\tau-1\right)_{n} 
  \left(\sigma+\tau-1\right)_{n}}{(s+1)_{n}\,n!}\,u^n\,,
  \nonumber \\
  \tau &=& 
  \frac{1}{2}\,\sqrt{(\gamma+s+1)^2-4 p/u}\,,
  \label{eq:cnsol1}  
  \\
  \sigma &=& \frac{s+3-\gamma}{2}\nonumber\,,
  \end{eqnarray}
with 
$(a)_n\equiv \Gamma(a+n)/\Gamma(a)$ being the Pochhammer symbol. The coefficient $d_{n_0}$ is given by
\begin{equation}
  d_{n_0}  = \frac{n_0!\,c_{n_0}}{u^{n_0}}\, 
  \frac{\prod_{i=1}^{n_0}\,(s+i)}{\prod_{k=1}^{n_0}((\sigma-2+k)^2-\tau^2)} \, .
  \label{eq:HRScn}
\end{equation}

Accepting such a parametrization one can follow different strategies: 
\begin{itemize}
  \item  
  Use the raw coefficients $c_n$ and/or $c_{n,hyp}$ fixed by the fitted values of the parameters
  in the loop order range $1 \le n \le 20$ as determined by the NSPT computation to
  investigate the perturbative series. This will be done in the next Section \ref{subsec:boosted_PT}.

  \item 
  Assume that the coefficients $c_{n,hyp}$, found as solution of (\ref{eq:HRSreceq}), belong
  to an infinite series and try to sum up the series on a finite lattice. This will be discussed 
  in the following.
\end{itemize}

The infinite series we want to compute is defined by
\begin{eqnarray}
  W^{(n_0)}_{NM,\infty}&=& 1+\sum_{n=1}^{n_0} \, 
  c_n\,g^{2n}+ \sum_{n=n_0+1}^\infty \, c_{n,hyp}\,g^{2n} 
  \nonumber \\
  &\equiv& 1+ \sum_{n=1}^\infty W_{NM,hyp}^{(n)} \, g^{2n} \, ,
  \label{eq:finsum1}
\end{eqnarray}
where the first $n_0$ coefficients $c_n \equiv W_{NM}^{(n)}$ are given by the NSPT measurements
and the $c_{n,hyp}$ are the solutions of (\ref{eq:HRSreceq}). 
For later use we have introduced the general coefficients $W_{NM,hyp}^{(n)}$.
The matching condition for
(\ref{eq:finsum1}) is that at $n_0$ we have $c_{n_0} =c_{n_0,hyp} $.
Introducing the hypergeometric function $_2F_1$ 
\begin{eqnarray}
  _2F_1\left(a, b; c;t \right) = \sum_{n=0}^\infty\, A_n \, t^n \equiv 
  \sum_{n=0}^\infty\, \frac{(a)_n\,(b)_n}{(c)_n\,n!} \, t^n  \, ,
  \label{eq:finsum111}
\end{eqnarray}
we get the closed expression
\begin{eqnarray}
  &&  W^{(n_0)}_{NM,\infty}= 1+ \sum_{n=1}^{n_0}\, \left( c_n - d_{n_0} A_n \,u^n\right)\,g^{2n} 
  \nonumber
  \\
  && + d_{n_0}\Big[ {_2F_1} \left(\sigma-\tau-1,\sigma+\tau-1;s+1;u\,g^2\right)
  \nonumber
  \\
  &&
  -1 \Big]  \,.
  \label{eq:finsum1res}
\end{eqnarray}
The result expressed in terms of 
$_2F_1\left(a, b; c;u\,g^2\right)$  has a branch cut discontinuity at the positive $g^2$-axis for $g^2 > 1/u$.
This means that the parameter $u$ in (\ref{eq:HRSratio}) (just as well as in (\ref{plaqfit})) determines the 
convergence radius: for $g^2 < 1/u$ the series can be summed up to $n=\infty$ without analytic continuation into the 
complex plane. All parameters $u, \gamma, p,s$ depend on the corresponding underlying data set. 
As discussed above (see also Figure~\ref{WratFig}) we will assume that the convergence radius is the same 
for all Wilson loop sizes which implies a common value for $u$.

We found that Wilson loops larger in size than the plaquette (e.g., $W_{21}, W_{31}, W_{22}$) 
give rise to ratios $r_n$ (for $n <5$) that show a pronounced oscillating behavior. 
Therefore, we restrict the fit of the ratio function (\ref{eq:HRSratio}) to the data for $n > n_0 =4$ only. 
The fit results are shown in Figure~\ref{fig:DSPlotsWNM} as thin lines.

It should be pointed out that fitting the parameters $(u,\gamma,p,s)$ in ansatz (\ref{eq:HRSratio}) to
the NSPT data is non-trivial. We have determined the optimal values by minimizing the function 
\begin{equation}
  \delta^2(u,\gamma,p,s) = \sum_{n=n_0+1}^{20} 
  \frac{ \left[r_n(u,\gamma,p,s) - r_n(\mathrm{NSPT})\right]^2}{\left[r_n(\mathrm{NSPT})\right]^2}\,,
  \label{eq:fitfun}
\end{equation}
where $r_n(\mathrm{NSPT})$ are the ratios computed from the corresponding NSPT data.
The most sensitive parameter in (\ref{eq:HRSratio}) is $s$. Therefore, we 
vary $s$ over a certain range $s_{\min} < s < s_{\max}$ by a small increment $\Delta_s$
as $s_0(k)=s_{\min} + k\,\Delta_s$ ($k$ - integer) and minimize
$\delta^2(u,\gamma,p,s_0(k))$ with respect to $(u,\gamma,p)$ at every  $s_0(k)$ which is held fixed.
The smallest of all minimized $\delta^2_{\min}(u,\gamma,p,s_0(k))$ defines the starting
set $(u_\star, \gamma_\star, p_\star, s_0(k_\star))$ for a final minimization fit - 
now with respect to all parameters $(u,\gamma,p,s)$. 
\begin{figure}[!htb]
  \begin{center}
        \includegraphics[scale=0.64,clip=true]{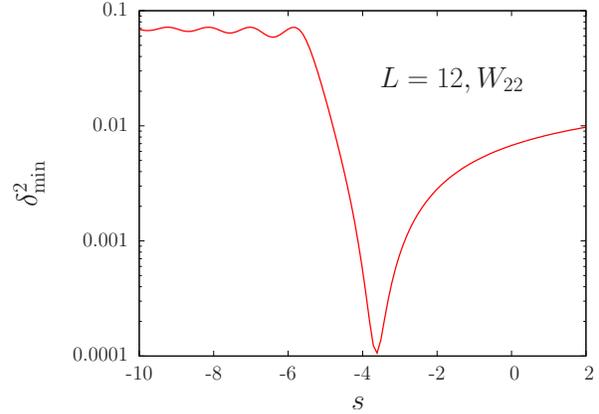}
  \end{center}
  \caption{$\delta^2_{\min}$ as function of parameter $s$ for $W_{22}$.}
  \label{fig:deltaW22}
\end{figure}
In Figure~\ref{fig:deltaW22} we show one example for $\delta^2_{\min}$ for $W_{22}$ with $n_0=4$.
One recognizes a couple of shallow local minima (besides the absolute one) where minimization
procedures could have been trapped. 
In Table \ref{tab:FitErrTab} 
\begin{table}[!htb]
  \begin{center}
    \begin{tabular}{|c|c|r@{.}l|r@{.}l|r@{.}l|r@{.}l|}
      \hline
      &&\multicolumn {1}{c}{ }&&\multicolumn {1}{c}{ }&&\multicolumn {1}{c}{ }&&\multicolumn {1}{c}{ }&\\[-0.8ex]
      $W_{NM}$ &   $\delta_{\min}^2$ 
               & \multicolumn{2}{c|}{u}
               & \multicolumn{2}{c|}{$\gamma$}
               & \multicolumn{2}{c|}{p}
               & \multicolumn{2}{c|}{s}
      \\[0.5ex] \hline 
      &&\multicolumn {1}{c}{ }&&\multicolumn {1}{c}{ }&&\multicolumn {1}{c}{ }&&\multicolumn {1}{c}{ }&\\[-0.8ex]
      $W_{11}$ & $8\cdot 10^{-6}$  & $0$    & $9694(4)$ & $1$ &  $13(5)$  & $1$ &   $5^{+1.3}_{-0.6}$   & $0$  &   $7^{+3.0}_{-1.8}$  \\
      $W_{21}$ & $1\cdot 10^{-5}$  & $0$   & $9694(5)$ & $1$ &  $ 02(4)$ & $1$ &  $ 6(5)$                   & $-1$ &  $4(8)$   \\
      $W_{31}$ & $2\cdot 10^{-5}$  & $0$   & $9694(6)$ & $0$ &   $91(4)$ & $1$ &  $ 7(2)$                   & $-3$ &  $3(2)$  \\
      $W_{22}$ & $4\cdot 10^{-5}$  & $0$   & $9694(9)$ & $0$ &   $82(4)$ & $1$ &   $9(2)$                   & $-3$ &  $9(1)$\\[0.8ex]
      \hline
    \end{tabular}
  \end{center}
  \caption{Minimal  value of $\delta_{\min}^2$ and resulting fit parameters.  The fit range in $n$ is $[5,20]$.}
  \label{tab:FitErrTab}
\end{table}
we give the results of our minimal fit function and 
the final fit parameters for various Wilson loops.
The given errors $(\Delta u,\Delta \gamma,\Delta p,\Delta s)$ are the extreme values 
within the error ellipsoid obtained from the relation
\bea
  &&\delta^2(u^\star+\Delta u,\gamma^\star+\Delta \gamma,p^\star+\Delta p,s^\star+\Delta s)
  \nonumber \\
  &=&  2\, \delta^2(u^\star,\gamma^\star,p^\star,s^\star)\,,
\eea
where  $(u^\star,\gamma^\star,p^\star,s^\star)$ are the best fit parameters.
For extrapolation of the perturbative series we use hypergeometric
fits in the interval $ [5,20]$. Fits to the coefficients in this range are excellent, 
with relative errors $\le 0.5\%$.
 
The hypergeometric fit still gives a fairly good description of
the data all the way down to $n=1$.  For most loop sizes a fit
from $n=1$ to $20$ describes the data  within $\sim 5\%$, except for 
the $2 \!\times\! 1$ loop, which has some errors $\approx 10\%$. 
Given that the coefficients vary through 4 orders of magnitude in this
interval, an error of $5$ or $10\%$ is still impressive. 
 
\begin{figure*}[!htb]
  \begin{center}
     \begin{tabular}{cc}
        \includegraphics[scale=0.32,clip=true,angle=270]{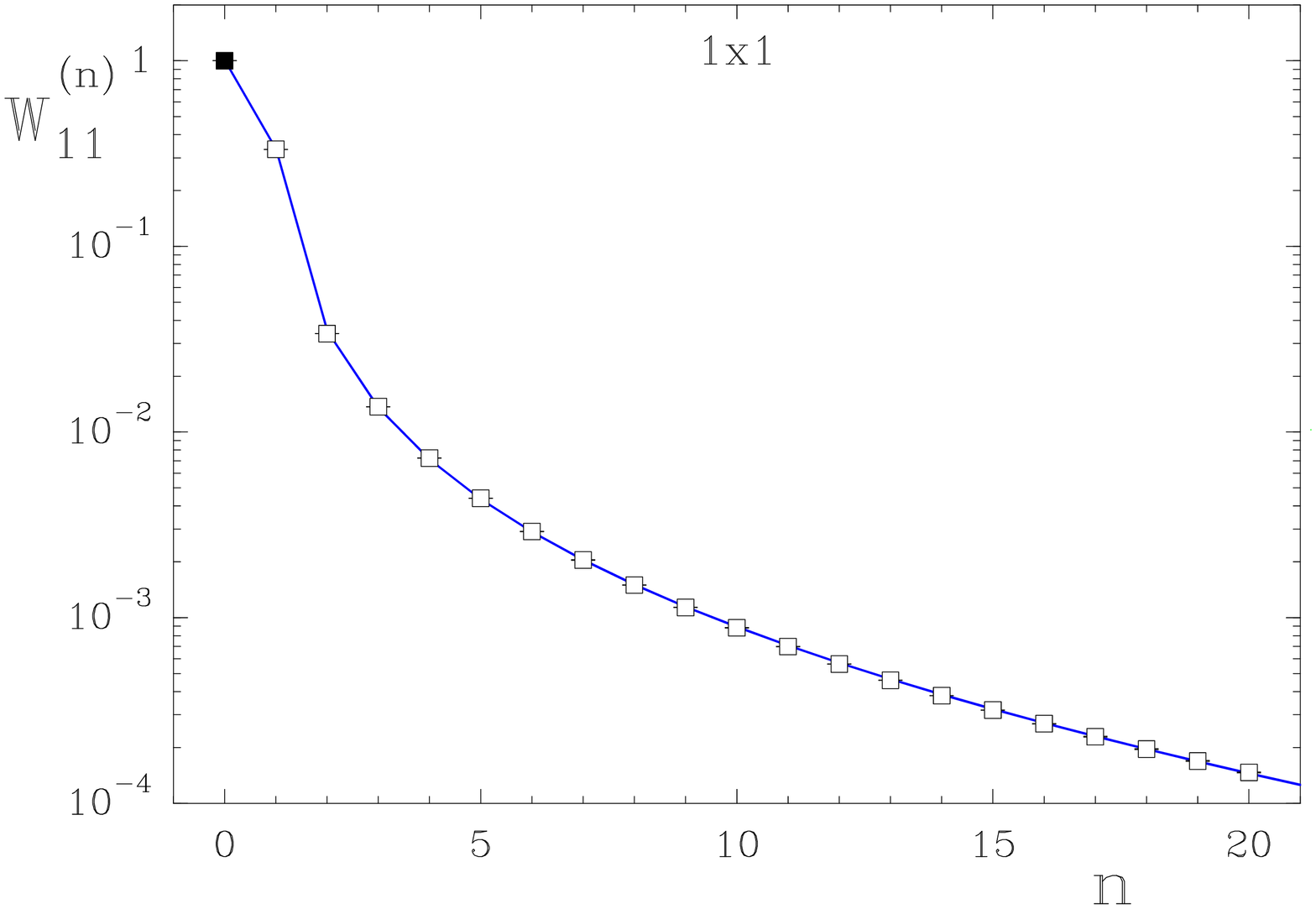}
        &
        \includegraphics[scale=0.32,clip=true,angle=270]{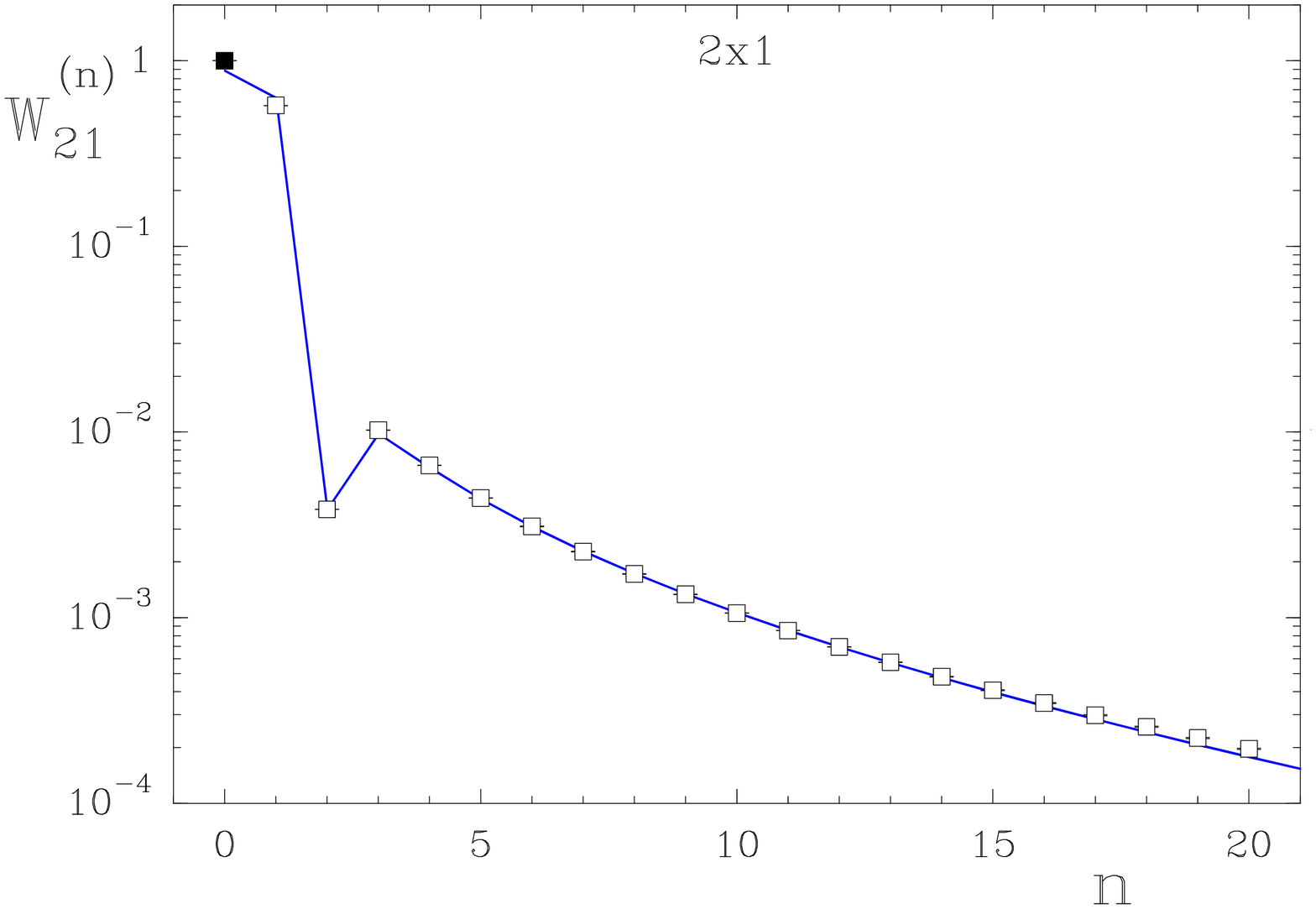}
        \\
        \includegraphics[scale=0.32,clip=true,angle=270]{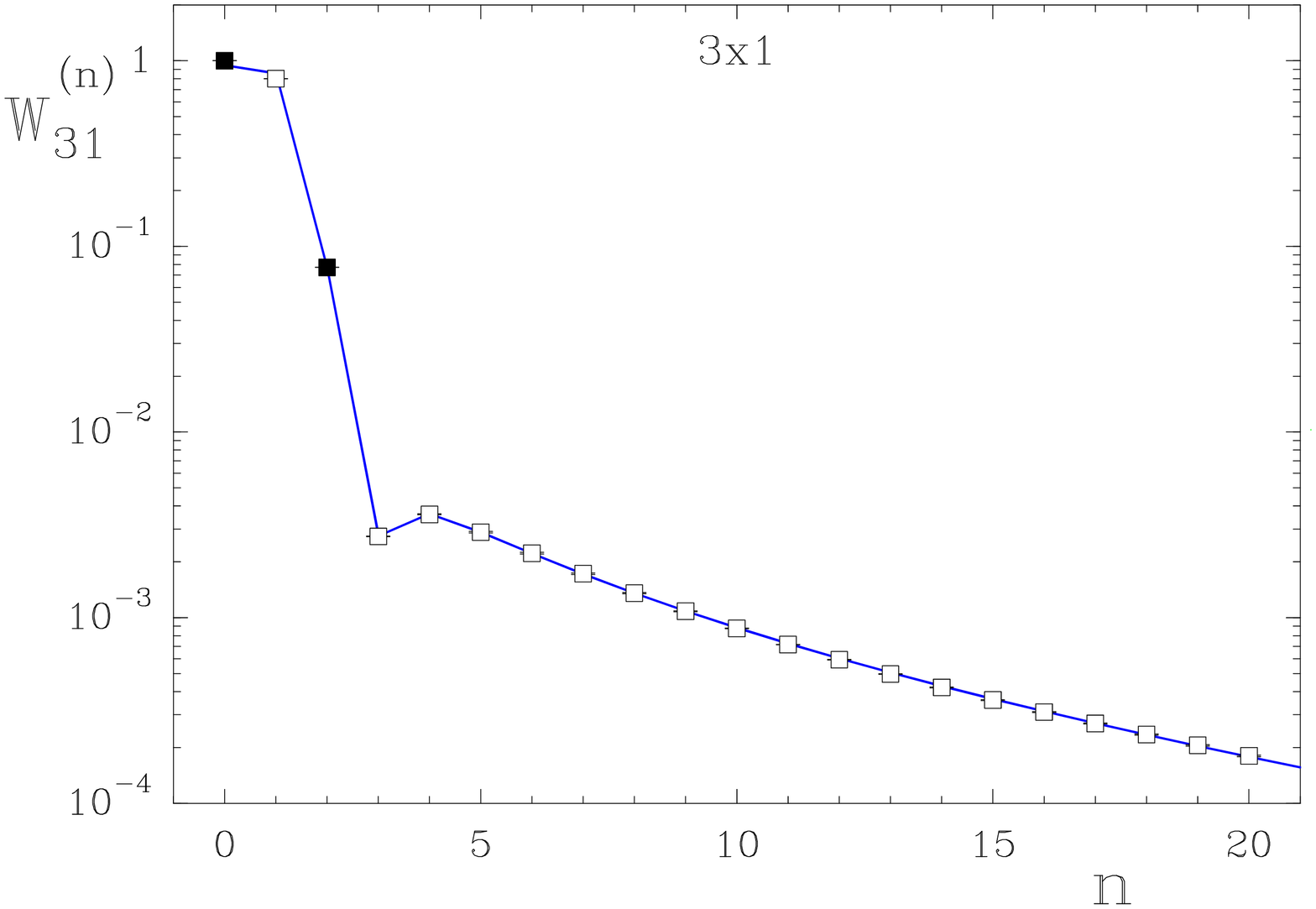}
        &
        \includegraphics[scale=0.32,clip=true,angle=270]{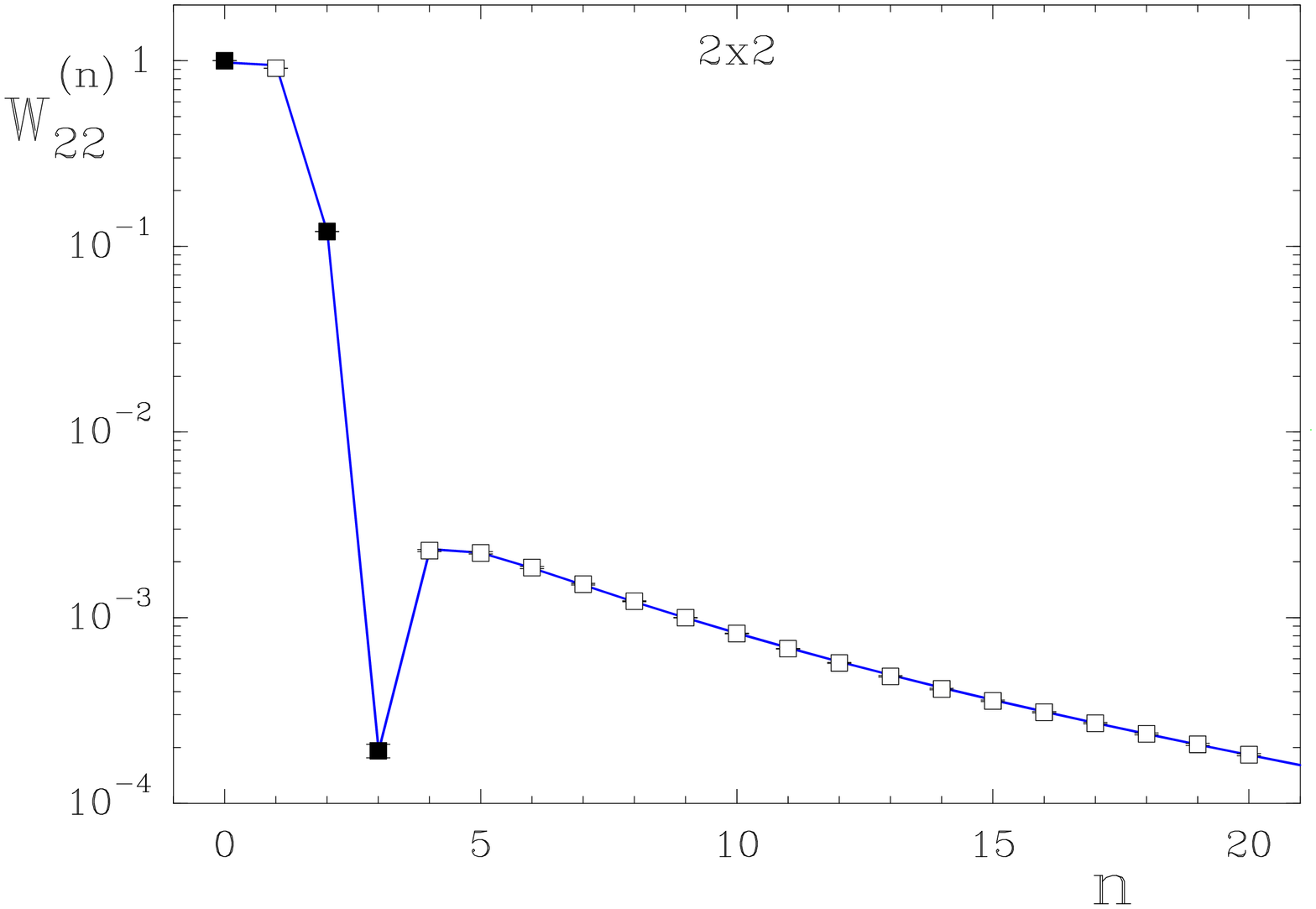}
     \end{tabular}
  \end{center}
  \caption{The hypergeometric fit to the
           coefficients $W_{NM}^{(n)}$ for the four small Wilson loops.
           Solid symbols represent positive terms, open symbols are
           negative terms. The blue line is an equal weight fit,
           including all points from $W_{NM}^{(0)}$ to $W_{NM}^{(20)}$.
           The agreement is remarkably good.}
  \label{wijfit}
\end{figure*}

\begin{figure*}[!htb]
  \begin{center}
    \begin{tabular}{cc}
       \includegraphics[scale=0.55,clip=true]{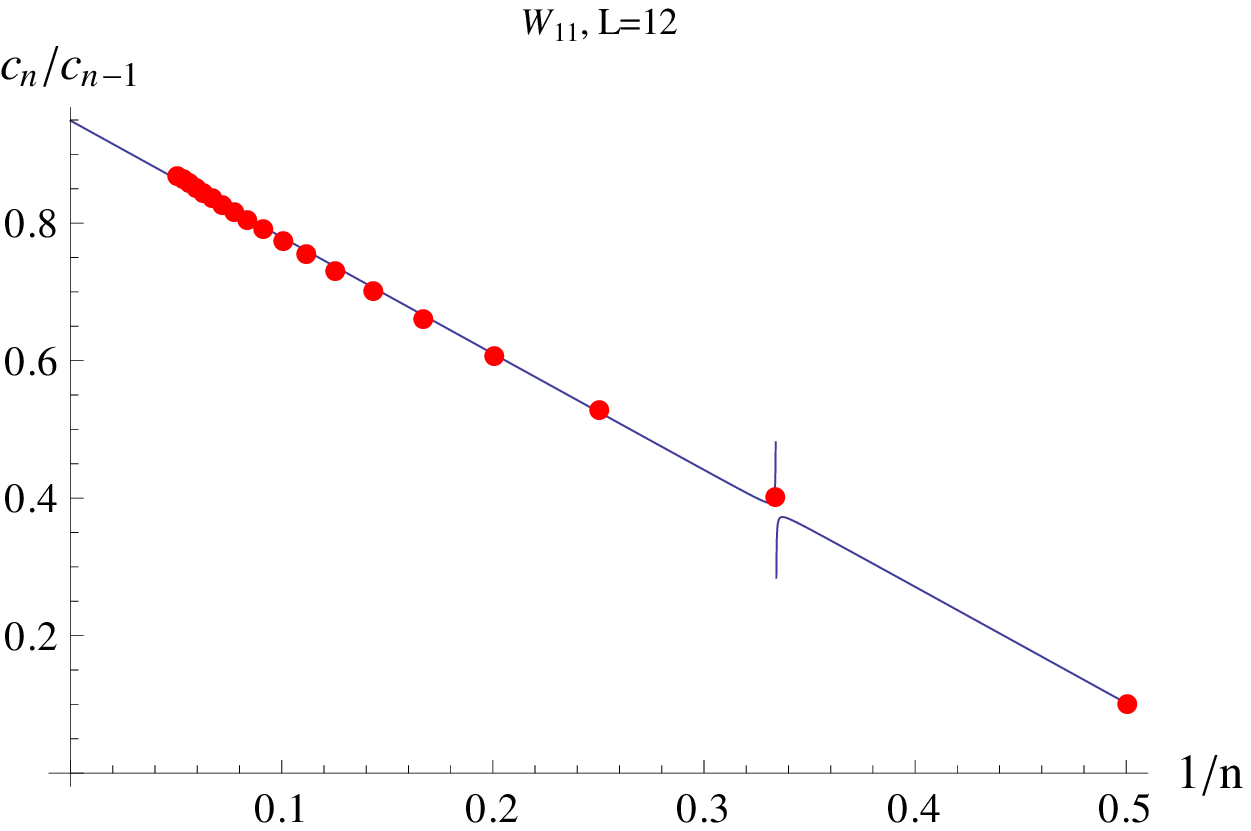}
       &
       \includegraphics[scale=0.55,clip=true]{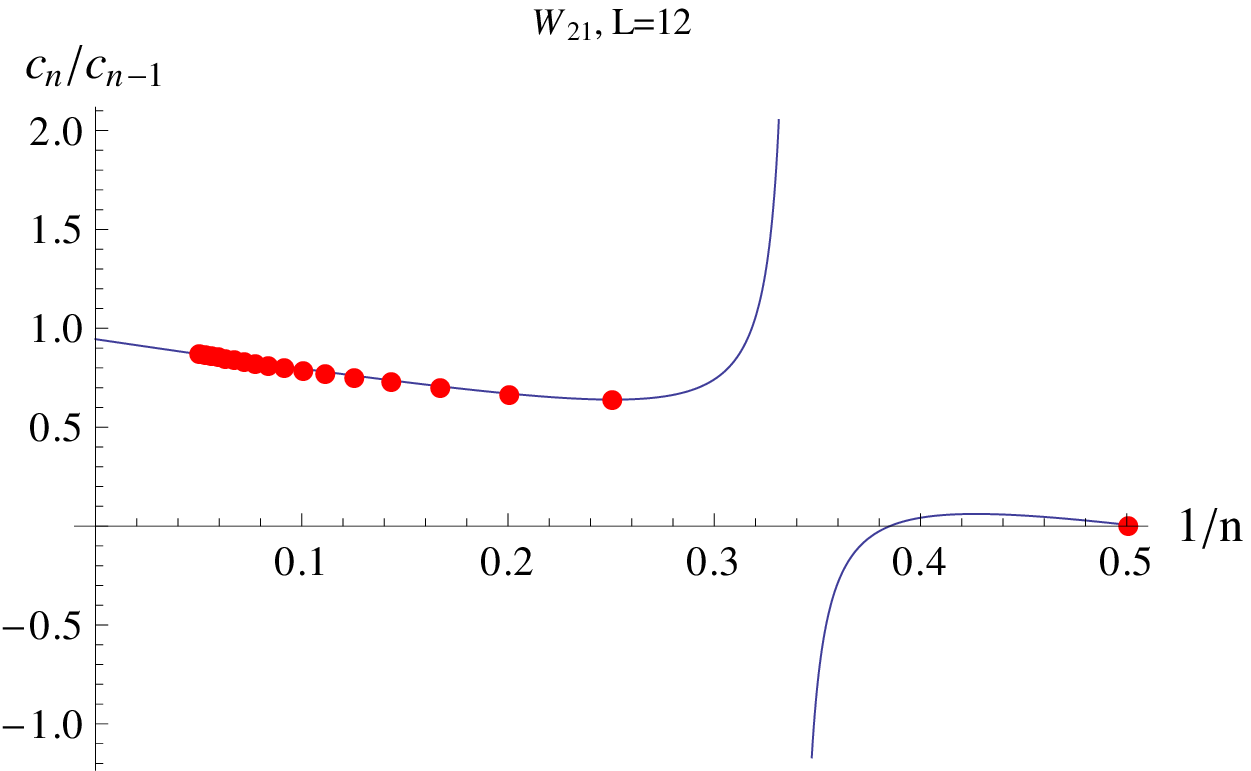}
       \\
       \includegraphics[scale=0.55,clip=true]{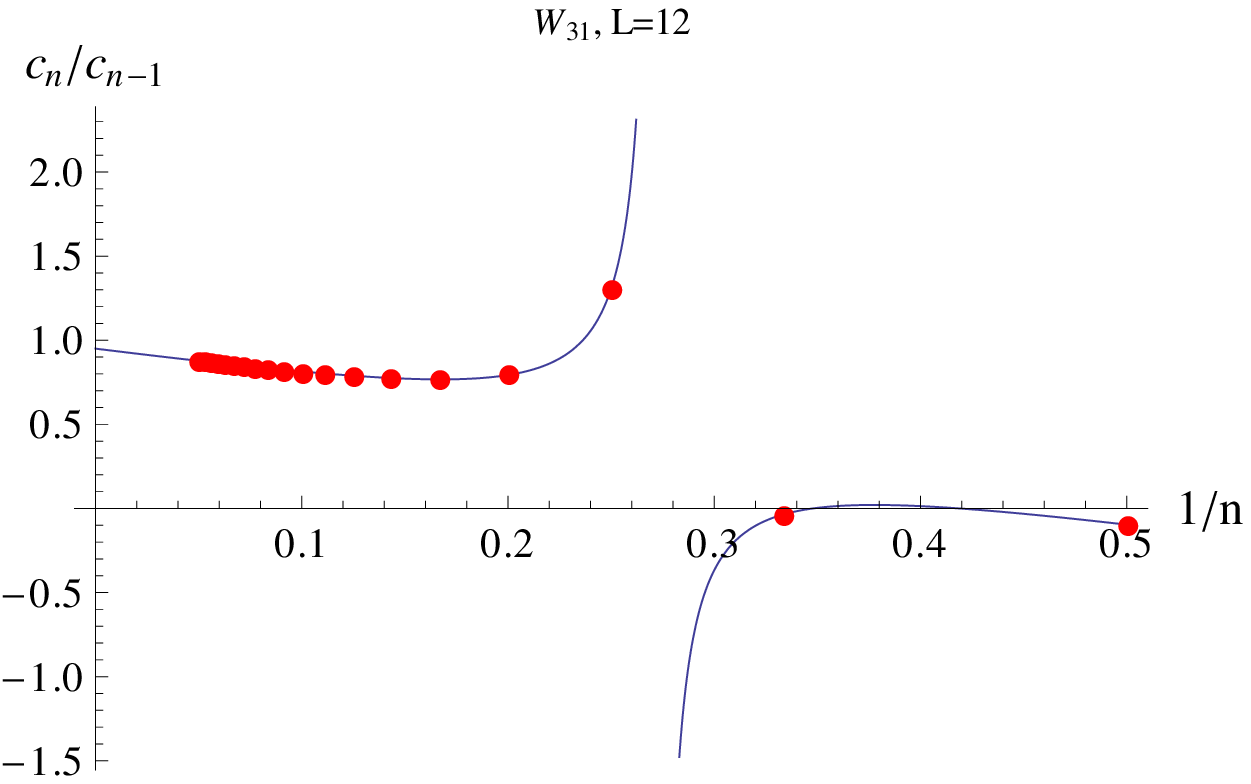}
       &
       \includegraphics[scale=0.55,clip=true]{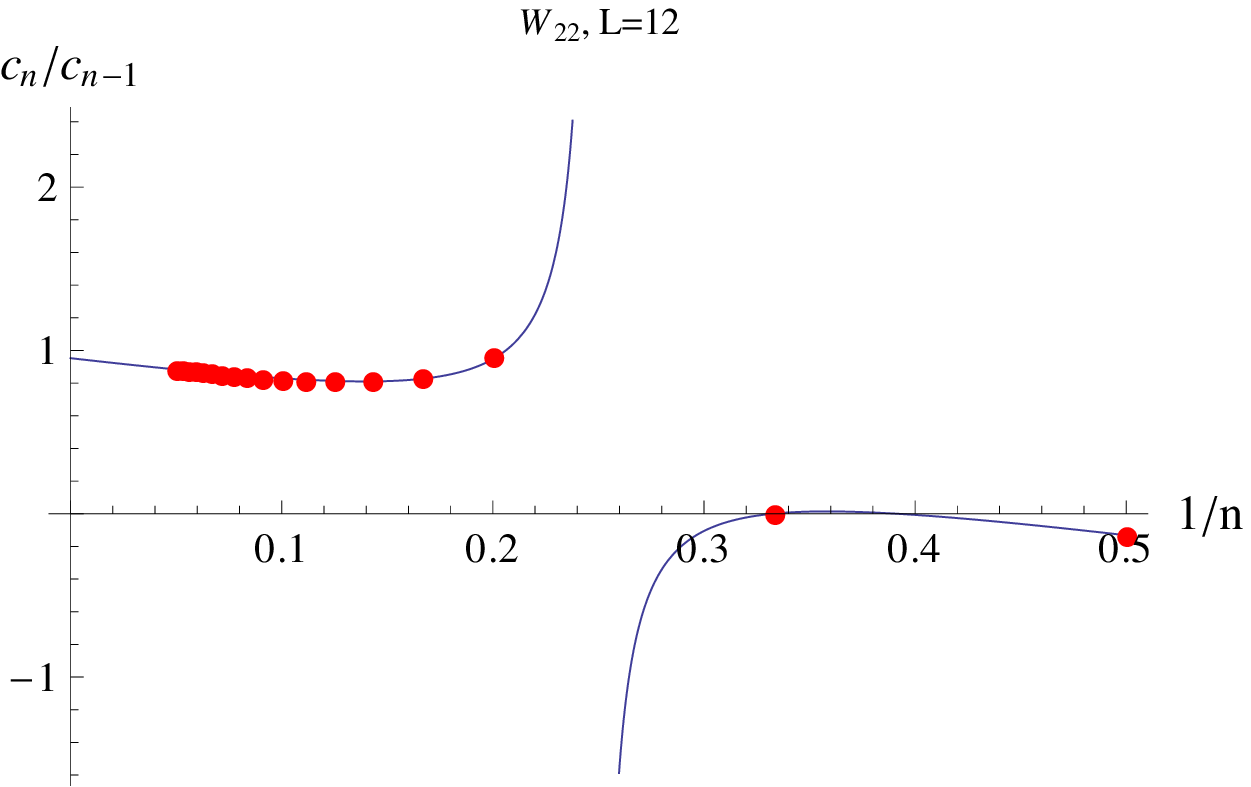}
    \end{tabular}
  \end{center}
  \caption{Domb-Sykes plots for $W_{NM}$ with fits to the parameters in the range $n \in [2,20]$.}
  \label{fig:CoffRats}
\end{figure*}

All the Wilson loops show rather similar behavior at large order $n$,
see Figure~\ref{wijfit}.
At small $n$ they look quite different from the plaquette,
with a mixture of positive and negative terms. 
It is interesting that there is often
a ``notch'' just before the asymptotic region begins, i.e. a
particularly steep drop to a small coefficient, followed by a
jump back up again. This is particularly dramatic in the
$2 \!\times\! 2$ Wilson loop, where the $n=3$ coefficient is about
600 times smaller than the $n=2$ coefficient. 
The notch gives rise to big changes in $r_n$, for example for the  $2 \!\times\! 2$ loop we have
\bea
  &&\dots, \,   r_2 = -0.1319, \, r_3 = +0.0016,
  \nonumber \\
  && \, r_4 = -11.98, \, r_5 = +0.9722, \dots
\eea
The notch corresponds to the singularity in the Domb-Sykes plot.
The anomalously large $r_n$ value occurs when $n$ is close to the 
pole at $n = -s$ in (\ref{eq:HRSratio}). 
This is demonstrated in Figure~\ref{fig:CoffRats}
where the fit to the parameters has been extended to the range $n \in [2,20]$.
Using this fit range one clearly recognizes the corresponding pole terms. 

Our analysis shows that we can reproduce our NSPT data up to order $n=20$ 
for Wilson loops of moderate size (at least the elongated ones) 
with this hypergeometric model sufficiently well. 
This means that we do not find any evidence for a factorial behavior which should result
in a behavior $r_n \sim n$. In Section \ref{subsec:nsptres} we showed that in the range $4 \le L \le 12$
the volume dependence of each individual perturbative coefficient is rather smooth
and already very weak at sizes like $L \approx 12$. 
So we do not expect a significant change extrapolating the results to infinite lattice
size.

Even beyond the apparent radius of convergence $(g_c^2=1/u, \beta_c \approx 5.82) $ the perturbative series still
has some information on the Wilson loops. In that case the terms in the series decrease
initially, before reaching a minimum and then growing. 
Summing the series up to the minimum term would give an approximation to the Wilson loop. 
The minimum term in the series can be estimated from the condition on the ratio 
of neighboring coefficients in (\ref{rn-def}) $r_{n_{\min}} \, g^2=1$.
The corresponding minimal number $n_{\min}$ in the summation is approximately 
(neglecting the parameters $s$ and $p$)
\begin{equation}
  n_{\min} \approx \frac{(1+\gamma) \, u g^2}{u g^2 -1} = 
  \frac{6u \, (1+\gamma)}{6 u - \beta} \approx \frac{12}{\beta_c - \beta}
  \label{eq:nmin}
\end{equation}
for $\beta < 6u$. So, at $\beta=5.7$ we would have to sum about 100 terms before reaching the
minimum (assuming, of course, that the hypergeometric form remains applicable),
and even at $\beta=5.2$ ($g^2=1.15$) we would still have about 20 decreasing 
terms before reaching the minimum term. To stay on the safe side, we have 
not used any data beyond the apparent convergence radius $g_c$ in our analysis of non-perturbative 
Wilson loops described below. We have restricted ourselves to $\beta \ge 5.85$, i.e. $g^2 \le 1.026$.

\subsection{Boosted perturbation theory}
\label{subsec:boosted_PT}

It is well-known that the bare lattice coupling $g$ is a bad expansion parameter 
by virtue of lattice artefacts like tadpoles. There is hope that, by redefining the bare coupling $g$ into a
boosted coupling $g_b$ and the corresponding rearrangement of the series, a better convergence behavior 
can be achieved~\cite{Lepage:1992xa}. For the case of perturbative Wilson loops 
this idea has been applied for the first time by Rakow~\cite{Rakow:2005yn}. 

Let us denote the perturbative Wilson loop summed up to order $n^\star$ using the bare coupling $g$ 
by 
\begin{equation}
  W_{NM}(g,n^\star)= 1 + \sum_{n=1}^{n^\star}\,W_{NM}^{(n)}\,g^{2n}
  \label{eq:Ppert}
\end{equation}
and call in the following any series in $g^2$ a ``naive series''.
We define the boosted coupling as 
\begin{equation}
  g^2_{b}=\frac{g^2}{W_{11}(g,n^\star)}\,.
  \label{eq:gboost}
\end{equation}
The corresponding ``boosted series'' for an arbitrary Wilson loop $W_{NM}$ is then given by
\begin{equation}
  W_{NM,b}(g_b,n^\star)=1+\sum_{n=1}^{n^\star}\,W_{NM,b}^{(n)}\,g^{2n}_{b}
  \label{eq:Pertboost}
\end{equation}
with coefficients $W_{NM,b}^{(n)}$ to be calculated from $W_{NM}^{(k)}$ and $W_{11}^{(l)}$ with $k,l \le n$.
Setting
\begin{equation}
   W_{NM}(g,n^\star)= W_{NM,b}(g_b,n^\star)
   \label{setting} 
\end{equation}
and inserting (\ref{eq:gboost}) into the right hand side of (\ref{setting}), 
we can compute the boosted coefficients $W_{NM,b}^{(n)}$ order by order.

It should be emphasized that the prescribed procedure is done by solving a hierarchical set of recursive 
equations. Especially for large loop orders $n$ these equations involve hundreds or thousands of terms.
Using the NSPT raw data $W_{NM}^{(n)}$ with their errors
gives rise to 
significant numerical instabilities in the boosted result for larger $n$. Therefore, it turned out to be
advantageous to use the  coefficients $W_{NM,hyp}^{(n)}$ (\ref{eq:finsum1})
as  input  for the recursive equations. Using that form up to
loop order $n \le 20$ means that we are smoothing the data of the naive series.
In addition we are in the position to extend the maximal loop order beyond $n=20$.
This leads to a stable numerical result for the boosted coefficients $W_{NM,b,hyp}^{(n)}$.
An additional improvement can be achieved by 
replacing the lowest order perturbative coefficients at $L=12$ by the corresponding coefficients 
of the infinite volume limit. 
In the Appendix we give those numbers for the one- and
two-loop coefficients obtained in the diagrammatic 
approach~\cite{Wohlert:1984hk,Alles:1998is,Athenodorou:2004xt}.

\begin{figure}[!htb]
  \begin{center}
    \includegraphics[scale=0.63,clip=true]{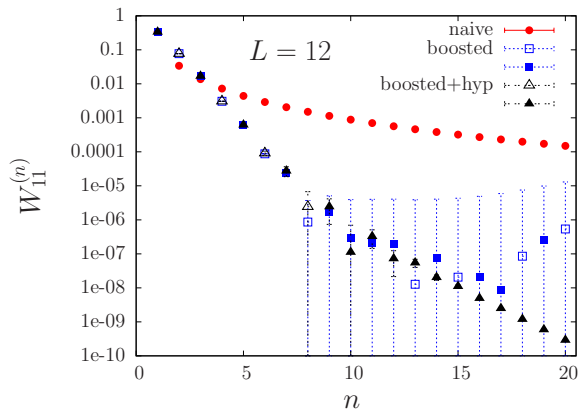}
  \end{center}
  \caption{Comparison of perturbative coefficients for the naive series ($W_{11}^{(n)}$), the boosted series 
           from NSPT raw data ($W_{11,b}^{(n)}$) and the boosted series using the hypergeometric model 
           ($W_{11,b,hyp}^{(n)}$). The boosted coupling (\ref{eq:gboost}) is used,
           positive/negative signs of the coefficients are given by open/full symbols.}
  \label{fig:W11CoeffCompare}
\end{figure}

In Figure~\ref{fig:W11CoeffCompare}
we compare the perturbative coefficients of the plaquette for the NSPT raw data $W_{11}^{(n)}$ with the
$W_{11,b}^{(n)}$ calculated from the raw data and the $W_{11,b,hyp}^{(n)}$.
The boosted coefficients obtained via the model show a smooth decreasing behavior with much
smaller errors than the boosted coefficients based on the raw data,
all the way down to the highest order $n=20$.
The superior result concerning the error is due to the fact that the errors
of the model-fitted coefficients are computed from the correlated
errors of the parameters $(u,\gamma,p,s)$ as discussed in the
preceding section. The errors of the boosted coefficients 
(when constructed from the raw data) are calculated with standard error propagation
through the set of recurrence equations involving thousands of terms.
Since the perturbative plaquette (as the non-perturbative plaquette) 
is less than one,  $W_{11}(g,n^\star) <1$, 
it is clear from (\ref{eq:gboost}) that $g^2_{b} > g^2$. 
On the other hand, we find $|W_{11,b}^{(n)}| \ll |W_{11}^{(n)}|$ 
for $n > 4$ as shown in Figure~\ref{fig:W11CoeffCompare}. We remark
that also the boosted coefficients 
are characterized by oscillating signs as function of the loop order $n$ for smaller $n$
(open vs. filled symbols).

The above mentioned numerical problems relating the boosted series to 
the naive series obtained directly from NSPT
would be less severe if we could start
from a coupling constant which was closer to $g_b$.   
Therefore in~\cite{Rakow:2005yn} 
one of us proposed a simulation with a shifted 
``reference'' coupling constant, $g_{ref}$. Instead of simulating NSPT
with the action~(\ref{eq:WGAction}), we could use the slightly modified action
\bea
   S_{ref}[U]&=& 6 \left( \frac{1}{g_{ref}^2} + \hat{r}_1 + \hat{r}_2 \, g_{ref}^2 \right) \times
  \nonumber \\
  && \sum_P
   \left[ 1 - \frac{1}{6} \mathrm{Tr} \left( U_P + U_P^{\dagger} \right) \right]
   \label{refAction}
\eea
where now $U_P$ is expanded as a power series in $g_{ref}$ rather than $g$.   
Physically, the action is still the usual plaquette action -- all we have
done is to redefine the coupling constant. 
This modified action leads to changes in the drift term of (\ref{eq:Langevin_I}). 
The advantage is that the simulation now gives
us a series for the plaquette in terms of the coupling $g_{ref}$, 
related to the bare coupling by
\begin{equation} 
  \frac{1}{g^2} =  \frac{1}{g_{ref}^2} + \hat{r}_1 + \hat{r}_2 \, g_{ref}^2 \;. 
\end{equation} 
If we choose the parameters $\hat{r}_1$ and $\hat{r}_2$ well, the new intermediate
coupling will be close to the boosted coupling, so the transformation
from $g_{ref} \to g_b$ will be numerically stable and will not introduce large
uncertainties as in the transformation from $g^2\to g^2_b$. 

In~\cite{Rakow:2005yn} simulations have been performed with
$\hat{r}_1 = 1/3, \hat{r}_2 = 0.033911$. These values were chosen 
such that $g_b^2 = g_{ref}^2 + O(g_{ref}^8)$, making the
transformation between the two couplings numerically robust. 
The resulting boosted series is shown in Table~\ref{reftab}. 
\begin{table*}[!htb] 
  \begin{center}
    \begin{tabular}{|c|r@{$.$}l|r@{$.$}l|r@{$.$}l|}\hline
        $n$ & \multicolumn{2}{c|}{$ W_{11,b}^{(n)} $ from (\ref{refAction})} &
        \multicolumn{2}{c|}{$ W_{11,b}^{(n)} $ from NSPT raw data} &
        \multicolumn{2}{c|}{ $ W_{11,b}^{(n)} $ from (\ref{eq:finsum1})}\\ \hline
        1 &$\; -0$&$333334(42) $ &
        \multicolumn{2}{c|}{ $-1/3 \qquad\qquad$} &  \multicolumn{2}{c|}{ $-1/3\qquad$} \\
        2 &$  0$&$077187(30) $&$ 0$&$0772001181(8) $&$ 0$&$0772001181(8)$  \\
        3 &$ -0$&$016817(10) $&$\quad -0$&$0168321(4) $&$ -0$&$0168321(4)$     \\
        4 &$  0$&$0030488(10) $&$ 0$&$0030612(3)   $&$ 0$&$0030612(3)$        \\
        5 &$ -0$&$0006101(14) $&$ -0$&$00061867(9) $&$ -0$&$000620(11)$       \\
        6 &$  0$&$0000831( 7) $&$ 0$&$000087(2)     $&$ 0$&$0000911(14)$      \\
        7 &$ -0$&$00002209(34) $&$ -0$&$000024(2)   $&$ -0$&$0000275(89)$     \\
        8 &$ -0$&$00000007(30) $&$ 0$&$0000009(28)   $&$ 0$&$0000024(43)$      \\
        9 &$ -0$&$00000138(11) $&$ -0$&$0000017(33)   $&$ -0$&$0000024(17)$     \\
        10&$ -0$&$00000042( 8) $&$ -0$&$00000029(360) $&$ -0$&$00000011(58)$    \\
        11&$ -0$&$000000201(12) $&$ -0$&$00000022(380) $&$ -0$&$00000033(18)$     \\
        12&$ -0$&$000000087(14) $&$ 0$&$000000012(3877) $&$ -0$&$000000073(51)$  \\ \hline
    \end{tabular}
  \end{center}
  \caption{Coefficients for the plaquette in boosted perturbation theory, 
           calculated using the modified action (\ref{refAction}) on a $12^4$ 
           lattice. They are compared to the corresponding coefficients from the
           NSPT raw data (second column) and the hypergeometric model data 
           (third column). The loop order $n$ given in the table is restricted by 
           the order used in \cite{Rakow:2005yn}. 
  \label{reftab}}
\end{table*} 
The results are compatible with those found by transforming both the naive 
series from the NSPT raw data and from the hypergeometric model, respectively, but the error
bars are now considerably reduced. In particular, the change in the behavior 
beyond $n=8$, from an alternating series to a single-sign series is
confirmed in this calculation. 
So far we have only applied this method to the
series describing the plaquette, but we expect that it would also be useful 
for the larger Wilson loops.

\begin{figure*}[!htb]
  \begin{center}
     \begin{tabular}{cc}
        \includegraphics[scale=0.63,clip=true]{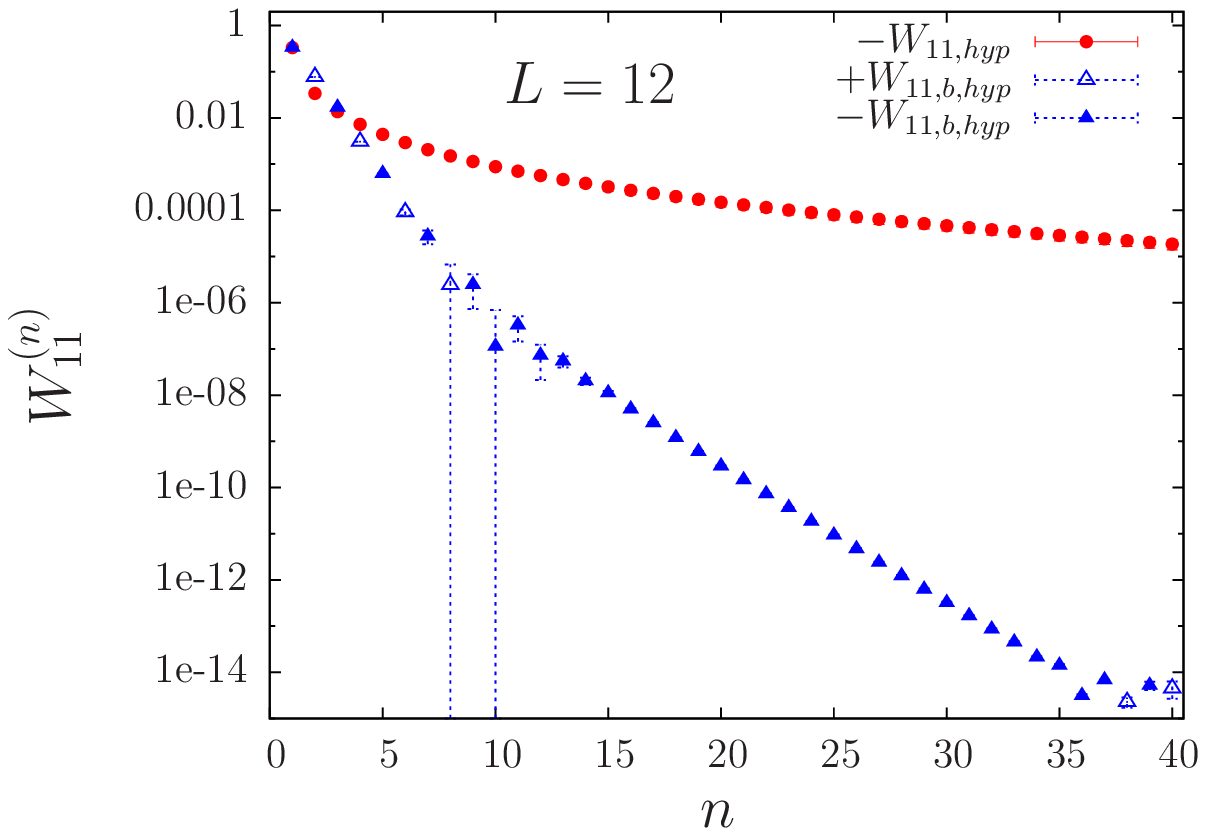}
        &
        \includegraphics[scale=0.63,clip=true]{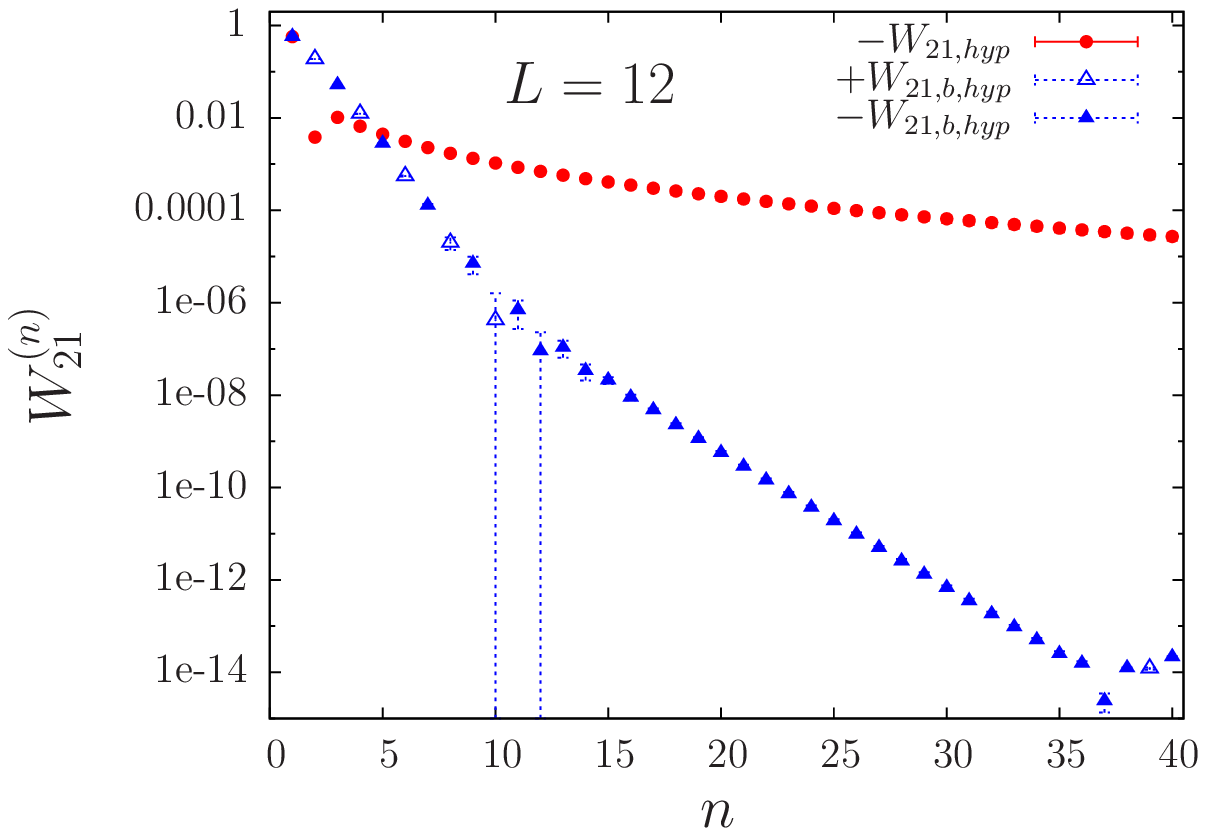}
        \\
        \includegraphics[scale=0.63,clip=true]{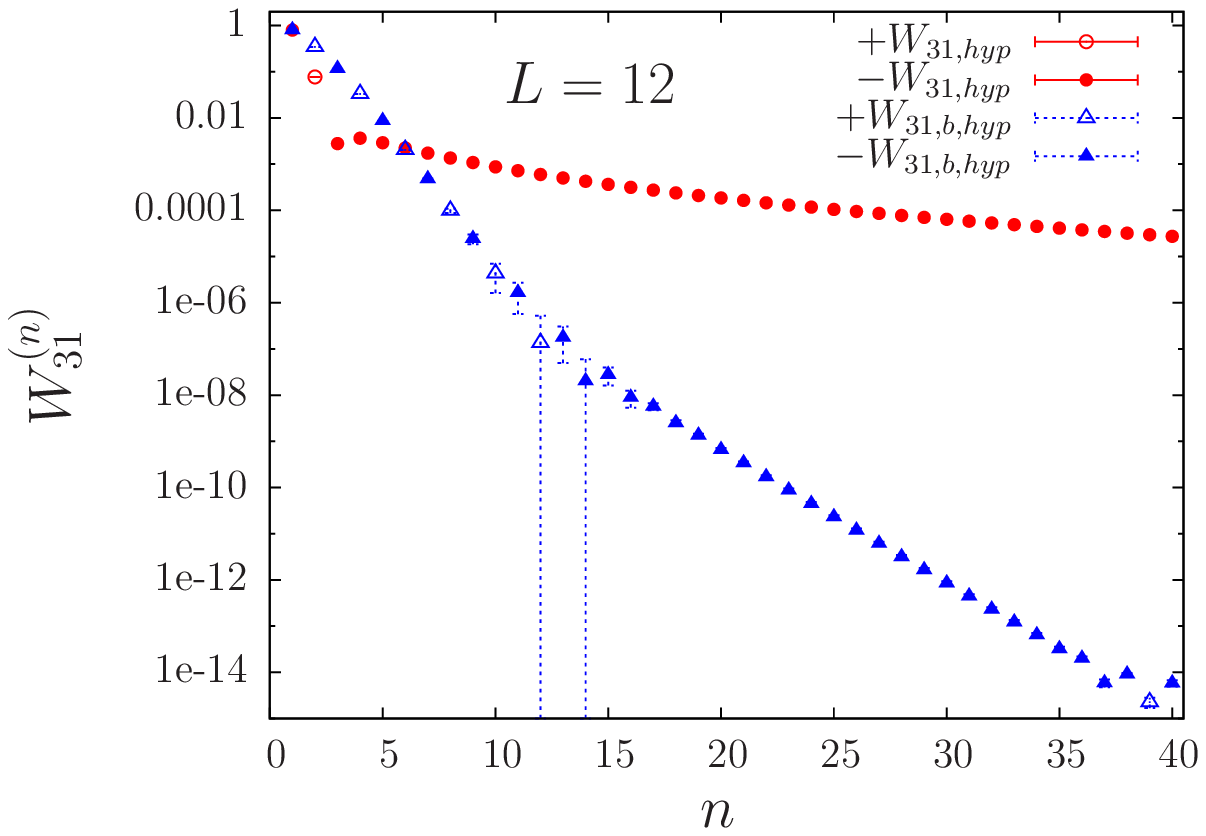}
        &
        \includegraphics[scale=0.63,clip=true]{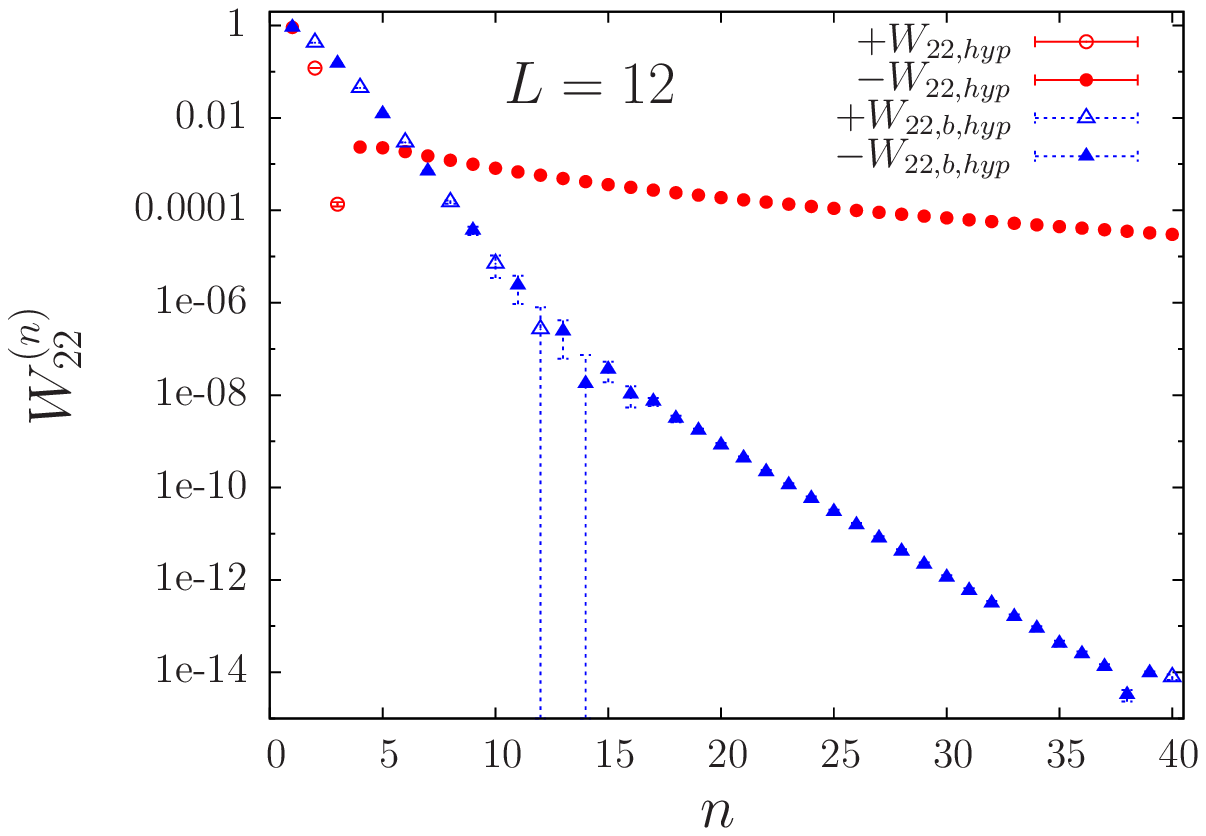}
     \end{tabular}
  \end{center}
  \caption{Coefficients for the naive and boosted series based on the hypergeometric model
           for $W_{11}$, $W_{21}$ , $W_{31}$ and $W_{22}$ as function of the loop order $n$.}
  \label{fig:WijCoeffCompare}
\end{figure*}

The successful hypergeometric model fit to the NSPT raw data (as presented
in the preceding section) and the very smooth behavior of the boosted 
coefficients based on the fit formula (\ref{eq:finsum1res})
allows us to extend the accessible loop order for the
coefficients  both in the naive and boosted series far beyond $n=20$ loops.
In Figure~\ref{fig:WijCoeffCompare} 
the corresponding coefficients for $W_{11}$, $W_{21}$, $W_{31}$ and $W_{22}$ 
are shown throughout the extended range of loop orders 
$n \le 40$ relying on the information contained in the set 
of smoothed data represented by the hypergeometric model.

\begin{figure*}[!htb]
  \begin{center}
     \begin{tabular}{cc}
        \includegraphics[scale=0.63,clip=true]{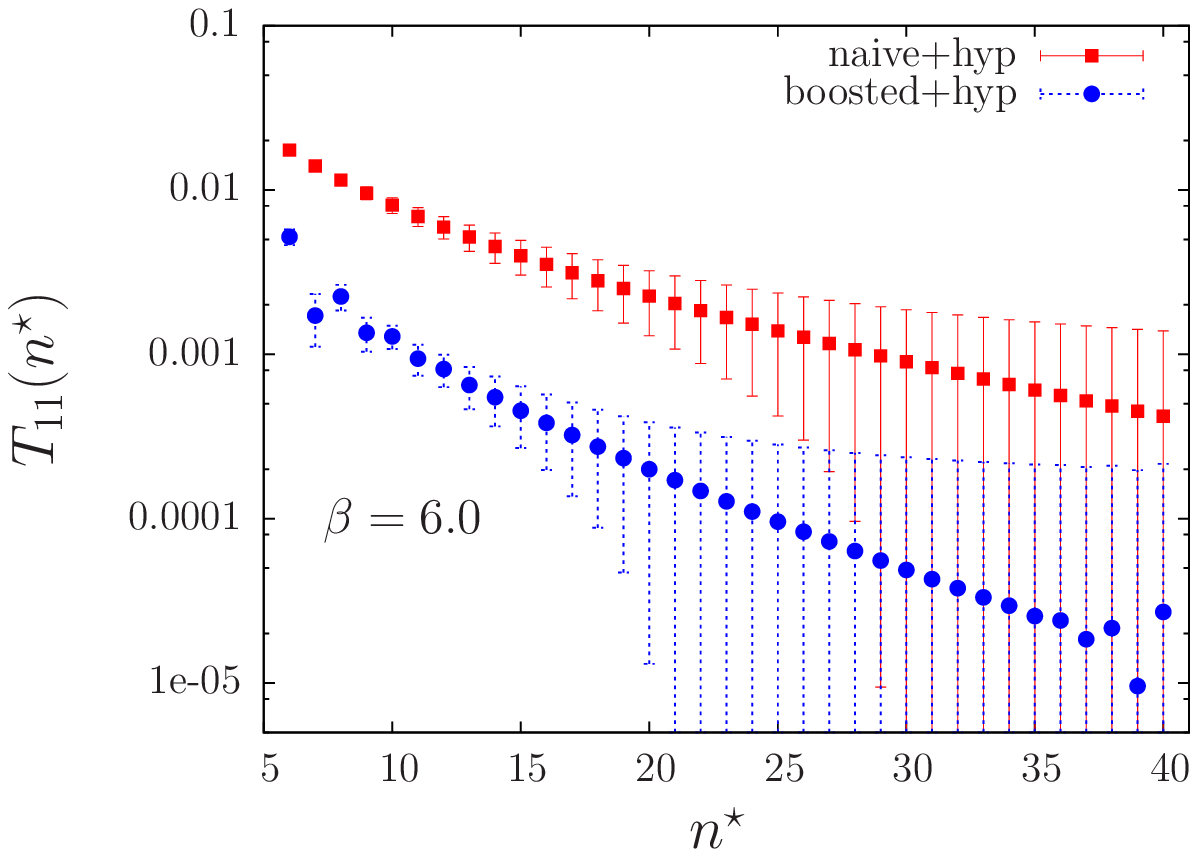}
        &
        \includegraphics[scale=0.63,clip=true]{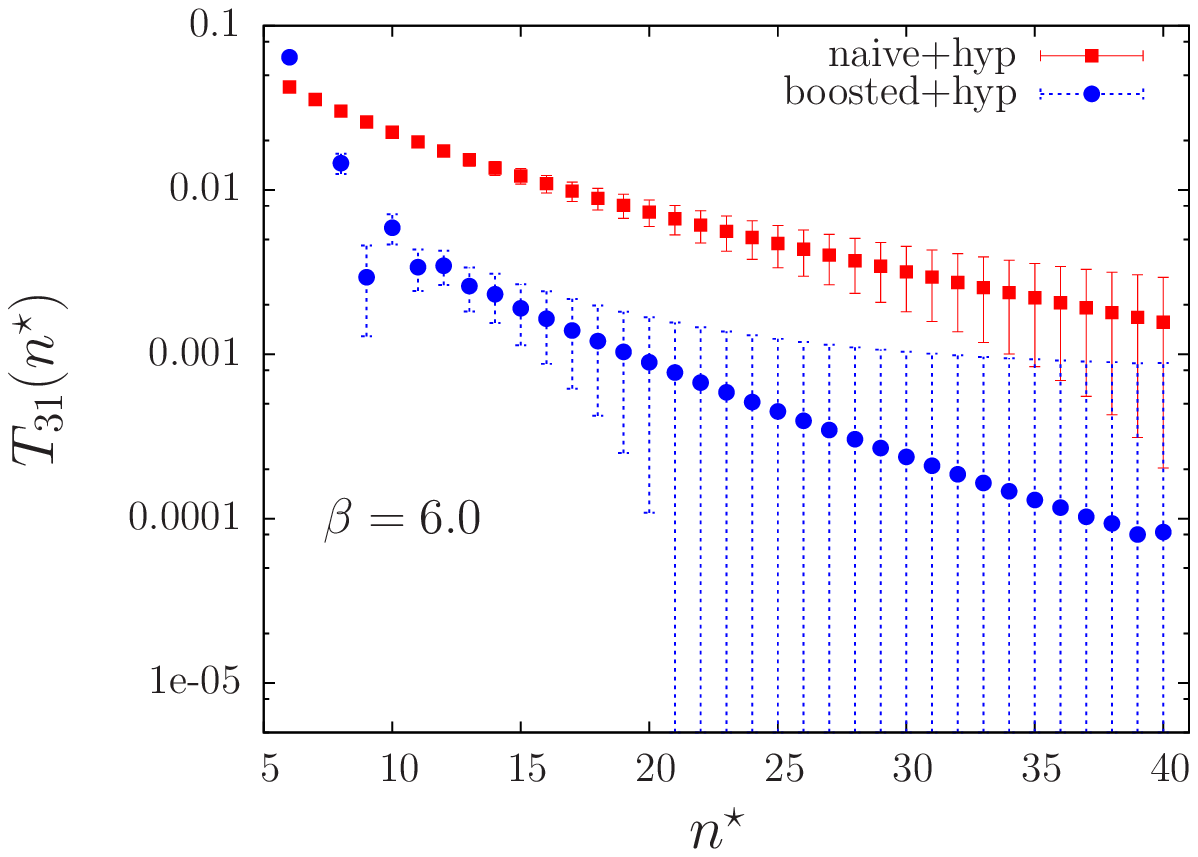}
     \end{tabular}
  \end{center}
  \caption{Truncation errors $T_{NM}(n^\star)$ (\ref{eq:delta}) for $W_{11}$ (left) and $W_{31}$ (right)
           at $L=12$ and $\beta=6$ using the naive and the boosted series on the basis of the hypergeometric model.
           The hypergeometric model values for the total sum are  
           $W^{(n_0=4)}_{11,\infty}=0.59409(8)$ and $W^{(n_0=4)}_{31,\infty}=0.25337(22)$.}
  \label{fig:WNMSumCompare}
\end{figure*}

In Figure~\ref{fig:WNMSumCompare} 
we compare the effect of truncating the sum at order
$n^\star$ for the naive and boosted series, both on the basis of the hypergeometric model. The corresponding truncation
error $T_{NM}(n^\star)$ is defined by
\begin{equation}
  T_{NM}(n^\star) = \frac{\Big|W_{NM}(n^\star)- W^{(n_0)}_{NM,\infty}\Big|}{W^{(n_0)}_{NM,\infty} }\,,
  \label{eq:delta}
\end{equation}
where $W_{NM}(n^\star)$ is either the naive ($W_{NM}(g,n^\star)$) or the  boosted ($W_{NM,b}(g_b,n^\star)$) truncated series.
As the asymptotic value $W^{(n_0)}_{NM,\infty}$ we take 
the hypergeometric sum (\ref{eq:finsum1}) with $n_0=4$ computed at the chosen $g^2=6/\beta=1$. 
Even though part of the decrease in the boosted coefficients is ``eaten" up by
the fact that $g^2_b(=1.6832) > g^2(=1)$, we see that the boosted series is clearly superior.
For example, for $W_{11}$ we have a truncation error $\sim 10^{-3}$ at $10^{th}$
order in the boosted series, but we would have to go nearly to the $30^{th}$
order in naive perturbation theory to achieve the same accuracy.

Figure~\ref{fig:WNMSumCompare} suggests that using the naive perturbative series 
for $W_{11,hyp}(g,n^\star)$ in (\ref{eq:gboost}) to compute $g_b^2$ for a given $g^2$ 
is a poor choice. A much better convergence towards the total perturbative plaquette
is obtained by using the coefficients $W_{11,b,hyp}^{(n)}$.
This suggests to define the  boosted coupling $g_b^2(g^2)$ by solving the implicit equation
\bea
  g^2_{b}=\frac{g^2}{W_{11,b,hyp}(g_b,n^\star)}
  \label{eq:gboostmod}
\eea
where
\bea
   W_{11,b,hyp}(g_b,n^\star)= 1+\sum_{n=1}^{n^\star}\,W_{11,b,hyp}^{(n)}\,g^{2n}_{b} \,.
\eea

\begin{figure}[!htb]
  \begin{center}
     \includegraphics[scale=0.63,clip=true]{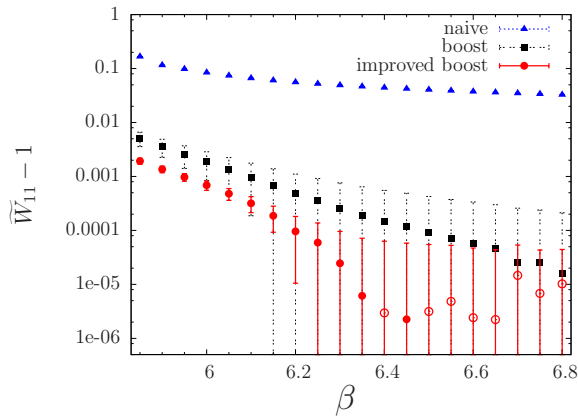}
  \end{center}
  \caption{$\widetilde{W}_{11}(\beta)-1$ as function of $\beta=6/g^2$ for $n^\star=20$:
           ``naive'' with $W_{11,PT}=W_{11}(g,20)$;
           ``boost'' with $W_{11,PT}=W_{11,b}(g_b,20)$  and $g_b^2$ computed from (\ref{eq:gboost});
           ``improved boost'' with $W_{11,PT}=W_{11,b}(g_b,20)$  and $g_b$ computed from (\ref{eq:gboostmod}).
           (Full/open symbols denote positive/negative numbers.) }
  \label{fig:DiffMCPT}
\end{figure}

One essential justification for choosing~(\ref{eq:gboostmod}) is the behavior of
the perturbative series of a Wilson loop for large $\beta$ (small $g^2$)
in comparison to the non-perturbative measurement: in this coupling range the
Wilson loop should be dominated by the perturbative content. 
We introduce the relative difference
\begin{equation}
  \widetilde{W}_{NM}(\beta)-1=\frac{W_{NM,PT}(\beta)-W_{NM,MC}(\beta)}{W_{NM,MC}(\beta)}\,.
  \label{dWNMPTMC}
\end{equation}
where the index ``PT'' stands for the perturbative value of the Wilson loop and  ``MC'' denotes the Monte Carlo result.
This quantity should tend to zero for large $\beta$. 
In Figure \ref{fig:DiffMCPT}
we plot $\widetilde{W}_{11}(\beta)-1$ as function of $\beta$.
The $\beta$ dependence  clearly shows that  the boosted coupling computed from~(\ref{eq:gboostmod})
gives the best behavior for the small $g^2$ where the plaquette from that perturbative series practically coincides
with the Monte Carlo value.  Wilson loops with larger loop sizes show
a similar behavior.

Note that the definitions of the boosted coupling using either (\ref{eq:gboostmod})
or (\ref{eq:gboost}) are calculated from a perturbative input exclusively.
Using boosting in standard Monte Carlo measurements, the boosted coupling $g^2_b$
is defined by dividing the bare coupling squared $g^2$ by the measured plaquette at given
$\beta$ value. 
Numerically, this coupling constant behaves in a similar way as that 
obtained from (\ref{eq:gboostmod}). 
This is another argument to use expression (\ref{eq:gboostmod}) as definition for the boosted 
coupling in perturbation theory.
 
\section{The non-perturbative part of Wilson loops}
\label{sec:GG}

\subsection{Reliability of high order lattice perturbation theory}
\label{subsec:GG1}

There is much debate to which extent high order lattice perturbation
theory can be trusted and how its results can be used to extract
physical quantities. In~\cite{DiRenzo:1996xd} the authors have investigated
the influence of the finite volume on the possibility to
find infrared renormalons. Using the steepest descent ($sd$)
method, they deduce an upper bound on the order of perturbation theory
$n^{sd}$ above which possible infrared effects are tamed for dimension four operators
\begin{equation}
  n < n^{sd} \approx 4\, \log L + c\,,
  \label{eq:nconstr}
\end{equation}
where $L$ is the lattice size. However, it is difficult to determine
the value of $c$ -- in~\cite{DiRenzo:1996xd}  it was
estimated as $c = \mathcal{O}(1)$. 

As shown in the preceding Section \ref{subsec:boosted_PT} we found that boosted
perturbation theory using the raw NSPT coefficients in the range
$1 \le n \le 12$ gives already reliable results for the summed series. Furthermore,
from the discussion at the end of Section \ref{subsec:nsptres} (see Figure \ref{fig:W11InfPlaquette})
we feel confident that the finite size effects are under control -- which would not be
the case if there are infrared effects.

On finite lattices one cannot expect renormalons because of hard 
ultraviolet ($k<1/a$) and infrared ($k \ge 2\pi/L\,a$) cut-offs.  
However, one might expect quadratic and quartic divergences.
For the plaquette $W_{11}$ one could write 
(see, e.g.~\cite{Martinelli:1996pk})
\begin{equation}
  W_{11}=C_1(a\,Q) \, \langle \mathds{1} \rangle + 
  C_2(a\,Q) \, a^4 \,\langle GG \rangle\,,
\end{equation}
with $\langle GG \rangle$ denoting a condensate of dimension four.
There could be a mixing between operators $\mathds{1}$  and $GG$
which would result in an $a^4$-contribution to $C_1$:
\begin{equation}
  C_1(a\,Q) = C_1^{0}(a\,Q) + C_1^{4}(a\,Q)\,(aQ)^4\,.
\end{equation}
The coefficients $C_1^{i}(a\,Q)$ themselves diverge at most as powers of
$\log(aQ)$. The existence of a quartic divergence would spoil the determination
of the condensate. This type of divergence is connected to a pole in the Borel transform of the corresponding, 
assumed divergent perturbative series with a factorial growth of the expansion coefficients~\cite{Beneke:1998ui}. 
We do not observe such a factorial growth up to loop order $n = 20$.
This is a fact, which we have to accept and 
appreciate theoretically~\cite{Suslov:2005zi,Zakharov:2010tx}.

\subsection{Ratios of Wilson loops}
\label{subsec:GG2}

A precise separation of the non-perturbative part of Wilson loops from the 
corresponding quantities measured on the lattice requires a perturbative 
computation to very high order. From the discussion in Section \ref{subsec:boosted_PT}
it is clear that boosted perturbation theory provides an optimal tool for that.
We use the version of boosting including the hypergeometric model to smooth the NSPT bare coefficients 
and go beyond loop order $n=20$. The boosted coupling is computed from (\ref{eq:gboostmod}) with $n^\star=40$.
Additionally we restrict ourselves to moderate loop sizes which ensures that the boosted
coefficients can be determined with sufficient accuracy.

Let us introduce generic ratios of powers of Wilson loops (together with their 
boosted perturbative expansion) as
\begin{equation}
  R^{k,m}_{NM,N^\prime M^\prime}=\frac{(W_{NM})^k}{(W_{N^\prime M^\prime})^m}
  =\sum_{n} \,[R^{k,m}_{NM,N^\prime M^\prime}]^{(n)} \, g_b^{2n}\,.
  \label{eq:ratioR1}
\end{equation}
In most of the following examples we restrict ourselves to reference loops
of size $N^{\prime}=M^\prime=1$ (plaquette) and integer powers $k,m>0$.
A generalization to larger  $N^\prime,M^\prime$ and also to non-integer powers $k$ and $m$ can
be easily performed.

We consider now the particular ratios $R^{1,2}_{21,11}$ and $R^{1,3}_{31,11}$. 
They fulfill the area relation
\begin{equation}
  k\,\times\,\mathcal{S}_{NM}=m\,\times\,\mathcal{S}_{N^\prime M^\prime}\,,
  \label{eq:ratioR2}
\end{equation} 
where $\mathcal{S}_{NM}$ is the area of the Wilson loop $W_{NM}$ -- in our case
of planar rectangular loops we have  $\mathcal{S}_{NM} = N \!\times\! M$.
{}From considerations of naturalness we would expect the convergence behavior 
of these types of ratios to be better than other ratios that are not constrained by the
area relation (\ref{eq:ratioR2}). 
We first compare the perturbative coefficients of these ratios
with the corresponding coefficients of Wilson loops $W_{NM}^{(n)}$.
\begin{figure*}[!htb]
  \begin{center}
     \begin{tabular}{cc}
        \includegraphics[scale=0.63,clip=true]{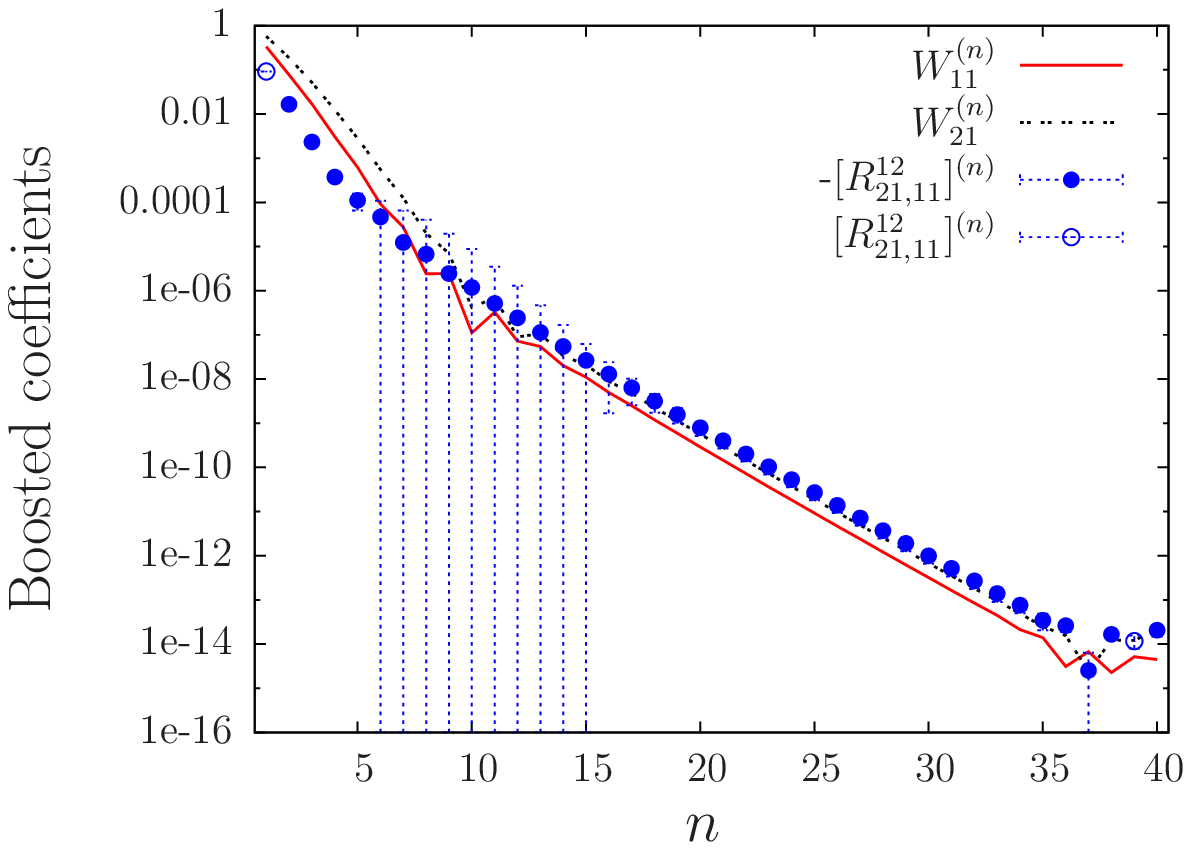}
        &
        \includegraphics[scale=0.63,clip=true]{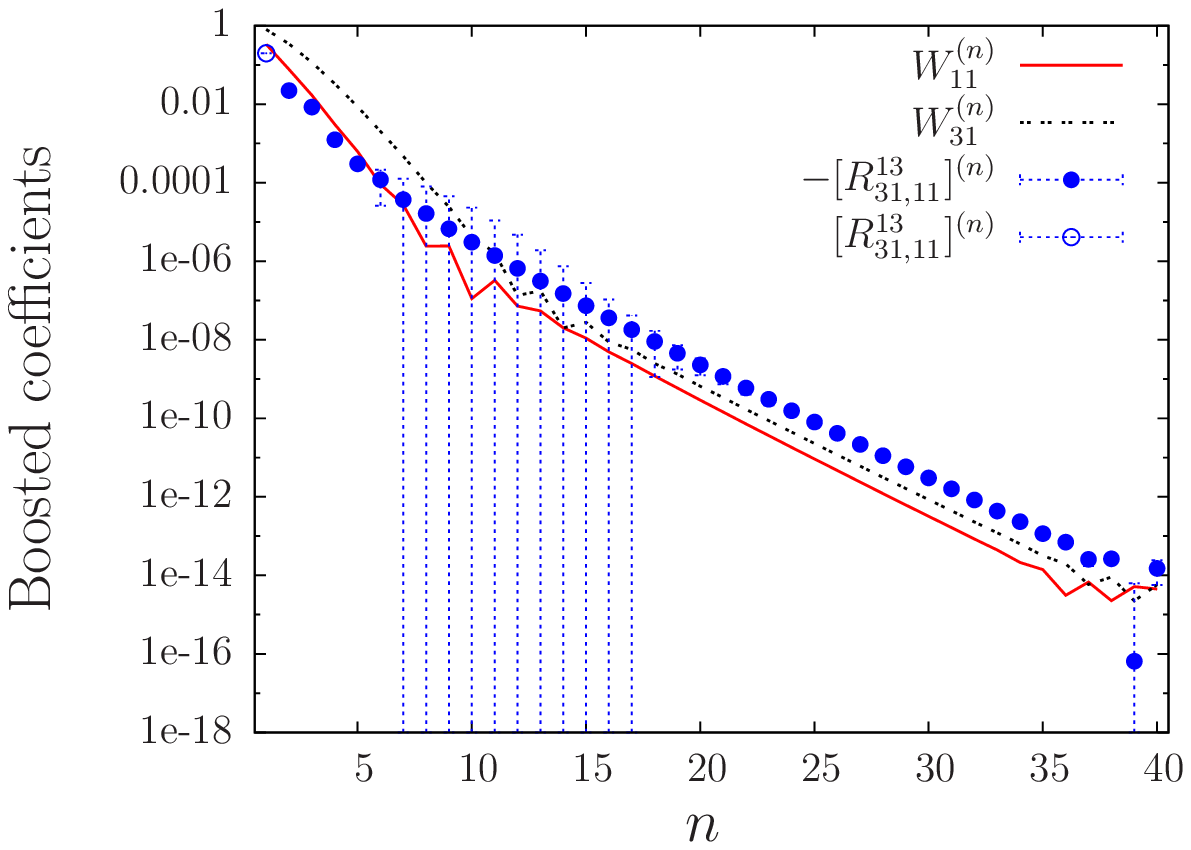}
     \end{tabular}
  \end{center}
  \caption{Boosted coefficients of ratios (left: $[R^{12}_{21,11}]^{(n)}$, right: $[R^{13}_{31,11}]^{(n)}$)
           defined in (\ref{eq:ratioR1}). The thin lines show the corresponding boosted coefficients
           for the $1\!\times\!1$ and $2\!\times\!1$ (left) and  $1\!\times\!1$ and $3\!\times\!1$(right) 
           Wilson loops, respectively.}
  \label{fig:RatioCoeffs1}
\end{figure*}
Figure~\ref{fig:RatioCoeffs1} shows that the coefficients of the ratios behave similar 
to the coefficients of the Wilson loops (shown for comparison as thin lines without errors) themselves.
 
Now we define the quantity
\begin{equation} 
  \Delta_\mathcal{A}=\mathcal{A}_{PT}-\mathcal{A}_{MC} \,,
  \label{eq:ratioR3}
\end{equation} 
where $\Delta_\mathcal{A}$ is then the non-perturbative value of the quantity $\mathcal{A}$
and the ratio
\begin{equation}
  \widetilde{\mathcal{A}}=\frac{\mathcal{A}_{PT}}{\mathcal{A}_{MC}} \,.
  \label{eq:ratioR40}
\end{equation}
In the case of Wilson loops $\Delta_\mathcal{A}>0$ and $\Delta_\mathcal{A}\ll \mathcal{A}_{PT}$.
Since we know the non-perturbative piece to be much smaller than 
the perturbative one we can expand $\widetilde{\mathcal{A}}$ 
in powers of $\Delta_\mathcal{A}$.
To first order we have 
\begin{equation}
  \widetilde{\mathcal{A}}\simeq 1+\frac{\Delta_\mathcal{A}}{\mathcal{A}_{PT}}\,.
  \label{eq:ratioR4}
\end{equation}
Applying this expansion taking in place of $\widetilde{\mathcal{A}}$ the ratios $\widetilde{R}$ 
for the $R$ introduced in (\ref{eq:ratioR1}) we have 
\begin{equation}
  \widetilde{R}^{k,m}_{NM,N^\prime M^\prime} \simeq 1+k\,\frac{\Delta_{W_{NM}}}{W_{NM,PT}}-
  m\,\frac{\Delta_{W_{N^\prime M^\prime}}}{W_{N^\prime M^\prime,PT}}\,.
  \label{eq:ratioR5}
\end{equation}

\begin{figure}[!htb]
  \begin{center}
     \includegraphics[scale=0.63,clip=true]{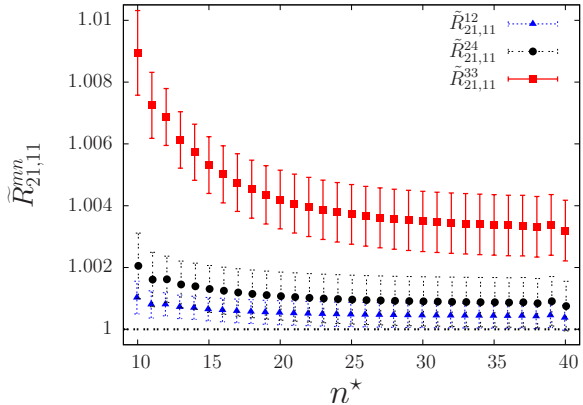}
  \end{center}
  \caption{$\widetilde{R}^{k,m}_{21,11}$ for $(k,m)=(1,2), (2,4)$ and $(3,3)$
           as function of loop order $n^\star$ up to which the ratio is summed up.}
  \label{fig:Ratio1}
\end{figure}

In Figure~\ref{fig:Ratio1} 
we show an example for some ratios $\widetilde{R}^{k,m}_{NM,N^\prime M^\prime}$
at $\beta=6$. We have used our own Monte Carlo measurements of Wilson loops generated at the same lattice size~\cite{ISP}.
One recognizes that for large $n^\star$ the ratios tend to $\widetilde{R}^{k,m}_{NM,N^\prime M^\prime} \simeq 1$.
For smaller powers $m$ and $k$ this behavior is more pronounced. 
Additionally, one finds that the ``non-natural" choice
$(k,m)=(3,3)$ leads to a significantly different behavior.  
Thus, Figure~\ref{fig:Ratio1} strongly suggests
to use powers $(k,m)$ which obey the area relation (\ref{eq:ratioR2}).

Using for $A$ in the $\widetilde A$ definition (\ref{eq:ratioR40}) the quantity 
${R}^{k,m}_{NM,N^\prime M^\prime}$ (\ref{eq:ratioR1}) one can easily
derive a formula to determine the ``deviation from perturbation theory'',
$\Delta_{W_{NM}}$, for a $N\!\times\!M$ Wilson loop as
\bea
  \label{eq:deltaWNM1}
  &\Delta_{W_{NM}}(W_{N^\prime M^\prime}) = 
  \\
   &
  \left[1-\exp \left( -\frac{d}{dk}\, 
   \log \left(\widetilde{R}^{k,m}_{NM,N^\prime M^\prime}  \right) \right)\right]\, W_{NM,PT}\,,
  \nonumber
\eea
where we made explicit the 
dependence of $\Delta_{W_{NM}}$ on the reference loop $W_{N^\prime M^\prime}$.
Values of $(N,M,k)$ and $(N^\prime M^\prime,m)$ are related  by (\ref{eq:ratioR2}).
Inserting the boosted perturbative series for $W_{NM,PT}$ and the Monte Carlo 
measured values $W_{NM,MC}$
for various values of the inverse coupling $\beta$ into (\ref{eq:deltaWNM1}) one obtains rather
easily the desired $a$-dependent non-perturbative part $\Delta_{W_{NM}}(a)$ of  $W_{NM}$ using
one's favorite known relation $\beta(a)$.

\subsection{Condensate of dimension four on the lattice}
\label{subsec:GG3}

One special case of the non-perturbative part of Wilson
loops is $\Delta_{W_{11}}=W_{11,PT}-W_{11,MC}$ which is directly connected
to the gluon condensate  introduced in~\cite{Shifman:1978bx}.
There is a commonly used relation between the Monte Carlo measured
plaquette and its perturbative counterpart
\begin{equation}
  W_{11,MC} = W_{11,PT} - a^4 \frac{\pi^2}{36}\left(\frac{-b_0 g^3}{\beta(g)}  \right) \langle 
  \frac{\alpha}{\pi} GG \rangle \,,
  \label{eq:GG1}
\end{equation}
which defines the gluon condensate $\langle \frac{\alpha}{\pi} GG \rangle$ on the
lattice\footnote{In (\ref{eq:GG1}) $\beta(g)$ denotes the standard $\beta$-function with $b_0$ being its
leading coefficient.}.
In contrast to (\ref{eq:deltaWNM1}), relation (\ref{eq:GG1}) allows us to determine
the gluon condensate from the $1 \!\times\! 1$  Wilson loop only.
An alternative could be to find $\Delta_{W_{11}}$  from (\ref{eq:deltaWNM1})
choosing a suitable reference Wilson loop.
As discussed in Section \ref{subsec:GG1} this is strictly valid only in the absence
of renormalon ambiguities which is assumed to be the case in the following.

In (\ref{eq:GG1}) it is assumed that there is only a single, 
non-perturbative quantity of dimension four contributing to the plaquette.
It has been speculated~\cite{Burgio:1997hc} that in the difference between the
perturbative and the lattice Monte Carlo plaquette also an $a^2$-contribution
might be present. That difference depends on $n^\star$
denoting the truncation of the perturbative series 
as expressed by the $n^\star$ dependence of the corresponding coefficients:
\bea
  \Delta_{W_{11}}(n^\star) &=& W_{11,PT}(n^\star) - W_{11,MC} 
  \nonumber \\
  &=&  c_2(n^\star) \,a^2+ c_4(n^\star)\, a^4\,.
  \label{eq:GG12}
\eea
In~\cite{Narison:2009ag}  Narison and Zakharov have presented arguments that a non-zero value of the coefficient 
$c_2(n^\star)$ is an artefact due to the truncation -- above some value of
$n^\star$ that coefficient should vanish.

For the estimate of the gluon condensate 
we are in the position to take the most precise perturbative
values available - in our computation these are the summed series based on hypergeometric functions ($n^\star\to \infty$)
given in (\ref{eq:finsum1res}) with the  parameters of Table \ref{tab:FitErrTab}.
So we can ask the question, whether there is a significant $a^2$-dependence 
for the non-perturbative
parts $\Delta_{W_{NM}}$ derived from (\ref{eq:deltaWNM1}) making a corresponding ansatz as in (\ref{eq:GG12}).

To find the dependence of the non-perturbative part on the lattice spacing $a$, we consider the lattice coupling region
$\beta_{\min} \le \beta \le \beta_{\max}$.
$\beta_{\min}=5.85$ is determined by the convergence radius  of the perturbative series. 
In the analysis we have used non-perturbative Wilson loops from the same lattice size as the largest NSPT lattice and  
have chosen $\beta_{\max}=6.3$.
To relate the different lattice couplings $\beta$ to $a/r_0$, where $r_0$ is the Sommer scale, we use~\cite{Necco:2001xg}.

\begin{figure*}[!htb]
  \begin{center}
     \begin{tabular}{cc}
        \includegraphics[scale=0.62,clip=true]{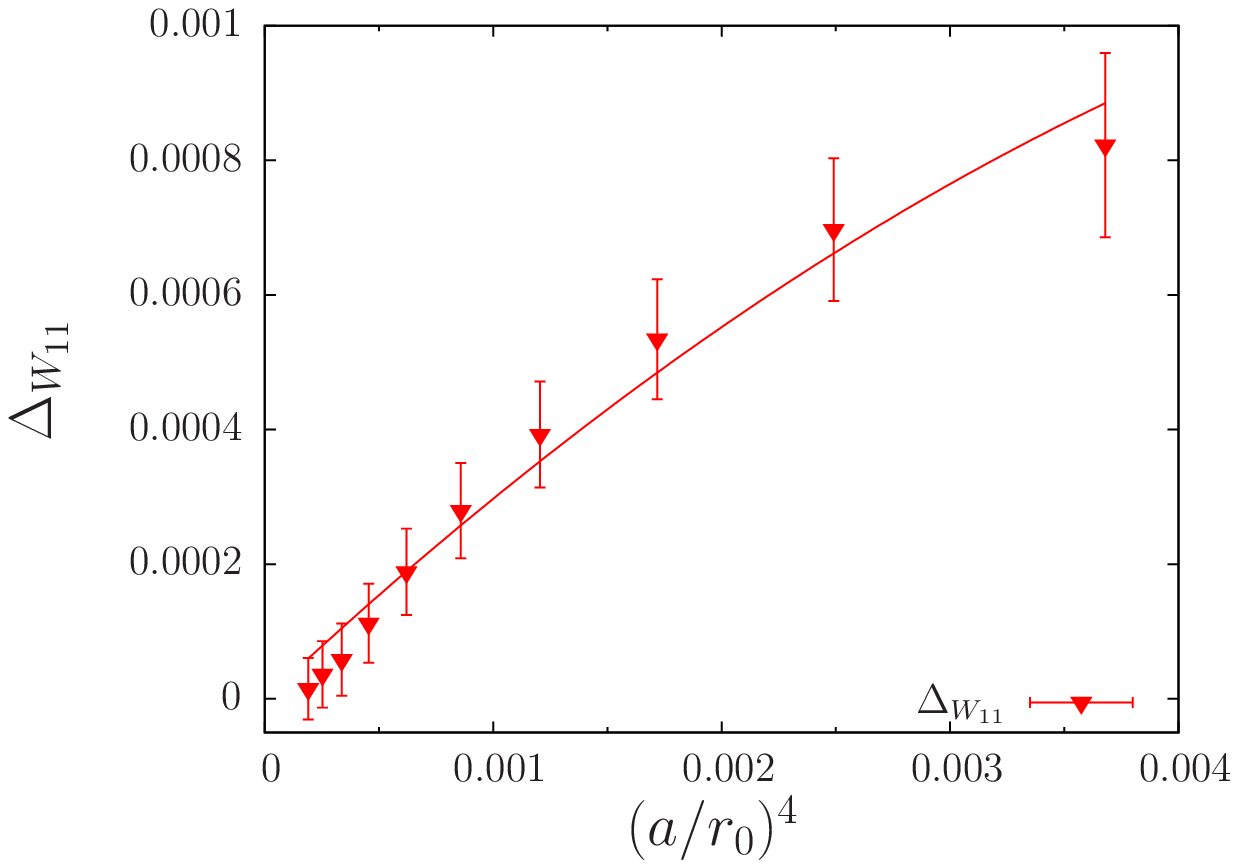}
        &
        \includegraphics[scale=0.62,clip=true]{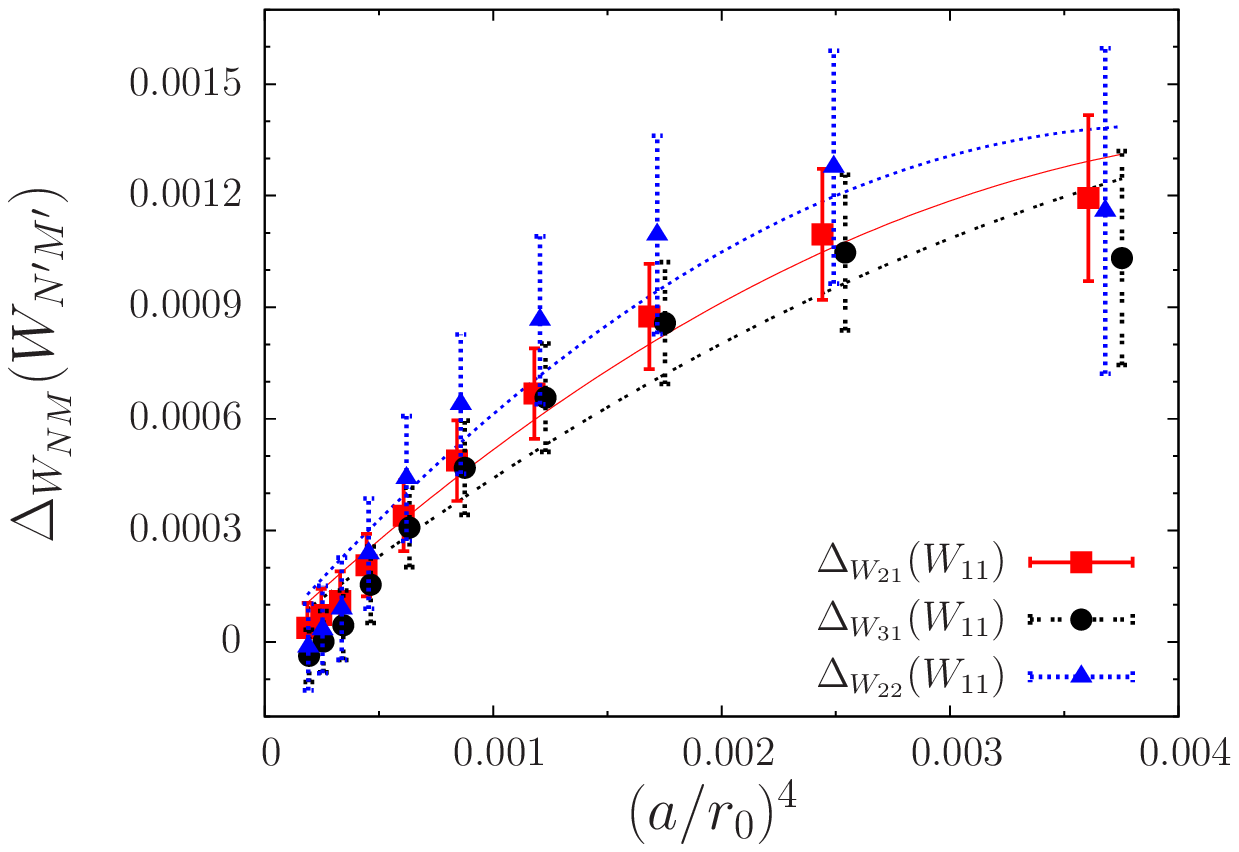}
   \end{tabular}
 \end{center}
  \caption{$\Delta_{W_{11}}$ (left) and $\Delta_{W_{NM}}$ (right) as function of $a^4$ 
           together with their corresponding fits assuming $a^4$ and $(a^4)^2$ contributions.}
  \label{fig:W11a4}
\end{figure*}

In the left of Figure \ref{fig:W11a4} we show $\Delta W_{11}(a)$ as function of $a^4$.
One observes that there is not much room for an
additional $a^2$-dependence. 
On the other hand, we find a significant bending for larger $a (g^2)$ which can
be parametrized as an $(a^4)^2$ correction term.
This might be a sign of breaking scaling on the coarsest lattices, 
or it could be the signature of  higher-dimensional condensates
considered in~\cite{Shifman:1980ui}. 
That correction is relatively small for  $\Delta_{W_{11}}$.
For larger Wilson loops we find this deviation from a pure $a^4$-dependence 
more pronounced as shown in the right of Figure \ref{fig:W11a4}.
We should mention that, 
using the summed perturbative series of the hypergeometric model,
the non-perturbative parts $\Delta_{W_{NM}}$ 
are independent of the choice of the reference loops (as indicated 
in (\ref{eq:deltaWNM1})) and also agree for the plaquette case
with the simple subtraction scheme (\ref{eq:GG1}).

\begin{figure}[!htb]
  \begin{center}
        \includegraphics[scale=0.63,clip=true]{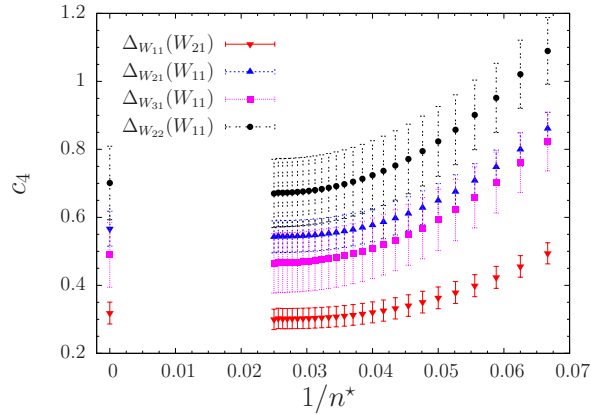}
  \end{center}
  \caption{Coefficient $c_4$ as function of the inverse loop order $1/n^\star$ for different Wilson loops.
           The data points at ``$1/n^\star=0$'' represent the series summed to infinity using the hypergeometric model.} 
  \label{fig:C4dWNM}
\end{figure}

In Figure \ref{fig:C4dWNM}
we plot $c_4(n^\star)$ for various Wilson loops.
One recognizes a pronounced plateau for $n^\star > 30$. In Table \ref{c4tab} we give the values
of the coefficients $c_4$ both for the boosted series summed up to $n^\star=40$ and
as obtained from the infinite series, respectively.
On dimensional grounds one would expect that $c_4$ would be approximately proportional to the 
square of the Wilson loop area~\cite{Shifman:1980ui}. 
From Table~\ref{c4tab} we do see an increase
 \begin{table}[!htb] 
   \begin{center}
      \begin{tabular}{|c|r|r|}\hline
         &  $c_4$ from boosting       & $c_4$ from the hypergeometric \\ 
         &  $(n^\star=40)$            &                         model \\ \hline 
         $W_{11}$  & $0.30(3)$        & $0.31(3)$      \\
         $W_{21}$  & $0.54(5)$        & $0.56(5)$   \\
         $W_{31}$  & $0.47(9)$        & $0.49(10)$        \\
         $W_{22}$  & $0.67(10)$       & $0.70(11)$          \\ \hline
      \end{tabular} 
   \end{center}
   \caption{Coefficients $c_4$ for the Wilson loops $W_{NM}$ obtained from boosted perturbation theory
   up to $n^\star=40$ and from the series summed to infinity using the hypergeometric model.} 
   \label{c4tab}
 \end{table} 
in $c_4$, but it is much slower than area squared (in fact the $a^4$ 
term in the $3 \!\times\! 1$ loop is smaller than the $2 \!\times\! 1$ loop, though
the error bars overlap). 

Introducing the Sommer scale $r_0$, 
a physical value for the condensate can be extracted from the coefficient $c_4$.
If we  approximate $\left(\frac{-b_0\,g^3}{\beta(g)}\right) \sim 1 $
in (\ref{eq:GG1}), we extract from $\Delta_{W_{11}}$ the gluon condensate 
as given in Table \ref{tab:GG}.
\begin{table}[!htb]
  \vspace{0.5cm}
  \begin{center}
    \begin{tabular}{|l |c |c |c|}
      \hline
      &&\\[-1.3ex]
      & $r_0^4\,\langle \frac{\alpha}{\pi}G\,G \rangle$ &  $\langle \frac{\alpha}{\pi}G\,G \rangle$ [GeV$^4$]\\[0.5ex]
      \hline
      &&\\[-1.3ex]
      $\Delta_{W_{11}}$                           & $1.16(12)$       & $0.028(3)$ \\[0.5ex]
      \hline
    \end{tabular}
  \end{center}
  \caption{Gluon condensate at $L=12$  ($r_0=0.5$ fm).}
  \label{tab:GG}
\end{table}
This value is slightly lower than the value $0.04(1)$~GeV$^4$ found 
in~\cite{Rakow:2005yn}. The main reason for the difference is that 
in~\cite{Rakow:2005yn} 
the boosted series was truncated at $n^\star = 12$, 
while in the present work we make an estimate of the contribution from higher 
terms in the boosted series.

\section{Summary}
\label{sec:summary}

In this paper we presented the result of NSPT calculations for Wilson loops 
of various sizes using the Wilson gauge action. Within the framework of NSPT we were
able to determine the perturbative coefficients of those loops up to loop order $n=20$ 
for different lattice sizes as numerically clear signals.

Up to that order we did not observe signs of a factorial $n$-dependence as expected for an asymptotic series.
Assuming that this behavior is not spoiled at larger $n$, we were able to describe the $n$ dependence of the series
by a simple recursion relating subsequent orders. Solving that relation, the sum over all orders has been represented
by a hypergeometric function. Its branch cut discontinuity defines a convergence radius of the series at positive $g^2$.

Using the naive perturbative series of the Wilson loops in the bare coupling squared $g^2=6/\beta$, the summed series 
up to $n^\star$ converges only slowly to some asymptotic value. 
This has led us to apply boosting -- a rearrangement of the perturbative series in terms of
the so-called boosted coupling as expansion parameter
where we expect that the summed series reaches a stable plateau already after moderate loop orders. 
For moderate Wilson loop sizes these plateaus have been found.
 
The transformation from the naive perturbative series to the boosted series
is numerically delicate, involving large cancellations. Simply transforming
the NSPT raw expansion coefficients leads to very noisy boosted coefficients beyond $n\approx 8$. 
To get around this problem we ``smoothed" the coefficients of the naive 
perturbative series using the presented hypergeometric model before calculating the boosted series. The resulting
``smoothed'' boosted coefficients are much more stable, and 
this strongly suggests that 
the observed rapid fall-off of the  boosted coefficients continues to large loop orders. 

We introduced ratios of powers of Wilson loops which 
then have been treated within boosted perturbation theory.
In many cases the truncation errors for these ratios are much smaller 
than the truncation errors for the Wilson loops themselves. 

The results of the boosted perturbative series are extremely close to the 
Monte Carlo values of the Wilson loops, 
the same applies to their ratios. For $\beta > 6$ 
($g^2 < 1$) the differences are typically in the third or fourth decimal
place. Looking at the small deviations 
between Monte Carlo results and boosted perturbation theory
allows for a determination of the non-perturbative 
parts of Wilson loops. 
We find that the dominant behavior of the non-perturbative
part scales like $a^4$. 

As a special case we have calculated the gluon condensate $\langle \frac{\alpha}{\pi}G\,G \rangle$
from the plaquette. The found number is somewhat larger than that
in the phenomenological SVZ sum rule approach~\cite{Shifman:1978bx} --
at least for our $12^4$ lattice.
Our number agrees within errors with the estimate 
$\langle \frac{\alpha}{\pi} GG\rangle= 0.024(8)$ GeV$^4$ presented by Narison in~\cite{Narison:1995tw} 
which is based on a study of heavy quarkonia mass splittings.

We have checked the regularly reappearing claim, that the Wilson
loop has, in addition to its
``canonical'' $a^4$ dependence, a significant part showing a $a^2$
power dependence.
Our results show that in the chosen
$\beta$--region the  non-perturbative parts of the Wilson loops $W_{NM}$  can be well
described by an $a^4$-ansatz with an $(a^4)^2$ correction term. 
For the difference between the perturbative and the lattice Monte Carlo plaquette 
$\Delta_{W_{11}}$ this correction is rather small.

If infinite or large order perturbation theory was to reflect the long 
distance properties of QCD, we would expect the Wilson loops to show 
an area-law behavior and the static potential to grow linearly with 
distance. As a result, the Borel transform would exhibit a pole
at $1/b_0=16 \, \pi^2/11$,
and the coefficients of the perturbative series should show a factorial
growth. 
(Then, for comparison, the gluon condensate would show up as a pole at $2/b_0 = 32 \, \pi^2/11$.)
Instead, we find
\begin{equation}
  W(R,T) \propto \frac{T}{R}
  \label{WL0}
\end{equation}
for $R=2,3,4$ and $T=5$, within a few per cent, and no sign of an infrared 
renormalon\footnote{We have nothing to add to~\cite{Bauer:2011ws} and to the argument of~\cite{Luke:1994xd}
that there is no physical significance to these ambiguities.}.
This result holds for all couplings within the radius of convergence of 
the perturbative series, $0  < g^2 \lesssim 1.1$.

\begin{figure}[!htb]
  \begin{center}
        \includegraphics[scale=0.60,clip=true]{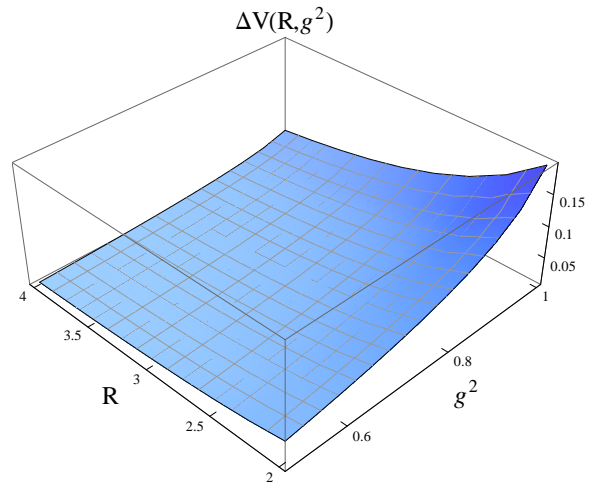}
  \end{center}
  \caption{The perturbative potential difference $\Delta V$ obtained from the perturbative Wilson 
           loops up to loop order 20 as function of
           the distance $R$ and $g^2$.} 
  \label{fig:VLG}
\end{figure}

In Figure~\ref{fig:VLG}
we show the potential difference $\Delta V$ as function of $R$ and $g^2$ calculated from 
the series variant of the Creutz ratio
\bea
  \Delta V(R) &=& V(R-1)-V(R) 
  \nonumber \\
   &=& \log \frac{W(R,T)\, W(R-1,T-1)}{W(R,T-1) \,W(R-1,T)} 
\eea
using the perturbative Wilson loops up to loop order 20.
For a linearly increasing potential one would expect
$\Delta V$ to be a constant proportional to the string tension. 
In fact,  $\Delta V$ decreases with $R$ for all $g^2$ within the radius of convergence consistent 
with the expected Coulomb behavior $1/(R(R-1))$.

A look at the $\beta$ function suggests, furthermore, that the
perturbative theory is separated from the strong coupling phase  
through a pole, similar to the supersymmetric Yang-Mills 
theory~\cite{Kogan:1995hn}, indicating that there is no 
direct contradiction with the strong coupling expansion. 
A similar result to (\ref{WL0}) was
found in Monte Carlo simulations of gauge-fixed non-compact 
lattice QCD~\cite{Patrascioiu:1981vq, Seiler:1983mz},
which as well take into account small fluctuations of the gauge fields only.

This leads us to conclude - on the basis of our present results, 
{\it nota bene} - that the
perturbative series carry no information on the confining properties 
of the theory and the non-trivial features of the QCD vacuum. 
The positive aspect of this result is that the 
perturbative tail can be cleanly separated from the Monte Carlo 
results for the plaquette.

\section*{Acknowledgements}
\vspace{-1mm}
This investigation has been supported partly by DFG under contract SCHI 422/8-1
and by the EU grant 227431 (Hadron Physics2).
R.M. is supported by the Research Executive Agency (REA) of the European Union 
under Grant Agreement PITN-GA-2009-238353 (ITN STRONGnet).
We thank the RCNP at Osaka university for providing computer resources.

\renewcommand{\thesection}{\Alpha{section}}
\setcounter{section}{0}
\section*{Appendix}

We present in Tables~\ref{tabWCL4} - \ref{tabWCL122} all considered rectangular perturbative Wilson loops of 
sizes $N\!\times\!M$ with $N,M=1,\dots,L/2$ for different sizes $L$ of the used hypercubic lattices $L^4$ in the form
\begin{equation}
     W_{NM}= 1 + \sum_{n=1}^{20} W_{NM}^{(n)} \, g^{2n} \,.
\end{equation}
The expansion coefficients $W_{NM}^{(n)}$ are the result of the extrapolation to zero Langevin step size using (\ref{eq:fitWNMeps}).
The reported errors are the fit errors from the extrapolation $\varepsilon\to0$.
The presented numbers for larger Wilson loops and higher loop orders are collected irrespective of possible problems with 
the signal to noise ratio at a given order $n$ as discussed in Section \ref{subsec:nsptres} and have to be taken with care.
In Table \ref{tabhypinf} we give  some perturbative Wilson loops as result of an infinite series
using the described hypergeometric model for various $\beta$ values at $L=12$.
In Table \ref{tab:c1c2LInf} we collect the values for known loop order coefficients in the infinite volume limit.
For $W_{11}$ the first three loop order coefficients are given in~\cite{Alles:1998is,Athenodorou:2004xt}
whereas for the larger Wilson loops only the first two loop
orders are known~\cite{Wohlert:1984hk}.
The first order coefficients can be computed to high precision.

\renewcommand{\thetable}{A\arabic{table}}
\setcounter{table}{0}
\begin{table*}[!htb]
{\scriptsize
\begin{center}
\vspace{0.5cm}
  \begin{tabular}{||c|r|r|r|r||}
\hline
n & $W_{11}^{(n)}$ & $W_{21}^{(n)}$ &$W_{22}^{(n)}$ \\ 
\hline
$1 $&$  -0.332147(22) $&$  -0.567064(34) $&$  -0.874683(122)$\\[0.7ex]
$2 $&$  -0.033411(15) $&$  -0.004571(25) $&$  0.104041(63)$\\[0.7ex]
$3 $&$  -0.013368(13) $&$  -0.010094(28) $&$  -0.000735(58)$\\[0.7ex]
$4 $&$  -0.006983(1) $&$  -0.006394(13) $&$  -0.002683(12)$\\[0.7ex]
$5 $&$  -0.004179(8) $&$  -0.004167(9) $&$  -0.002284(1)$\\[0.7ex]
$6 $&$  -0.002719(6) $&$  -0.002859(8) $&$  -0.001777(9)$\\[0.7ex]
$7 $&$  -0.001872(6) $&$  -0.002041(8) $&$  -0.001368(1)$\\[0.7ex]
$8 $&$  -0.001342(5) $&$  -0.001503(8) $&$  -0.001063(9)$\\[0.7ex]
$9 $&$  -0.000992(5) $&$  -0.001134(7) $&$  -0.000834(8)$\\[0.7ex]
$10 $&$  -0.000752(4) $&$  -0.000874(6) $&$  -0.000663(7)$\\[0.7ex]
$11 $&$  -0.000581(4) $&$  -0.000684(5) $&$  -0.000534(6)$\\[0.7ex]
$12 $&$  -0.000456(4) $&$  -0.000544(5) $&$  -0.000433(6)$\\[0.7ex]
$13 $&$  -0.000363(3) $&$  -0.000437(4) $&$  -0.000355(6)$\\[0.7ex]
$14 $&$  -0.000292(3) $&$  -0.000355(4) $&$  -0.000293(6)$\\[0.7ex]
$15 $&$  -0.000238(3) $&$  -0.000291(4) $&$  -0.000243(5)$\\[0.7ex]
$16 $&$  -0.000195(2) $&$  -0.000240(4) $&$  -0.000204(5)$\\[0.7ex]
$17 $&$  -0.000161(2) $&$  -0.000200(3) $&$  -0.000171(5)$\\[0.7ex]
$18 $&$  -0.000134(2) $&$  -0.000167(3) $&$  -0.000145(5)$\\[0.7ex]
$19 $&$  -0.000112(2) $&$  -0.000141(3) $&$  -0.000123(4)$\\[0.7ex]
$20 $&$  -0.000094(2) $&$  -0.000119(3) $&$  -0.000105(4)$\\
\hline
\end{tabular}
 \end{center}
 }
  \caption{Perturbative coefficients for $L=4$.}
\label{tabWCL4}
\end{table*}
\begin{table*}[!htb]
{\scriptsize
\begin{center}
\vspace{0.5cm}
  \begin{tabular}{||c|r|r|r|r|r|r||}
\hline
n & $W_{11}^{(n)}$ & $W_{21}^{(n)}$ &$W_{22}^{(n)}$ &$W_{31}^{(n)}$ &$W_{32}^{(n)}$&$W_{33}^{(n)}$\\ 
\hline
$1 $&$  -0.333112(15) $&$  -0.573644(44) $&$  -0.907518(112) $&$  -0.798086(71) $&$  -1.193307(174) $&$  -1.500876(291)$\\[0.7ex]
$2 $&$  -0.033829(6) $&$  -0.003938(26) $&$  0.118297(96) $&$  0.075949(52) $&$  0.313019(171) $&$  0.609060(335)$\\[0.7ex]
$3 $&$  -0.013641(4) $&$  -0.010199(9) $&$  0.000024(16) $&$  -0.002820(12) $&$  -0.005402(37) $&$  -0.050954(108)$\\[0.7ex]
$4 $&$  -0.007202(2) $&$  -0.006571(5) $&$  -0.002375(9) $&$  -0.003622(3) $&$  -0.000139(18) $&$  -0.000273(38)$\\[0.7ex]
$5 $&$  -0.004366(3) $&$  -0.004366(6) $&$  -0.002227(12) $&$  -0.002878(6) $&$  -0.000573(8) $&$  -0.000083(34)$\\[0.7ex]
$6 $&$  -0.002881(4) $&$  -0.003047(7) $&$  -0.001813(12) $&$  -0.002190(8) $&$  -0.000684(1) $&$  -0.000138(7)$\\[0.7ex]
$7 $&$  -0.002014(4) $&$  -0.002214(7) $&$  -0.001440(12) $&$  -0.001675(8) $&$  -0.000629(13) $&$  -0.000147(12)$\\[0.7ex]
$8 $&$  -0.001467(4) $&$  -0.001661(7) $&$  -0.001151(12) $&$  -0.001303(8) $&$  -0.000551(13) $&$  -0.000156(1)$\\[0.7ex]
$9 $&$  -0.001103(4) $&$  -0.001278(6) $&$  -0.000927(1) $&$  -0.001028(7) $&$  -0.000473(1) $&$  -0.000157(8)$\\[0.7ex]
$10 $&$  -0.000850(3) $&$  -0.001004(5) $&$  -0.000755(8) $&$  -0.000824(6) $&$  -0.000404(8) $&$  -0.000150(7)$\\[0.7ex]
$11 $&$  -0.000669(3) $&$  -0.000802(4) $&$  -0.000622(6) $&$  -0.000670(5) $&$  -0.000346(6) $&$  -0.000139(6)$\\[0.7ex]
$12 $&$  -0.000535(3) $&$  -0.000650(3) $&$  -0.000518(4) $&$  -0.000551(4) $&$  -0.000298(4) $&$  -0.000126(5)$\\[0.7ex]
$13 $&$  -0.000434(2) $&$  -0.000533(3) $&$  -0.000435(3) $&$  -0.000458(3) $&$  -0.000257(2) $&$  -0.000114(5)$\\[0.7ex]
$14 $&$  -0.000356(2) $&$  -0.000442(2) $&$  -0.000368(2) $&$  -0.000384(2) $&$  -0.000222(2) $&$  -0.000102(4)$\\[0.7ex]
$15 $&$  -0.000295(2) $&$  -0.000370(2) $&$  -0.000313(2) $&$  -0.000324(2) $&$  -0.000192(2) $&$  -0.000091(4)$\\[0.7ex]
$16 $&$  -0.000247(2) $&$  -0.000312(2) $&$  -0.000268(2) $&$  -0.000276(2) $&$  -0.000167(2) $&$  -0.000080(3)$\\[0.7ex]
$17 $&$  -0.000208(2) $&$  -0.000265(2) $&$  -0.000231(2) $&$  -0.000236(2) $&$  -0.000145(2) $&$  -0.000071(3)$\\[0.7ex]
$18 $&$  -0.000177(2) $&$  -0.000227(2) $&$  -0.000200(2) $&$  -0.000203(2) $&$  -0.000127(2) $&$  -0.000063(2)$\\[0.7ex]
$19 $&$  -0.000151(2) $&$  -0.000195(2) $&$  -0.000173(2) $&$  -0.000175(2) $&$  -0.000111(2) $&$  -0.000056(2)$\\[0.7ex]
$20 $&$  -0.000130(1) $&$  -0.000169(2) $&$  -0.000151(2) $&$  -0.000152(2) $&$  -0.000098(2) $&$  -0.000050(2)$\\
\hline
\end{tabular}
 \end{center}
 }
  \caption{Perturbative coefficients for $L=6$.}
\label{tabWCL6}
\end{table*}
\begin{table*}[!htb]
{\scriptsize
\begin{center}
  \begin{tabular}{||c|r|r|r|r|r||}
\hline
n & $W_{11}^{(n)}$ & $W_{21}^{(n)}$ &$W_{22}^{(n)}$ &$W_{31}^{(n)}$ &$W_{41}^{(n)}$ \\ 
\hline
$1 $&$  -0.333236(8) $&$  -0.574473(16) $&$  -0.911469(27) $&$  -0.800665(29) $&$  -1.023410(49)$\\[0.7ex]
$2 $&$  -0.033852(5) $&$  -0.003818(8) $&$  0.119976(19) $&$  0.076987(16) $&$  0.206839(26)$\\[0.7ex]
$3 $&$  -0.013670(3) $&$  -0.010214(4) $&$  0.000196(7) $&$  -0.002770(7) $&$  -0.002536(14)$\\[0.7ex]
$4 $&$  -0.007229(3) $&$  -0.006594(4) $&$  -0.002321(9) $&$  -0.003628(5) $&$  -0.001501(7)$\\[0.7ex]
$5 $&$  -0.004389(2) $&$  -0.004397(4) $&$  -0.002243(5) $&$  -0.002892(6) $&$  -0.001525(6)$\\[0.7ex]
$6 $&$  -0.002903(2) $&$  -0.003080(3) $&$  -0.001845(5) $&$  -0.002209(5) $&$  -0.001309(6)$\\[0.7ex]
$7 $&$  -0.002034(2) $&$  -0.002246(2) $&$  -0.001478(5) $&$  -0.001697(3) $&$  -0.001076(4)$\\[0.7ex]
$8 $&$  -0.001487(1) $&$  -0.001693(2) $&$  -0.001194(6) $&$  -0.001328(3) $&$  -0.000880(3)$\\[0.7ex]
$9 $&$  -0.001122(1) $&$  -0.001310(2) $&$  -0.000973(7) $&$  -0.001057(3) $&$  -0.000725(3)$\\[0.7ex]
$10 $&$  -0.000869(1) $&$  -0.001035(3) $&$  -0.000800(7) $&$  -0.000854(3) $&$  -0.000601(3)$\\[0.7ex]
$11 $&$  -0.000687(1) $&$  -0.000832(3) $&$  -0.000664(6) $&$  -0.000700(3) $&$  -0.000502(3)$\\[0.7ex]
$12 $&$  -0.000553(1) $&$  -0.000678(3) $&$  -0.000555(6) $&$  -0.000579(3) $&$  -0.000423(3)$\\[0.7ex]
$13 $&$  -0.000451(2) $&$  -0.000560(3) $&$  -0.000468(5) $&$  -0.000484(3) $&$  -0.000358(3)$\\[0.7ex]
$14 $&$  -0.000372(2) $&$  -0.000467(3) $&$  -0.000398(5) $&$  -0.000408(3) $&$  -0.000306(3)$\\[0.7ex]
$15 $&$  -0.000310(2) $&$  -0.000393(3) $&$  -0.000340(5) $&$  -0.000346(3) $&$  -0.000262(3)$\\[0.7ex]
$16 $&$  -0.000261(2) $&$  -0.000333(3) $&$  -0.000292(5) $&$  -0.000296(3) $&$  -0.000226(3)$\\[0.7ex]
$17 $&$  -0.000221(2) $&$  -0.000284(3) $&$  -0.000252(4) $&$  -0.000254(3) $&$  -0.000195(3)$\\[0.7ex]
$18 $&$  -0.000189(1) $&$  -0.000244(2) $&$  -0.000219(4) $&$  -0.000220(3) $&$  -0.000170(3)$\\[0.7ex]
$19 $&$  -0.000162(1) $&$  -0.000211(2) $&$  -0.000191(4) $&$  -0.000191(3) $&$  -0.000148(3)$\\[0.7ex]
$20 $&$  -0.000140(1) $&$  -0.000183(2) $&$  -0.000167(3) $&$  -0.000167(2) $&$  -0.000130(2)$\\
\hline
\end{tabular}
 \end{center}
 }
  \caption{Perturbative coefficients for $L=8$.}
\label{tabWCL8}
\end{table*}
\begin{table*}[!htb]
{\scriptsize
\begin{center}
  \begin{tabular}{||c|r|r|r|r|r||}
\hline
n & $W_{32}^{(n)}$ & $W_{33}^{(n)}$ &$W_{42}^{(n)}$ &$W_{43}^{(n)}$ &$W_{44}^{(n)}$ \\ 
\hline
$1 $&$  -1.204201(52) $&$  -1.528486(114) $&$  -1.485430(97) $&$  -1.830535(174) $&$  -2.140917(228)$\\[0.7ex]
$2 $&$  0.320661(27) $&$  0.636544(74) $&$  0.595785(62) $&$  1.028662(168) $&$  1.524356(276)$\\[0.7ex]
$3 $&$  -0.005468(14) $&$  -0.055098(20) $&$  -0.048959(22) $&$  -0.174438(58) $&$  -0.396169(154)$\\[0.7ex]
$4 $&$  -0.000135(19) $&$  -0.000547(45) $&$  -0.000495(31) $&$  0.002744(73) $&$  0.025146(116)$\\[0.7ex]
$5 $&$  -0.000592(1) $&$  -0.000131(15) $&$  -0.000219(24) $&$  -0.000108(45) $&$  0.000200(94)$\\[0.7ex]
$6 $&$  -0.000687(12) $&$  -0.000159(28) $&$  -0.000216(20) $&$  -0.000068(44) $&$  0.000011(96)$\\[0.7ex]
$7 $&$  -0.000652(1) $&$  -0.000208(13) $&$  -0.000238(13) $&$  -0.000082(17) $&$  -0.000067(33)$\\[0.7ex]
$8 $&$  -0.000588(1) $&$  -0.000214(17) $&$  -0.000246(11) $&$  -0.000089(14) $&$  -0.000057(28)$\\[0.7ex]
$9 $&$  -0.000514(1) $&$  -0.000196(16) $&$  -0.000233(11) $&$  -0.000077(15) $&$  -0.000027(26)$\\[0.7ex]
$10 $&$  -0.000443(9) $&$  -0.000174(13) $&$  -0.000209(9) $&$  -0.000063(12) $&$  -0.000011(14)$\\[0.7ex]
$11 $&$  -0.000380(7) $&$  -0.000153(1) $&$  -0.000183(7) $&$  -0.000054(8) $&$  -0.000010(7)$\\[0.7ex]
$12 $&$  -0.000326(6) $&$  -0.000136(8) $&$  -0.000160(6) $&$  -0.000048(7) $&$  -0.000013(5)$\\[0.7ex]
$13 $&$  -0.000281(6) $&$  -0.000120(7) $&$  -0.000141(5) $&$  -0.000043(6) $&$  -0.000015(4)$\\[0.7ex]
$14 $&$  -0.000243(5) $&$  -0.000107(7) $&$  -0.000124(5) $&$  -0.000039(5) $&$  -0.000015(4)$\\[0.7ex]
$15 $&$  -0.000210(5) $&$  -0.000095(6) $&$  -0.000109(5) $&$  -0.000035(5) $&$  -0.000013(4)$\\[0.7ex]
$16 $&$  -0.000183(5) $&$  -0.000084(5) $&$  -0.000096(5) $&$  -0.000032(5) $&$  -0.000011(4)$\\[0.7ex]
$17 $&$  -0.000160(5) $&$  -0.000075(5) $&$  -0.000086(5) $&$  -0.000029(4) $&$  -0.000010(4)$\\[0.7ex]
$18 $&$  -0.000141(4) $&$  -0.000068(4) $&$  -0.000076(4) $&$  -0.000027(4) $&$  -0.000009(3)$\\[0.7ex]
$19 $&$  -0.000124(4) $&$  -0.000061(4) $&$  -0.000068(4) $&$  -0.000025(3) $&$  -0.000008(3)$\\[0.7ex]
$20 $&$  -0.000110(4) $&$  -0.000055(4) $&$  -0.000061(4) $&$  -0.000023(3) $&$  -0.000007(2)$\\
\hline
\end{tabular}
 \end{center}
 }
  \caption{Perturbative coefficients for $L=8$ (continued).}
\label{tabWCL81}
\end{table*}
\begin{table*}[!htb]
{\scriptsize
\begin{center}
  \begin{tabular}{||c|r|r|r|r|r|r|r||}
\hline
n & $W_{11}^{(n)}$ & $W_{21}^{(n)}$ &$W_{22}^{(n)}$ &$W_{31}^{(n)}$ &$W_{41}^{(n)}$ &$W_{51}^{(n)}$  &$W_{61}^{(n)}$\\ 
\hline
$1 $&$  -0.333320(4) $&$  -0.574758(4) $&$  -0.912636(19) $&$  -0.801260(5) $&$  -1.024718(1) $&$  -1.247323(13) $&$  -1.469522(15)$\\[0.7ex]
$2 $&$  -0.033898(1) $&$  -0.003835(2) $&$  0.120423(2) $&$  0.077139(9) $&$  0.207624(22) $&$  0.387381(18) $&$  0.616194(13)$\\[0.7ex]
$3 $&$  -0.013698(3) $&$  -0.010247(5) $&$  0.000136(15) $&$  -0.002788(4) $&$  -0.002610(3) $&$  -0.020698(3) $&$  -0.067876(11)$\\[0.7ex]
$4 $&$  -0.007251(3) $&$  -0.006625(7) $&$  -0.002337(13) $&$  -0.003640(8) $&$  -0.001503(7) $&$  -0.000967(6) $&$  -0.000342(11)$\\[0.7ex]
$5 $&$  -0.004410(3) $&$  -0.004425(6) $&$  -0.002255(11) $&$  -0.002914(9) $&$  -0.001539(1) $&$  -0.000784(1) $&$  -0.000475(6)$\\[0.7ex]
$6 $&$  -0.002922(3) $&$  -0.003106(6) $&$  -0.001861(9) $&$  -0.002233(7) $&$  -0.001326(1) $&$  -0.000724(12) $&$  -0.000406(15)$\\[0.7ex]
$7 $&$  -0.002052(3) $&$  -0.002272(4) $&$  -0.001503(7) $&$  -0.001726(5) $&$  -0.001101(5) $&$  -0.000645(6) $&$  -0.000373(8)$\\[0.7ex]
$8 $&$  -0.001504(2) $&$  -0.001718(3) $&$  -0.001217(4) $&$  -0.001355(3) $&$  -0.000906(3) $&$  -0.000557(4) $&$  -0.000334(6)$\\[0.7ex]
$9 $&$  -0.001138(2) $&$  -0.001333(2) $&$  -0.000994(2) $&$  -0.001082(2) $&$  -0.000748(1) $&$  -0.000475(2) $&$  -0.000289(3)$\\[0.7ex]
$10 $&$  -0.000884(1) $&$  -0.001056(2) $&$  -0.000820(2) $&$  -0.000876(2) $&$  -0.000621(1) $&$  -0.000403(2) $&$  -0.000251(3)$\\[0.7ex]
$11 $&$  -0.000700(1) $&$  -0.000851(1) $&$  -0.000683(3) $&$  -0.000719(2) $&$  -0.000519(2) $&$  -0.000344(2) $&$  -0.000218(3)$\\[0.7ex]
$12 $&$  -0.000565(1) $&$  -0.000696(2) $&$  -0.000574(4) $&$  -0.000597(3) $&$  -0.000438(3) $&$  -0.000295(4) $&$  -0.000191(4)$\\[0.7ex]
$13 $&$  -0.000462(1) $&$  -0.000577(2) $&$  -0.000487(4) $&$  -0.000502(3) $&$  -0.000373(4) $&$  -0.000256(4) $&$  -0.000168(4)$\\[0.7ex]
$14 $&$  -0.000383(1) $&$  -0.000484(2) $&$  -0.000418(4) $&$  -0.000426(3) $&$  -0.000321(4) $&$  -0.000223(4) $&$  -0.000149(3)$\\[0.7ex]
$15 $&$  -0.000320(1) $&$  -0.000409(2) $&$  -0.000361(4) $&$  -0.000364(3) $&$  -0.000278(3) $&$  -0.000196(3) $&$  -0.000132(2)$\\[0.7ex]
$16 $&$  -0.000271(1) $&$  -0.000350(2) $&$  -0.000314(3) $&$  -0.000315(2) $&$  -0.000243(2) $&$  -0.000173(2) $&$  -0.000117(1)$\\[0.7ex]
$17 $&$  -0.000231(1) $&$  -0.000301(2) $&$  -0.000275(2) $&$  -0.000274(2) $&$  -0.000213(2) $&$  -0.000153(2) $&$  -0.000105(1)$\\[0.7ex]
$18 $&$  -0.000199(1) $&$  -0.000261(2) $&$  -0.000242(2) $&$  -0.000239(2) $&$  -0.000188(2) $&$  -0.000136(2) $&$  -0.000094(2)$\\[0.7ex]
$19 $&$  -0.000172(1) $&$  -0.000228(1) $&$  -0.000213(3) $&$  -0.000210(2) $&$  -0.000166(2) $&$  -0.000122(3) $&$  -0.000085(2)$\\[0.7ex]
$20 $&$  -0.000150(1) $&$  -0.000200(1) $&$  -0.000189(3) $&$  -0.000185(2) $&$  -0.000147(3) $&$  -0.000109(3) $&$  -0.000077(2)$\\
\hline
\end{tabular}
 \end{center}

 }
  \caption{Perturbative coefficients for $L=12$.}
\label{tabWCL12}
\end{table*}
\begin{table*}[!htb]
{\scriptsize
\begin{center}
  \begin{tabular}{||c|r|r|r|r|r|r|r||}
\hline
n & $W_{32}^{(n)}$ & $W_{33}^{(n)}$ &$W_{42}^{(n)}$ &$W_{43}^{(n)}$ &$W_{44}^{(n)}$ &$W_{52}^{(n)}$ &$W_{53}^{(n)}$ \\ 
\hline
$1 $&$  -1.207005(31) $&$  -1.535522(52) $&$  -1.491384(41) $&$  -1.845142(72) $&$  -2.170005(100) $&$  -2.148586(117) $&$  -0.000077(2)$\\[0.7ex]
$2 $&$  0.322694(4) $&$  0.643882(9) $&$  0.601963(23) $&$  1.048051(37) $&$  1.571598(94) $&$  1.538376(110) $&$  -0.000077(2)$\\[0.7ex]
$3 $&$  -0.005740(18) $&$  -0.056823(24) $&$  -0.050320(9) $&$  -0.181032(13) $&$  -0.418636(81) $&$  -0.404144(11) $&$  -0.000077(2)$\\[0.7ex]
$4 $&$  -0.000112(19) $&$  -0.000446(44) $&$  -0.000514(12) $&$  0.003334(36) $&$  0.028597(70) $&$  0.027039(50) $&$  -0.000077(2)$\\[0.7ex]
$5 $&$  -0.000592(11) $&$  -0.000136(15) $&$  -0.000182(2) $&$  -0.000224(12) $&$  -0.000196(13) $&$  -0.000207(18) $&$  -0.000077(2)$\\[0.7ex]
$6 $&$  -0.000685(9) $&$  -0.000113(15) $&$  -0.000197(8) $&$  -0.000059(18) $&$  -0.000019(16) $&$  0.000003(30) $&$  -0.000077(2)$\\[0.7ex]
$7 $&$  -0.000663(7) $&$  -0.000178(12) $&$  -0.000241(6) $&$  -0.000074(13) $&$  -0.000064(18) $&$  -0.000054(20) $&$  -0.000077(2)$\\[0.7ex]
$8 $&$  -0.000598(4) $&$  -0.000196(5) $&$  -0.000248(4) $&$  -0.000074(11) $&$  -0.000031(14) $&$  -0.000043(20) $&$  -0.000077(2)$\\[0.7ex]
$9 $&$  -0.000523(2) $&$  -0.000192(5) $&$  -0.000234(5) $&$  -0.000067(9) $&$  -0.000024(4) $&$  -0.000027(11) $&$  -0.000077(2)$\\[0.7ex]
$10 $&$  -0.000453(3) $&$  -0.000177(3) $&$  -0.000210(3) $&$  -0.000059(1) $&$  -0.000013(8) $&$  -0.000015(8) $&$  -0.000077(2)$\\[0.7ex]
$11 $&$  -0.000391(5) $&$  -0.000161(7) $&$  -0.000188(6) $&$  -0.000056(8) $&$  -0.000015(11) $&$  -0.000018(14) $&$  -0.000077(2)$\\[0.7ex]
$12 $&$  -0.000339(6) $&$  -0.000148(9) $&$  -0.000170(7) $&$  -0.000058(11) $&$  -0.000023(7) $&$  -0.000025(12) $&$  -0.000077(2)$\\[0.7ex]
$13 $&$  -0.000297(6) $&$  -0.000137(9) $&$  -0.000155(7) $&$  -0.000059(8) $&$  -0.000029(5) $&$  -0.000028(7) $&$  -0.000077(2)$\\[0.7ex]
$14 $&$  -0.000261(5) $&$  -0.000126(7) $&$  -0.000141(5) $&$  -0.000058(5) $&$  -0.000030(5) $&$  -0.000029(4) $&$  -0.000077(2)$\\[0.7ex]
$15 $&$  -0.000230(4) $&$  -0.000116(5) $&$  -0.000128(3) $&$  -0.000055(4) $&$  -0.000026(8) $&$  -0.000026(6) $&$  -0.000077(2)$\\[0.7ex]
$16 $&$  -0.000204(3) $&$  -0.000106(5) $&$  -0.000116(3) $&$  -0.000051(6) $&$  -0.000023(11) $&$  -0.000023(8) $&$  -0.000077(2)$\\[0.7ex]
$17 $&$  -0.000182(3) $&$  -0.000096(7) $&$  -0.000105(4) $&$  -0.000047(8) $&$  -0.000020(12) $&$  -0.000021(9) $&$  -0.000077(2)$\\[0.7ex]
$18 $&$  -0.000162(4) $&$  -0.000087(8) $&$  -0.000095(5) $&$  -0.000043(9) $&$  -0.000018(11) $&$  -0.000020(9) $&$  -0.000077(2)$\\[0.7ex]
$19 $&$  -0.000145(5) $&$  -0.000078(9) $&$  -0.000086(6) $&$  -0.000040(9) $&$  -0.000017(9) $&$  -0.000020(8) $&$  -0.000077(2)$\\[0.7ex]
$20 $&$  -0.000130(5) $&$  -0.000071(9) $&$  -0.000077(6) $&$  -0.000037(8) $&$  -0.000016(6) $&$  -0.000018(6) $&$  -0.000077(2)$\\
\hline
\end{tabular}
 \end{center}
 }
  \caption{Perturbative coefficients for $L=12$ (continued).}
\label{tabWCL121}
\end{table*}
\begin{table*}[!htb]
{\scriptsize
\begin{center}
  \begin{tabular}{||c|r|r|r|r|r|r|r||}
\hline
n & $W_{54}^{(n)}$ & $W_{55}^{(n)}$ &$W_{62}^{(n)}$ &$W_{63}^{(n)}$ &$W_{64}^{(n)}$ &$W_{65}^{(n)}$&$W_{66}^{(n)}$ \\ 
\hline
$1 $&$  -2.484945(156) $&$  -2.807271(225) $&$  -2.052506(78) $&$  -2.448879(142) $&$  -2.794661(184) $&$  -3.121967(258) $&$  -3.438947(291)$\\[0.7ex]
$2 $&$  2.181680(181) $&$  2.906812(327) $&$  1.391688(52) $&$  2.114904(186) $&$  2.879504(270) $&$  3.716609(450) $&$  4.631615(573)$\\[0.7ex]
$3 $&$  -0.790329(77) $&$  -1.342663(185) $&$  -0.341621(19) $&$  -0.751605(84) $&$  -1.322744(141) $&$  -2.086833(280) $&$  -3.069039(366)$\\[0.7ex]
$4 $&$  0.101235(36) $&$  0.260769(172) $&$  0.020658(33) $&$  0.093854(31) $&$  0.255213(42) $&$  0.548424(285) $&$  1.025063(600)$\\[0.7ex]
$5 $&$  -0.002457(18) $&$  -0.015589(148) $&$  -0.000052(25) $&$  -0.002038(23) $&$  -0.015007(49) $&$  -0.056854(283) $&$  -0.156497(515)$\\[0.7ex]
$6 $&$  0.000160(35) $&$  0.000420(19) $&$  -0.000080(18) $&$  0.000078(14) $&$  0.000253(48) $&$  0.001978(79) $&$  0.008403(118)$\\[0.7ex]
$7 $&$  -0.000078(29) $&$  -0.000175(44) $&$  -0.000076(11) $&$  -0.000051(12) $&$  -0.000048(49) $&$  -0.000388(108) $&$  -0.000301(267)$\\[0.7ex]
$8 $&$  -0.000020(16) $&$  -0.000011(8) $&$  -0.000059(11) $&$  -0.000045(23) $&$  -0.000034(33) $&$  0.000034(26) $&$  -0.000051(119)$\\[0.7ex]
$9 $&$  -0.000008(6) $&$  0.000015(11) $&$  -0.000049(9) $&$  -0.000022(1) $&$  -0.000009(25) $&$  -0.000004(50) $&$  0.000008(124)$\\[0.7ex]
$10 $&$  0.000007(12) $&$  0.000030(16) $&$  -0.000037(4) $&$  0.000003(1) $&$  0.000015(17) $&$  0.000056(16) $&$  0.000093(31)$\\[0.7ex]
$11 $&$  -0.000005(12) $&$  0.000008(1) $&$  -0.000035(3) $&$  0.000002(17) $&$  -0.000004(1) $&$  -0.000014(7) $&$  -0.000066(26)$\\[0.7ex]
$12 $&$  -0.000016(6) $&$  -0.000002(2) $&$  -0.000037(5) $&$  -0.000005(11) $&$  -0.000011(4) $&$  -0.000022(9) $&$  -0.000049(5)$\\[0.7ex]
$13 $&$  -0.000020(4) $&$  -0.000006(1) $&$  -0.000038(3) $&$  -0.000012(5) $&$  -0.000013(11) $&$  -0.000010(9) $&$  -0.000026(17)$\\[0.7ex]
$14 $&$  -0.000019(6) $&$  -0.000008(3) $&$  -0.000038(1) $&$  -0.000016(5) $&$  -0.000009(11) $&$  -0.000004(8) $&$  0.000002(22)$\\[0.7ex]
$15 $&$  -0.000015(8) $&$  -0.000007(6) $&$  -0.000036(2) $&$  -0.000016(5) $&$  -0.000004(9) $&$  -0.000001(7) $&$  0.000012(11)$\\[0.7ex]
$16 $&$  -0.000010(1) $&$  -0.000005(8) $&$  -0.000033(3) $&$  -0.000015(6) $&$  -0.000002(7) $&$  0.000000(6) $&$  0.000014(3)$\\[0.7ex]
$17 $&$  -0.000008(1) $&$  -0.000004(9) $&$  -0.000030(5) $&$  -0.000013(7) $&$  -0.000001(6) $&$  0.000000(5) $&$  0.000009(5)$\\[0.7ex]
$18 $&$  -0.000007(8) $&$  -0.000003(7) $&$  -0.000028(5) $&$  -0.000012(6) $&$  0.000000(5) $&$  -0.000001(4) $&$  0.000002(4)$\\[0.7ex]
$19 $&$  -0.000007(6) $&$  -0.000003(5) $&$  -0.000025(5) $&$  -0.000011(5) $&$  0.000000(3) $&$  -0.000002(3) $&$  -0.000002(3)$\\[0.7ex]
$20 $&$  -0.000006(4) $&$  -0.000002(3) $&$  -0.000023(4) $&$  -0.000010(4) $&$  0.000000(2) $&$  -0.000002(2) $&$  -0.000004(5)$\\
\hline
\end{tabular}
 \end{center}
 }
  \caption{Perturbative coefficients for $L=12$ (continued).}
\label{tabWCL122}
\end{table*}
\begin{table*}[!htb]
{\scriptsize
\begin{center}
  \begin{tabular}{||c|r|r|r|r||}
\hline
$\beta$ & $W_{11}^{\infty}$ & $W_{21}^{\infty}$& $W_{31}^{\infty}$ &$W_{22}^{\infty}$ \\ 
\hline
$5.85$ & $ 0.57595(14)$ & $ 0.36021(22)$ & $ 0.22936(28)$& $ 0.16659(41)$\\[0.7ex]
$5.9$ & $ 0.58254(11)$ & $ 0.36901(16)$ & $ 0.23814(21)$& $ 0.17557(29)$\\[0.7ex]
$5.95$ & $ 0.588518(92)$ & $ 0.37692(13)$ & $ 0.24602(17)$& $ 0.18354(24)$\\[0.7ex]
$6$ & $ 0.594092(80)$ & $ 0.38429(11)$ & $ 0.25337(14)$& $ 0.19095(20)$\\[0.7ex]
$6.05$ & $ 0.599358(71)$ & $ 0.39125(10)$ & $ 0.26034(12)$& $ 0.19797(17)$\\[0.7ex]
$6.1$ & $ 0.604372(63)$ & $ 0.397894(90)$ & $ 0.26702(11)$& $ 0.20469(15)$\\[0.7ex]
$6.15$ & $ 0.609172(57)$ & $ 0.404260(81)$ & $ 0.273454(99)$& $ 0.21118(14)$\\[0.7ex]
$6.2$ & $ 0.613784(52)$ & $ 0.410391(75)$ & $ 0.279676(89)$& $ 0.21745(12)$\\[0.7ex]
$6.25$ & $ 0.618228(48)$ & $ 0.416313(67)$ & $ 0.285714(81)$& $ 0.22355(11)$\\[0.7ex]
$6.3$ & $ 0.622521(44)$ & $ 0.422047(62)$ & $ 0.291587(74)$& $ 0.22949(10)$\\[0.7ex]
$6.35$ & $ 0.626675(41)$ & $ 0.427612(57)$ & $ 0.297310(68)$& $ 0.235295(94)$\\[0.7ex]
$6.4$ & $ 0.630703(38)$ & $ 0.433020(53)$ & $ 0.302895(63)$& $ 0.240961(87)$\\[0.7ex]
$6.45$ & $ 0.634612(35)$ & $ 0.438283(49)$ & $ 0.308354(58)$& $ 0.246508(80)$\\[0.7ex]
$6.5$ & $ 0.638412(33)$ & $ 0.443412(45)$ & $ 0.313694(54)$& $ 0.251941(75)$\\[0.7ex]
$6.55$ & $ 0.642108(31)$ & $ 0.448415(44)$ & $ 0.318922(50)$& $ 0.257270(70)$\\[0.7ex]
$6.6$ & $ 0.645708(29)$ & $ 0.453299(40)$ & $ 0.324046(47)$& $ 0.262499(65)$\\[0.7ex]
$6.65$ & $ 0.649216(27)$ & $ 0.458071(37)$ & $ 0.329071(44)$& $ 0.267634(61)$\\[0.7ex]
$6.7$ & $ 0.652637(25)$ & $ 0.462737(35)$ & $ 0.334001(41)$& $ 0.272680(57)$\\[0.7ex]
$6.75$ & $ 0.655977(23)$ & $ 0.467302(33)$ & $ 0.338842(39)$& $ 0.277641(54)$\\[0.7ex]
$6.8$ & $ 0.659239(22)$ & $ 0.471771(31)$ & $ 0.343596(36)$& $ 0.282521(50)$\\[0.7ex]
\hline
\end{tabular}
 \end{center}
 }
  \caption{Summed series of perturbative Wilson loops at $L=12$ using the described hypergeometric model
           as function of $\beta$.}
\label{tabhypinf}
\end{table*}
\begin{table}[!htb]
  {\scriptsize
  \begin{center}
    \begin{tabular}{|l |c |c |c|}
    \hline
     &&&     \\[-1.3ex]
    $W_{NM}$  & $W_{NM,\infty}^{(1)}$ &  $W_{NM,\infty}^{(2)}$&  $W_{NM,\infty}^{(3)}$\\[0.5ex]
    \hline
    &&&\\[-1.3ex]
    $W_{11}$   \cite{Alles:1998is,Athenodorou:2004xt}& $-1/3$           & $-0.0339109931(3)$ & $-0.0137063(2)$\\[0.5ex]
    $W_{21}$   \cite{Wohlert:1984hk}                 & $-0.57483367 $   & $-0.003857(17)$    &\\[0.5ex]
    $W_{31}$   \cite{Wohlert:1984hk}                 & $-0.80146372 $   & $0.07717(5) $      & \\[0.5ex]
    $W_{22}$   \cite{Wohlert:1984hk}                 & $-0.91287436 $   & $0.12040(7) $      & \\[0.5ex]
    \hline 
    \end{tabular}
  \end{center}
  }
  \caption{Coefficients of lowest loop orders in the infinite volume limit.}
  \label{tab:c1c2LInf}
\end{table}

\cleardoublepage

\end{document}